\documentclass[twocolumn,trackchanges]{aastex62}
\usepackage{hyperref}

\graphicspath{{./}{figures/}}

\received{October 29, 2019}
\revised{May 17, 2020}
\accepted{May 17, 2020}

\shorttitle{Time-delay determination of MgII line for the quasar HE 0413-4031}
\shortauthors{Zaja\v{c}ek et al.}

\newcommand{\angstrom}{\mbox{\normalfont\AA}}

\begin{document}

\title{Time-delay measurement of MgII broad line response for the highly-accreting quasar HE 0413-4031: Implications for the MgII-based radius-luminosity relation}

\correspondingauthor{Michal Zaja\v{c}ek}
\email{zajacek@cft.edu.pl}

\author[0000-0001-6450-1187]{Michal Zaja\v{c}ek}
\affiliation{Center for Theoretical Physics, Polish Academy of Sciences, Al. Lotnik\'ow 32/46, 02-668 Warsaw, Poland}

\author[0000-0001-5848-4333]{Bo\.{z}ena Czerny}
\affiliation{Center for Theoretical Physics, Polish Academy of Sciences, Al. Lotnik\'ow 32/46, 02-668 Warsaw, Poland}

\author[0000-0002-7843-7689]{Mary Loli Martinez--Aldama}
\affiliation{Center for Theoretical Physics, Polish Academy of Sciences, Al. Lotnik\'ow 32/46, 02-668 Warsaw, Poland}

\author[0000-0002-0297-3346]{Mateusz Ra\l{}owski}
\affiliation{Center for Theoretical Physics, Polish Academy of Sciences, Al. Lotnik\'ow 32/46, 02-668 Warsaw, Poland}

\author{Aleksandra Olejak}
\affiliation{Center for Theoretical Physics, Polish Academy of Sciences, Al. Lotnik\'ow 32/46, 02-668 Warsaw, Poland}

\author[0000-0002-5854-7426]{Swayamtrupta Panda}
\affiliation{Center for Theoretical Physics, Polish Academy of Sciences, Al. Lotnik\'ow 32/46, 02-668 Warsaw, Poland}
\affiliation{Nicolaus Copernicus Astronomical Center, Polish Academy of Sciences, ul. Bartycka 18, 00-716 Warsaw, Poland}

\author[0000-0002-2005-9136]{Krzysztof Hryniewicz}
\affiliation{Nicolaus Copernicus Astronomical Center, Polish Academy of Sciences, ul. Bartycka 18, 00-716 Warsaw, Poland}

\author[0000-0003-2656-6726]{Marzena \'Sniegowska}
\affiliation{Center for Theoretical Physics, Polish Academy of Sciences, Al. Lotnik\'ow 32/46, 02-668 Warsaw, Poland}

\author[0000-0002-7604-9594]{Mohammad-Hassan Naddaf}
\affiliation{Center for Theoretical Physics, Polish Academy of Sciences, Al. Lotnik\'ow 32/46, 02-668 Warsaw, Poland}

\author{Wojtek Pych}
\affiliation{Nicolaus Copernicus Astronomical Center, Polish Academy of Sciences, ul. Bartycka 18, 00-716 Warsaw, Poland}

\author[0000-0002-9443-4138]{Grzegorz Pietrzy\'nski}
\affiliation{Nicolaus Copernicus Astronomical Center, Polish Academy of Sciences, ul. Bartycka 18, 00-716 Warsaw, Poland}

\author[0000-0001-9704-690X]{C. Sobrino Figaredo}
\affiliation{Astronomisches Institut - Ruhr Universitaet Bochum, Germany}

\author[0000-0002-5918-8656]{Martin Haas}
\affiliation{Astronomisches Institut - Ruhr Universitaet Bochum, Germany}

\author{Justyna \'Sredzi\'nska}
\affiliation{Nicolaus Copernicus Astronomical Center, Polish Academy of Sciences, ul. Bartycka 18, 00-716 Warsaw, Poland}

\author{Magdalena Krupa}
\affiliation{Astronomical Observatory of the Jagiellonian University, Orla 171, 30-244 Cracow, Poland}

\author{Agnieszka Kurcz}
\affiliation{Astronomical Observatory of the Jagiellonian University, Orla 171, 30-244 Cracow, Poland}
 
\author[0000-0001-5207-5619]{Andrzej Udalski}
\affiliation{Astronomical Observatory, University of Warsaw, Al. Ujazdowskie 4,  00-478 Warsaw, Poland}

\author[0000-0002-3125-9088]{Marek Gorski}
\affiliation{Departamento de Astronomia, Universidad de Concepcion, Casilla 160-C, Chile}

\author[0000-0003-4745-3923]{Marek Sarna}
\affiliation{Nicolaus Copernicus Astronomical Center, Polish Academy of Sciences, ul. Bartycka 18, 00-716 Warsaw, Poland}



\begin{abstract}

We present the monitoring of the AGN continuum and MgII broad line emission for the quasar HE 0413-4031 ($z=1.38$) based on the six-year monitoring by the South African Large Telescope (SALT). We managed to estimate a time-delay of $302.6^{+28.7}_{-33.1}$ days in the rest frame of the source using seven different methods: interpolated cross-correlation function (ICCF), discrete correlation function (DCF), $z$-transformed DCF, JAVELIN, two estimators of data regularity (Von Neumann, Bartels), and $\chi^2$ method. This time-delay is below the value expected from the standard radius-luminosity relation. However, based on the monochromatic luminosity of the source and the SED modelling, we interpret this departure as the shortening of the time-delay due to the higher accretion rate of the source, with the inferred Eddington ratio of $\sim 0.4$. The MgII line luminosity of HE 0413-4031 responds to the continuum variability as $L_{\rm line}\propto L_{\rm cont}^{0.43\pm 0.10}$, which is consistent with the light-travel distance of the location of MgII emission at $R_{\rm out} \sim 10^{18}\,{\rm cm}$. Using the data of 10 other quasars, we confirm the radius-luminosity relation for broad MgII line, which was previously determined for broad H$\beta$ line for lower-redshift sources. In addition, we detect a general departure of higher-accreting quasars from this relation in analogy to H$\beta$ sample. After the accretion-rate correction of the light-travel distance, the MgII-based radius-luminosity relation has a small scatter of only $0.10$ dex.  

\end{abstract}



\section{Introduction} \label{sec:intro}

Reverberation mapping of Active Galactic Nuclei (AGN) is a leading method to study the spatial scale as well as the structure of the Broad Line Region \citep[hereafter BLR; see][]{1982ApJ...255..419B,2004AN....325..248P,2009NewAR..53..140G,2019arXiv190800742C}.
The method is very time consuming since it requires tens or even hundreds of spectra, covering well the characteristic timescales in a given object. Collected data allow for the measurement of the time delay of a chosen emission line with respect to the continuum. Assuming light travel time of light propagation, we thus obtain the characteristic size of the BLR.
Subsequent discovery of the relation between the size of the BLR and source absolute monochromatic luminosity \citep{kaspi2000,peterson2004,bentz2013} opened a way to measure black hole masses in large quasar samples using just a single-epoch spectrum \citep[e.g.][]{collin2006,vestergaard2006,shen2011,mejia2018}.

The radius - luminosity relation based on the monitoring of broad H$\beta$ component is relatively well studied in the case of lower redshift sources, including nearby quasars (below $\sim 0.9$; \citealt{kaspi2000,bentz2013,grier2017}). For larger redshifts, H$\beta$ moves to the infrared bands, and in the optical band the spectrum is dominated by UV emission lines. Sources with redshifts in the range of $\sim 0.4$ to 1.5, when the spectrum is observed in the optical band, become dominated by MgII line, and at higher redshifts CIV moves into the optical band. Thus, the single-spectrum methods are used to scale the H$\beta$ and the other line properties (mostly systematic differences in the line widths) to be able to cover a large spectral range. Direct reverberation measurements in other lines than H$\beta$ are still rare. In the current paper, we show a new reverberation measurement done in MgII line for the redshift larger than one.  

MgII line seems to be suitable for the black hole mass measurements since together with H$\beta$ it belongs to Low Ionization Lines \citep{collin1988}, and thus should originate close to the accretion disk where the motion of the emitting material is largely influenced by the potential of the central black hole. Therefore the motion of the MgII emitting material is expected to be quasi-Keplerian, i.e. the velocity field is dominantly Keplerian with a certain turbulent component. In analogy to H$\beta$ broad component, MgII line is virialized \citep{2013A&A...555A..89M}, while high ionization lines exhibit clear line profile asymmetries that imply the outflowing motion and the importance of the radiation force.

On the other hand, monitoring of MgII is more difficult since the line in many sources has very low variability amplitude \citep{goad1999,woo2008,zhu2017,guo2019} and/or the timescales at larger redshifts and for more massive (and luminous) quasars are considerably longer. Also the width of MgII line is narrower than H$\beta$ broad component, which indicates the position at larger light-travel distances \citep{2013A&A...555A..89M}. In addition, MgII line emitting gas is evidenced to respond to non-thermal radiation from jets, which may further complicate the reverberation mapping for radio-loud and $\gamma$-ray emitting sources \citep{2013ApJ...763L..36L,2020ApJ...891...68C}.

Successful determination of the line time delay has been achieved only for 10 sources so far \citep{metzroth2006,2016ApJ...818...30S,grier2017,lira2018,czerny2019}, but it nevertheless allowed for the preliminary construction of the radius-luminosity relation based on MgII line \citep{czerny2019} with the slope close to $R\propto L^{0.5}$ \citep{2019arXiv190905572P} being consistent with the relation for the H$\beta$ line \citep{bentz2013}. The key measurement towards larger luminosities came from the bright quasar CTS C30.10 (z=0.90052, $\log{[L_{3000}\,({\rm erg\,s^{-1}})]} = 46.023$) monitored for 6 years with the South African Large Telescope (SALT). The source CT252, for which the reverberation mapping was also performed in MgII line \citep{lira2018}, alongside CIII] and CIV monitoring, so far had the largest redshift of $z=1.890$ among MgII sources.

In this paper, we show the results for the quasar HE 0413-4031
also monitored with the SALT, but brighter ($\log{[L_{3000}\,({\rm erg\,s^{-1}})]}=46.741$) and located at the redshift of $z=1.389$ (according to NED\footnote{https://ned.ipac.caltech.edu/}). This quasar found as part of the Hamburg/ESO survey for bright QSOs \citep{wisotzki2000}. Apart from the quasar spectrum, the source is also radio-loud and belongs to the flat-spectrum radio quasars (FSRQ) - blazars \citep{2016ApJS..224...26M}. In fact, according to the NED database, the radio spectral slope at lower radio frequencies between $0.843\,{\rm GHz}$ and $5\,{\rm GHz}$, is inverted with $\alpha=0.68$\footnote{We use the flux-frequency convention $F_{\nu}\propto \nu^{+\alpha}$.}, which indicates a compact self-absorbed radio core. From this, we estimate the flux density at $1.4\,{\rm GHz}$, $F_{1.4}\approx 21\,{\rm mJy}$, which implies the monochromatic luminosity per frequency of $L_{1.4}\approx 2.5 \times 10^{26}\,{\rm W\,Hz^{-1}}>10^{24}\,{\rm W\,Hz^{-1}}$, based on which HE 0413-4031 can be classified as radio-loud AGN \citep{2016A&ARv..24...10T}. 

In the analysis, we determine the rest-frame time-delay of the MgII line using different statistically robust methods --  interpolated cross-correlation function (ICCF), discrete correlation function (DCF), $z$-transformed DCF, JAVELIN, two estimators of data regularity (Von Neumann, Bartels), and $\chi^2$ method. The determined rest-frame time-delay of $\sim 303$ days turns out to be smaller than the time-delay predicted from the expected radius-luminosity relation, where the radius of the BLR is proportional to the square-root of the monochromatic luminosity. Since HE 0413-4031 is a quasar with the accretion rate close to the Eddington limit, which is inferred from the detailed SED fitting, we demonstrate that the shortening of the measured time-delay is due to the accretion-rate effect in analogy to the H$\beta$-based radius-luminosity relation \citep{Mart_nez_Aldama_2019}.  

The paper is structured as follows. In Section~\ref{sec:obs}, we present the observational analysis including both spectroscopy and photometry. Subsequently, in Section~\ref{sec_results_spectroscopy}, we analyze the mean spectrum, rms spectrum, spectral fits of individual observations, and the variability properties of light curves. The focus of the paper is on the time-delay determination of the MgII broad line emission with respect to the continuum using different statistical methods, which is presented in detail in Section~\ref{sec:result}. In Section~\ref{sec_MgII_RL_relation}, we present the preliminary virial black hole mass and Eddington ratio, and using other measurements of MgII time-delay, we construct a MgII-based radius-luminosity relation and demonstrate that the departure of the sources depends on their accretion rate, which leads to the significant time-delay shortening for the highly-accreting quasar HE 0413-4031. In the discussion part in Section~\ref{sec:discussion}, we analyze the response of the MgII line with respect to the continuum variability, which is related to the intrinsic Baldwin effect, we perform the SED fitting, and we discuss the source classification along the quasar main sequence, taking into account its radio properties. Finally, we summarize the main conclusions in Section~\ref{sec:conclusions}.

\section{Observations} \label{sec:obs}

The quasar HE 0413-4031, located at redshift $z = 1.389 $ according to the NED database, is a very bright source: \citet{veron_cetty2001} report the V mag of $M_{V}=16.5$ mag. 
Its position on the sky (04h 15m 14s; -40$^{\circ}$ 23' 41") made it a very good target for the spectroscopic monitoring with the SALT. The source has been monitored since 21 Jan 2013 till August 8, 2019. The spectroscopic and photometric data are summarized in Section~\ref{sec_photometry_spectroscopy} in Tables~\ref{tab:spectr}, \ref{sec:phot1}, and \ref{sec:phot2}. 

\subsection{Spectroscopy}
\label{sec:spectr}

The quasar was observed using the Robert Stobie Spectrograph on SALT (RSS; \citealt{burgh2003,kobulnicky2003,smith2006}). A slit spectroscopy mode was used, with the slit width of 2''. Adopted medium resolution grating PG1300 and the grating angle of 28.625, with the filter PC04600, gave a configuration of a spectral resolution of 1523 at 7370 \AA. The same configuration has been used in all 25 observations, covering more than six years. A single exposure usually lasted about 820 s, and two exposures were taken during each observation. All observations were performed in service mode. The observation dates are given in Table~\ref{tab:spectr}.

The basic reduction of the raw data was done by the SALT staff by applying a semi-automatic pipeline being a part of the SALT PyRAF package. At the next stage the two images were combined with the aim to remove the cosmic rays as well as to increase the signal to noise ratio. The wavelength calibration was performed using the calibration lamp exposures taken after the source observation. In most observations argon lamp has been used. We additionally checked the calibration using the OI sky line 6863.955 \AA~ since in our observations of another quasar with SALT telescope the lamp calibration was not very accurate at early years of monitoring. However, for HE 0413-4031 the differences between the lamp calibration and the sky line position were at a level of a fraction of an Angstroem.

Due to the specific design of the SALT telescope, correcting for vignetting is an important issue. For that purpose we used an ESO standard star LTT 4364 (white dwarf, with practically no spectral features in the interesting spectral range) which was observed with SALT in the same configuration as the quasar. By analytic parametrization of the ratio of ESO and SALT spectrum of the star, we obtained a correction to the spectral shape of a quasar in the observed wavelength range from 6342 to 6969 \AA~ in the observed frame.  Formally, the part of the spectrum up to 8600 \AA~ is available, apart from two gaps, but the correction of the spectral shape by the comparison star is not satisfactory in this spectral range.
Absolute calibration of the SALT spectra was performed using the supplementary photometry.

\subsection{Photometry}
\label{sec:photometry}

Spectroscopic observations were accompanied by denser photometric monitoring. For a significant part of our campaign, high quality data were collected as part of the OGLE-IV survey done with the 1.3m Warsaw telescope at the Las Campanas Observatory, Chile. Monitoring was performed in V band, with the exposure time 240 s, and the typical error was about 0.005 mag. 

We also obtained photometric measurements from the SALT telescope at the same night as the spectroscopic observations were performed, whenever the instrument SALTICAM was available. We used the images obtained in g band, usually two exposures were made, with the exposure time 45 s. Since SALT instrument is not suitable for highly accurate photometric observations, the typical error of this photometry is of the order of 0.012 mag. Since SALTICAM data were collected in a different band than OGLE, we allowed for a grey shift of the SALTICAM set using the periods when the two monitorings overlapped.

Finally, in the period between December 3, 2017 and March 24, 2019, we also performed short denser monitoring with the 40 cm Bochum Monitoring Telescope (BMT) based at the Universitaetssternwarte Bochum, near Cerro Armazones in Chile \footnote{\url{http://www. astro.ruhr-uni-bochum.de/astro/oca}}. This monitoring was done in two bands, B and V, but for the purpose of this work only V band lightcurve has been used. This data set is not entirely consistent with the OGLE + SALTICAM data, there appears to be a slight offset by 0.171 mag, when comparing the earliest BMT point with the last OGLE point. We corrected the magnitude of all the BMT points by this offset, i.e. increasing their magnitudes by $0.171$ mag to match the first BMT point with the nearest OGLE point. For comparison, we performed time-delay measurements with this data subset included or not included in the photometric lightcurve. The photometric data points are listed in Tables~\ref{sec:phot1} and \ref{sec:phot2}.

\subsection{Spectroscopic data fitting}
\label{sec:spec_fit}

We use the same approach to the modelling of the MgII region as in \citet{czerny2019}. Because of the potential problems with the remaining vignetting effect, we concentrate only on the relatively narrow spectral band, from 2700 to 2900 \AA, in the quasar rest frame. We allow for the following components: (i) power law component of arbitrary slope and normalization, representing the continuum emission from the accretion disk (ii) FeII pseudo-continuum modeled using theoretical templates of \citet{bruhweiler2008}, folded with a Gaussian of the width representing the kinematic velocity of the FeII emitter, and (iii) MgII line itself.
We test also other FeII templates for completeness, but we discuss this issue separately, in Appendix~\ref{sec_appendix_FeII_templates}.

In our model, MgII line is parametrized in general by two separate kinematic components, each modeled assuming a Gaussian or a Lorentzian shape. The amplitudes, the width and the separation are the model parameters. Each kinematic component in turn is modeled as a doublet, and the ratio within the doublet components (varying from 1 to 2) depends on the optical depth of the emitting cloud.  

The additional parameter is the source redshift, since the determination of the redshift in NED database is not accurate enough for our data. Since we do not have an independent measure of the redshift from narrow emission lines, we assume that FeII and one of the MgII components represent the source rest frame.

All model parameters are fitted together, we do not fit first the continuum, since in the presence of the FeII pseudo-continuum there is no clear continuum-dominated region and fitting all components at the same time is more appropriate. However, we differentiate between the global parameters and the parameters when modelling individual spectra. We first created a composite spectrum by averaging all observations, and for such an average spectrum we determined the redshift, the best FeII template and the FeII smearing velocity, and the best value of the doublet ratio, and these values were later kept fixed when individual data sets were modeled.

We calculate the equivalent width (EW) of the lines with respect to the power law component, within the limits where the model was applied (i.e. integrating between 2700 and 2900 \AA). Calculation is done from the model, by numerical integration, and EW(MgII) contains both kinematic and doublet components.

The reported errors of the fit parameters, including the errors of EWs of MgII and FeII were determined by construction of the error contours, that is computations for an adopted range of the parameter of interest, with all other parameters allowed to vary. This leads in general to asymmetric errors around the best fit value. For the requested accuracy, we adopted the $\chi^2$ increase by 2.706, appropriate for one parameter of interest which represents 90\% confidence level (Statistical significance 0.1)\footnote{https://heasarc.gsfc.nasa.gov/xanadu/xspec/manual/XSerror.html}.

\subsection{Spectroscopic flux calibration and MgII absolute luminosity}

The approach to data fitting outlined in Sect.~\ref{sec:spec_fit} allows only to derive EW of the MgII and FeII lines. However, computations of time delay require the knowledge of the continuum lightcurve and the line luminosity lightcurve. 

A continuum light curve is provided by the photometric monitoring, and we use this photometry to calibrate the SALT spectra and to determine the MgII line flux. 

Since we have three instruments providing us with the photometry, and they are of a different quality, as explained in Sect.~\ref{sec:photometry}, we first perform the interpolation of the photometry datapoints at the epochs for which the EW of MgII is available using the weighted least-squares linear B-spline interpolation, using the inverse of photometry uncertainties as the corresponding weights in the spline interpolation algorithm. 

Having established the photometric flux at the time of the spectroscopic measurements still does not allow to obtain the calibrated spectrum easily. The V band does not overlap with the wavelength covered by our spectroscopy (see Sect.~\ref{sec:spectr}), and  for the redshift of our source ($z = 1.389$) corresponds to the restframe wavelength of 2304 \AA. Therefore, we have to interpolate between V band and the median of our fitting band, 2800 \AA~ rest frame. Since the measurement of the continuum slope in our narrow wavelength range is not very precise, and the slope changes between observations, the use of this slope could introduce an unnecessary scatter into the line flux calibration. Therefore we decided to use the broad band quasar continuum spectrum of \citet{zheng1997}
\footnote{Downloaded from \url{https://archive.stsci.edu/prepds/composite_quasar/}} as a template, and we assumed that the ratio between the flux at 2304 \AA~ (corresponding to V band in our quasar) and a continuum at 2800 \AA~ in HE 0413-4031 is always the same as in the template. 

This gave us a relation between the V magnitude and the 2800 \AA~ continuum $\nu F_{\nu}$ flux at 2800 \AA, $F_{2800}$:
\begin{equation}
\log{F_{2800}} = -0.4V - 8.234 ~~~~ [{\rm erg~~ s^{-1} cm^{-2}}]
\end{equation}

This is of course an approximation but the amplitude of the flux variations in our source is not very large so the issue of a spectrum becoming bluer when the quasar is brighter \citep[see e.g.][and the references therein]{ulrich1997,wilhite2005,kokubo2015} should not be critical. 

\section{Results: spectroscopy}
\label{sec_results_spectroscopy}

\subsection{Mean spectrum}
\label{sec_mean_spectrum}

We first combined the SALT spectra in order to establish the global source parameters which will be fixed later in the analysis of all 25 spectra. 

\begin{figure}
\centering
\includegraphics[width=0.48\textwidth]{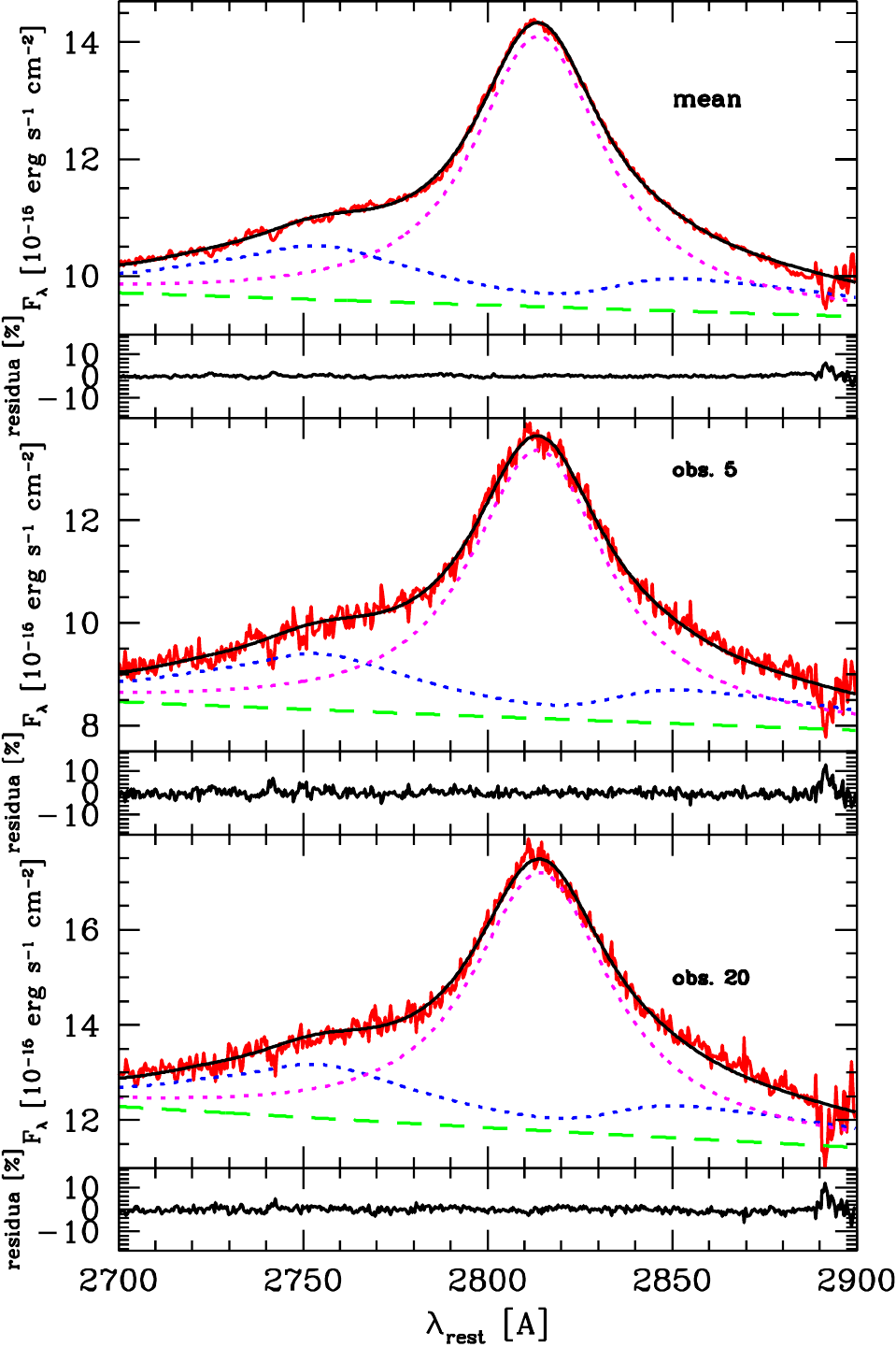}
\caption{{\bf Top: } Flux-calibrated mean spectrum of HE 0413-4031. Data is shown with a red line, and black line gives the best fit model. Remaining lines show the spectral components: power law (green dashed line), FeII pseudo-continuum (blue dotted line), and MgII total flux (two doublet components, magenta dotted line). The lower panel shows the residua, which are noticeable only close to 2900 \AA, where the sky line contamination was the strongest and the background subtraction did not fully correct for this effect. {\bf Middle:} Flux-calibrated spectrum of HE 0413-4031 for the minimum phase (epoch 5). The lines represent the same spectral components as in the top panel. {\bf Bottom:} Flux-calibrated spectrum of HE 0413-4031 for the maximum phase (epoch 20). The lines represent the same spectral components as in the top panel. The epochs are listed in Table~\ref{tab:spectr}.}
\label{fig:mean}
\end{figure}

The mean spectrum is shown in Fig.~\ref{fig:mean} (top panel). For comparison, in Fig.~\ref{fig:mean} (middle panel) we also show the spectrum in the early epoch (\#5) when the quasar was close to the minimum flux density as well as the spectrum from the later epoch close to the maximum flux density (\#20, bottom panel). The MgII line shape in this quasar looks simple, immediately suggesting that HE 0413-4031 belongs to class A quasars \citep{sulentic2000}. We checked that assuming just a single kinematic component of Lorentzian shape for MgII is sufficient, and adding the second component does not improve the $\chi^2$ significantly. The best fit for two-component model allows for 0.2 \% contribution from the second kinematic component, which is very broad (11 140 km s$^{-1}$), and the total $\chi^2$ for such a fit is better than in a single-component fit only by 1.0.
The source is thus a typical representative of a class A sources. We also checked that indeed a Lorentzian shape offers much better representation of the line shape than the Gaussian. If we assume a single kinematic component with a Gaussian shape, the reduced $\chi^2$ of the best fit is 16.0 per degree of freedom. If we allow for two Gaussian kinematic components, one with no shift with respect to FeII and the second one at arbitrary position, the fit improves (reduced $\chi^2 = 7.7$) but still does not match the one with a single Lorentzian shape, despite the higher number of parameters. We note that in the case of the two-Gaussian fit the component bound to FeII dominates (contains 57 \% of the line flux) which happened at the expense of the overall FeII contribution which dropped down by a factor of 3 in comparison with a single Lorentzian fit. However, such a fit is not favored by the data. The single-component Lorentzian profile typical of Population A sources \citep{sulentic2000} arises due to the turbulence in the line-emitting clouds and the broadening of the line is due to the rotation \citep{2011Natur.470..366K,2012MNRAS.426.3086G,2013A&A...549A.100K,2013A&A...558A..26K}.

Since in our full model one of the kinetic components was set at zero rest frame velocity, together with FeII, while the second kinematic component has an arbitrary shift in velocity space with respect to them, we eliminated the first kinematic component, leaving the second one, which allows us for the flexibility of the shift between MgII and FeII. This model is also later used to fit individual spectra.

We tested several templates of the FeII from the \citet{bruhweiler2008}, and the best fit was provided by the d12-m20-20-5.dat model which assumes the cloud number density $10^{12}$ cm$^{-3}$, the turbulent velocity $20$ km s$^{-1}$, and the hydrogen ionizing photon flux $10^{-20.5}$ cm$^{-2}$ s$^{-1}$. The same template was favored for the quasar CTS C30.10 also monitored by SALT \citep{czerny2019}. It is not surprising, since recent modelling of the quasar main sequence also suggest values of that order for the local BLR cloud density and the turbulent velocity \citep{panda2018,panda2019}. The best fit half-width of the Gaussian used for template convolution was 1200 km s$^{-1}$.

We calculated a grid of models for different redshift and different ratio of the doublet, and these two quantities are strongly coupled. We determined the best fit redshift as $z = 1.37648$, and the doublet ratio 1.9. This is a value quite close to the optically thin case, 2:1 ratio. 

These parameters: the choice of the FeII template, template smearing velocity, redshift and doublet ratio were later assumed to be the same in all fits of the individual spectra, while the FeII amplitude, MgII amplitude, line width and line shift, and the power law parameters were allowed to vary from observation to observation.


The best fit FWHM of the MgII line in the mean spectrum is $4380^{+14}_{-15}$ km s$^{-1}$, formally just above the line dividing the class A and class B source \citep{sulentic2000,marziani2018}. However, some trend with the mass in this division is expected, since for Seyfert galaxies the dividing line between Narrow Line Seyfert 1 galaxies and Seyfert galaxies is at 2000 km s$^{-1}$ \citep{osterbrock1985}, instead of at 4000 km s$^{-1}$, as in quasars. HE 0413-4031 is still more massive and brighter than most quasars in SDSS catalogs \citep{shen2011,paris2017}. Since we fit the single Lorentzian shape, we cannot derive the line dispersion $\sigma$ from the fit, since the Lorentzian shape corresponds to the limit of FWHM/$\sigma \rightarrow 0$.  We can, however, determine the line dispersion numerically since the FWHM/$\sigma$ ratio is an important parameter \citep{collin2006}. Therefore, we subtracted the fitted FeII and the remaining underlying continuum, and integrated the line profile. We obtained $\sigma = 2849$ km s$^{-1}$, and FWHM/$\sigma = 1.54$, which confirms that the source belongs to Population 2 of \citet{collin2006}, or class A of \citet{sulentic2000}.

In the mean spectrum, the EW(MgII) is $27.45^{+0.12}_{-0.10}$ \AA, a bit below the average value for the MgII from Large Bright Quasar Sample (42 \AA, \citealt{forster2001}). 

The most interesting part is the shift we detect between the MgII line and the FeII pseudo-continuum. This shift is by 15.1 \AA, or equivalently, 1620 km s$^{-1}$, and the MgII line is redshifted with respect to FeII. It may also be that FeII is blue-shifted with respect to MgII, however we cannot distinguish between these cases. \citet{kovacevic15} in their study observed redshifts, not blueshifts of the FeII. In addition, the conclusion about the relative shift strongly depends on the combination of the Fe II template used and the adopted redshift, as we discuss in Appendix~\ref{sec_appendix_FeII_templates}.

Unfortunately, we are unable to establish the proper position of the rest frame for our SALT observation. We failed to identify the narrow [NeV]3426.85\AA ~line which is relatively strong in the quasar spectra\footnote{http://classic.sdss.org/dr6/algorithms/linestable.html}, but this search did not yield a reliable identification.

\subsection{Determination of the mean and rms spectrum}

For constructing the mean and the rms spectra, we follow the standard procedure as explained by \citet{peterson2004}. The mean spectrum is calculated using the following relation
\begin{equation}
    \overline{F(\lambda)}=\frac{1}{N}\sum_{i=1}^{N} F_{i}(\lambda)\,,
    \label{eq_mean_spectrum}
\end{equation}
where $F_{i}(\lambda)$ are individual spectra. For studying variability phenomena, we also construct an rms spectrum using
\begin{equation}
   S(\lambda)=\left\{\frac{1}{N-1} \sum_{i=1}^{N}[F_{i}(\lambda)-\overline{F(\lambda)}]^2 \right\}^{1/2}\,. 
   \label{eq_rms_spectrum}
\end{equation}

\begin{figure}
\centering
\includegraphics[width=0.48\textwidth]{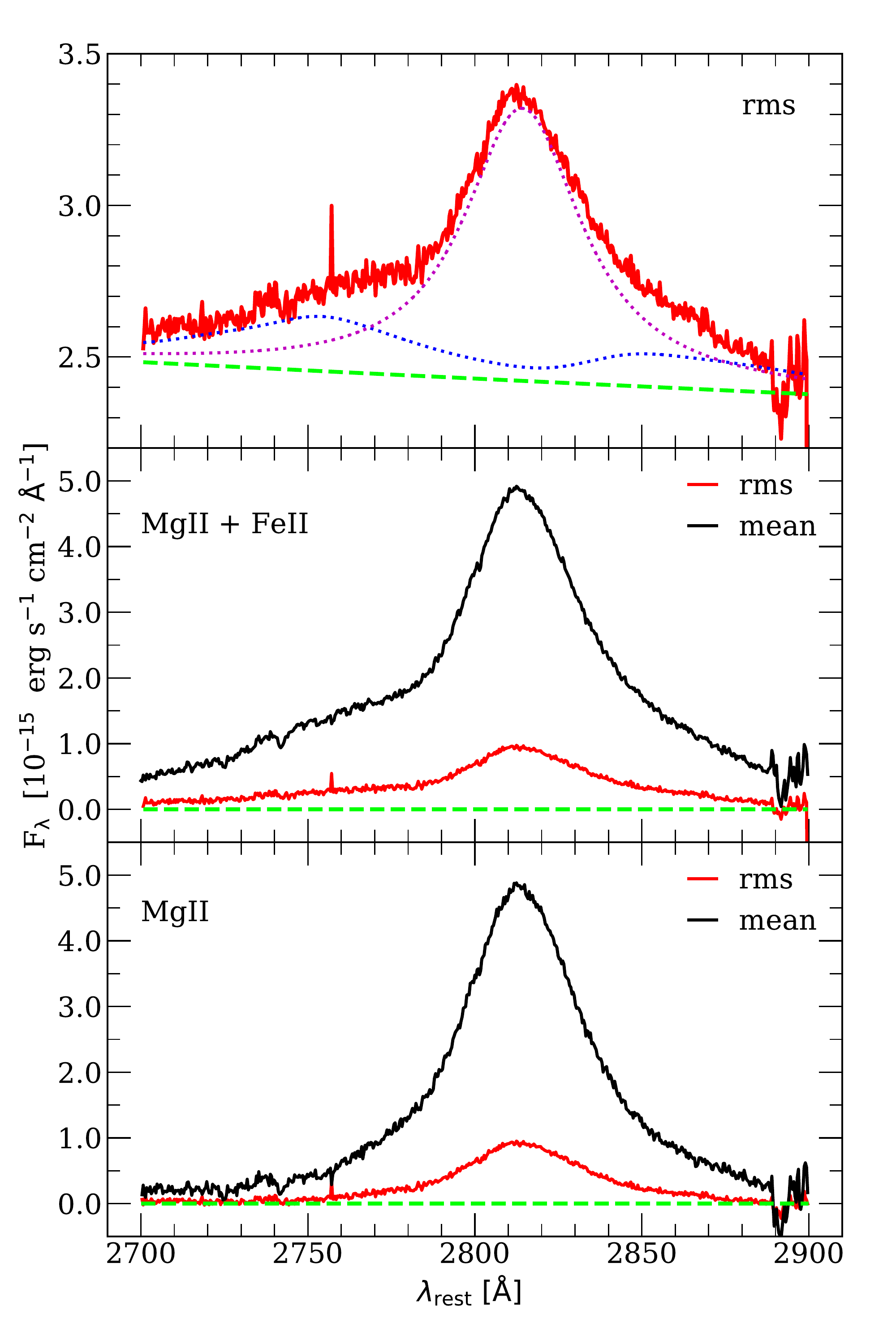}
\caption{Decomposition of the flux-calibrated rms spectrum (top panel) and the comparison of the rms (red) and mean (black) spectra at the zero-flux level (represented by a dashed green line), first with the continuum subtracted (MgII+FeII; middle panel) and subsequently, with the FeII emission subtracted (MgII; bottom panel). From the rms spectra it is clear, that the core of MgII line is the most variable, with a much smaller variability in the wings.}
\label{fig:rms2}
\end{figure}

The flux calibrated mean and root-mean-square (rms) spectra are shown in the top panels of Figures~\ref{fig:mean} and \ref{fig:rms2}, respectively. For the flux calibration we used the composite quasar spectra created by \citet{2001AJ....122..549V}. For the continuum, they proposed a power--law with an index of $\alpha_\lambda=-1.56$. This continuum was normalized for each spectrum according to the V magnitudes reported in Table~\ref{tab:Vmag}, which were simply converted to flux units in order to get the flux normalization. 

The rms spectrum was estimated following Eq.~\ref{eq_rms_spectrum}. Mean and rms profiles look similar, but to check this more quantitatively, we fitted the rms spectrum in the same way as we fitted the mean spectrum. The result is shown in Figure~\ref{fig:rms2}, upper panel. The line is still well fitted when we use a single Lorentzian model. The FWHM in rms spectrum is 4337 km s$^{-1}$, only marginally narrower than the FWHM of the mean spectrum (4380 km s$^{-1}$). When the mean and rms spectra are compared at the zero-flux level, i.e. with the continuum subtracted, we see that the core of MgII is most variable with the wings having a much smaller variability which could be attributed to FeII emission, see the central panel of Fig.~\ref{fig:rms2}. If we subtract the FeII pseudo-continuum from the rms and mean spectra, the only variable part of the MgII emission is at the core of the line, see the bottom panel of Fig.~\ref{fig:rms2}. The EW(MgII) if measured in the rms spectrum is 21.01 \AA, lower than in the mean spectrum (27.45 \AA), and also EW(FeII) is lower than in the mean spectrum (8.32 \AA~ instead of 10.13 \AA), which results from the enhanced role of the continuum power-law. The consistency of the rms and the mean spectrum fits also supports the single-component Lorentzian fit of the line shape.


\subsection{Spectral fits of individual observations}

\begin{figure}[tbh]
    \centering
    \includegraphics[width=0.5\textwidth]{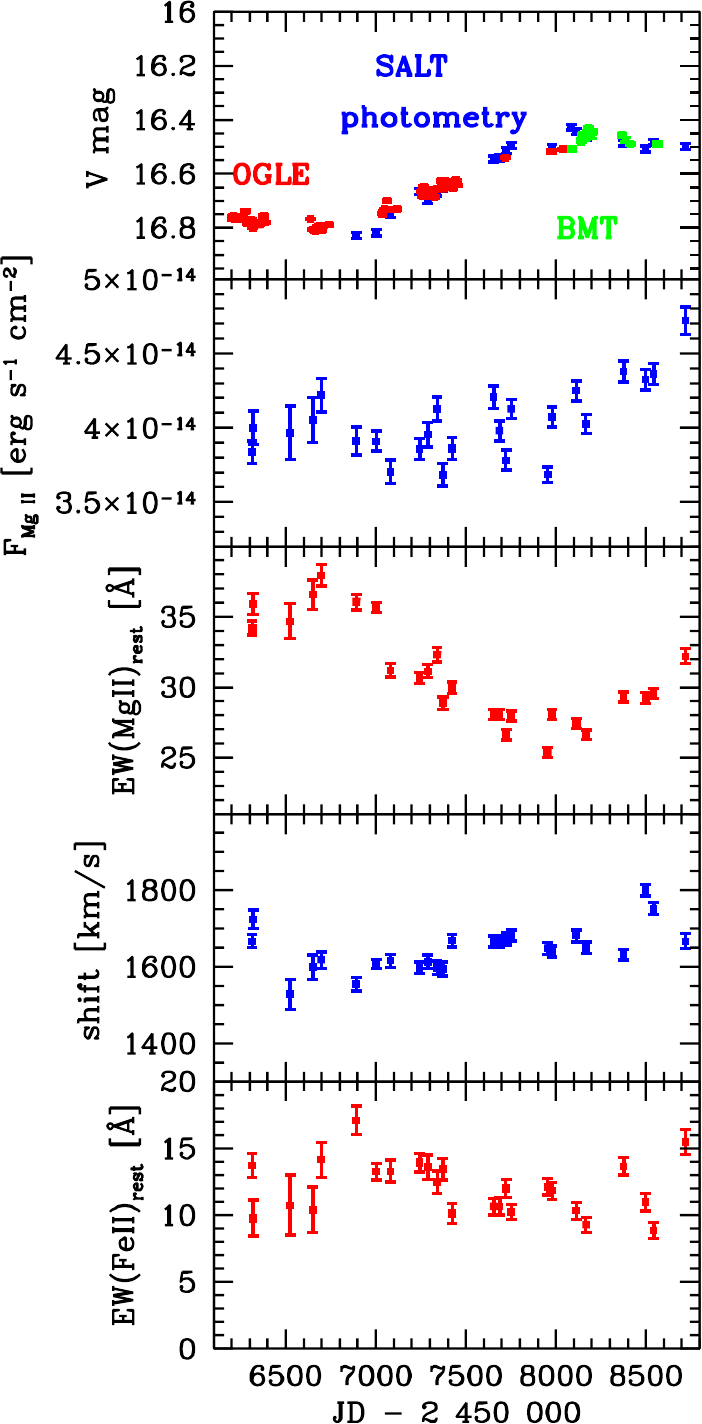}
    \caption{Temporal evolution of photometric and spectroscopic characteristics from the monitoring of the quasar HE 0413-4031. From the top to the bottom panels: photometric light curve ($V$-band magnitude) from OGLE, SALT, and BMT (colour-coded; BMT data were shifted by 0.171 mag up to correct for the systematic offset with respect to the OGLE data); MgII line-emission light curve (in ${\rm erg\,s^{-1}\,cm^{-2}}$); the equivalent width of MgII line in \AA; velocity shift of MgII line with respect to the FeII line in km/s; the equivalent width of FeII line in \AA. The time is expressed in JD-2450000.}
    \label{fig_light_curves_EW}
\end{figure}

For each of the spectra, the EW(MgII), FWHM(MgII), EW(FeII), the shift between MgII and FeII, and the power spectrum parameters were determined. The results are given in Table~\ref{tab:spectr}, and Fig.~\ref{fig_light_curves_EW} visually show the evolution of these properties with time.

The mean shift of the MgII and FeII lines, calculated from the individual spectra is 1582 km s$^{-1}$, somewhat smaller than obtained from the mean spectrum. Variations from one spectrum to the other are at the level of 82 km s$^{-1}$ (dispersion), larger than individual errors. If we fit a linear trend, we see a systematic increase of the MgII and FeII separation by 109 km s$^{-1}$ in six years but it is not much larger than the dispersion in the measurements; however, it seems formally significant if we use the individual measurement errors given in Table~\ref{tab:spectr}. The corresponding acceleration 18 km s$^{-1}$ yr$^{-1}$ is much smaller than the large value  of $104 \pm 14$ km s$^{-1}$yr$^{-1}$ found for the quasar HE 0435-4312 using also SALT instrument \citep{sredzinska2017}.

The averaged FWHM is 4390.8 km s$^{-1}$, the dispersion is 200 km s$^{-1}$, again slightly larger than typical measurement error but no interesting trends could be noticed. Thus, we observe some small variations in the line shape, but they are indeed marginal, consistent with the fact that rms spectrum is similar to the mean shape of the line. 

\subsection{Light curves: variability and linear trends}

The continuum photometric lightcurve and the MgII lightcurve are presented in Fig.~\ref{fig_light_curves_EW}. The continuum shows mostly slow but a noticeable variation. A single brightening trend dominates for most of the monitoring period, replaced with some dimming during the last 1.5 years. The overall variability level of the continuum is $F_{\rm var} = 13.0$ \%, if BMT telescope is included, and 10.4\%, if these data are not taken into account. Here we use the standard definition of the excess variance,
\begin{equation}
F_{\rm var} = \sum_{i=1}^{N} {(x_i -\overline{x})^2 - \delta x_i^2  \over N \overline{x}^2 }\,,
\end{equation}
where $\overline{x}$ is the average value, and $\delta x_i$ is the individual measurement error.
Since this linear trend seems suggestive, we also checked the shorter timescale variability by fitting first a linear trend to the lightcurve in the log space (i.e. when using magnitudes), and then we subtracted this trend from the original lightcurve. We did this only for the data without BMT. The $F_{var}$ dropped from 10.4~\% down to 7.4~\%.

The MgII line variability is lower, $F_{\rm var} = 7.2 \%$, and $F_{\rm var} = 7.1 \%$, depending whether BMT telescope data were or were not used for MgII calibration, respectively. It is interesting to note, however, that the level of variability in MgII and the continuum are comparable if the long term trend was subtracted from the data.  

We also determine FeII lightcurve, and the variability level of FeII seems higher, at the level of 14.7\% if BMT data are neglected, and 14.9\% if the BMT data are included. However, the measurement errors are large due to the coupling between the continuum and FeII pseudo-continuum.

\begin{figure*}[tbh]
    \centering
    \includegraphics[width=0.49\textwidth]{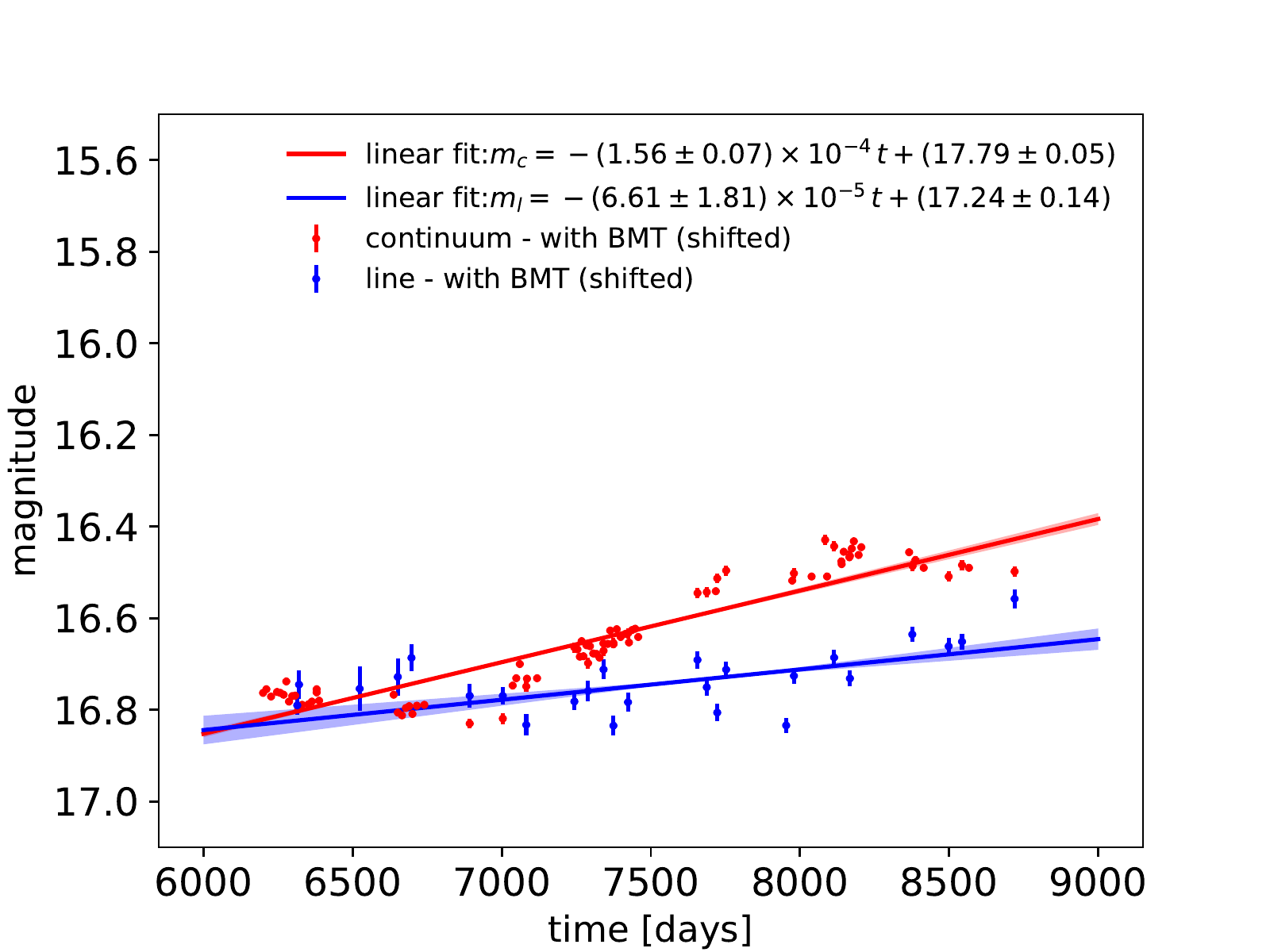}  \includegraphics[width=0.49\textwidth]{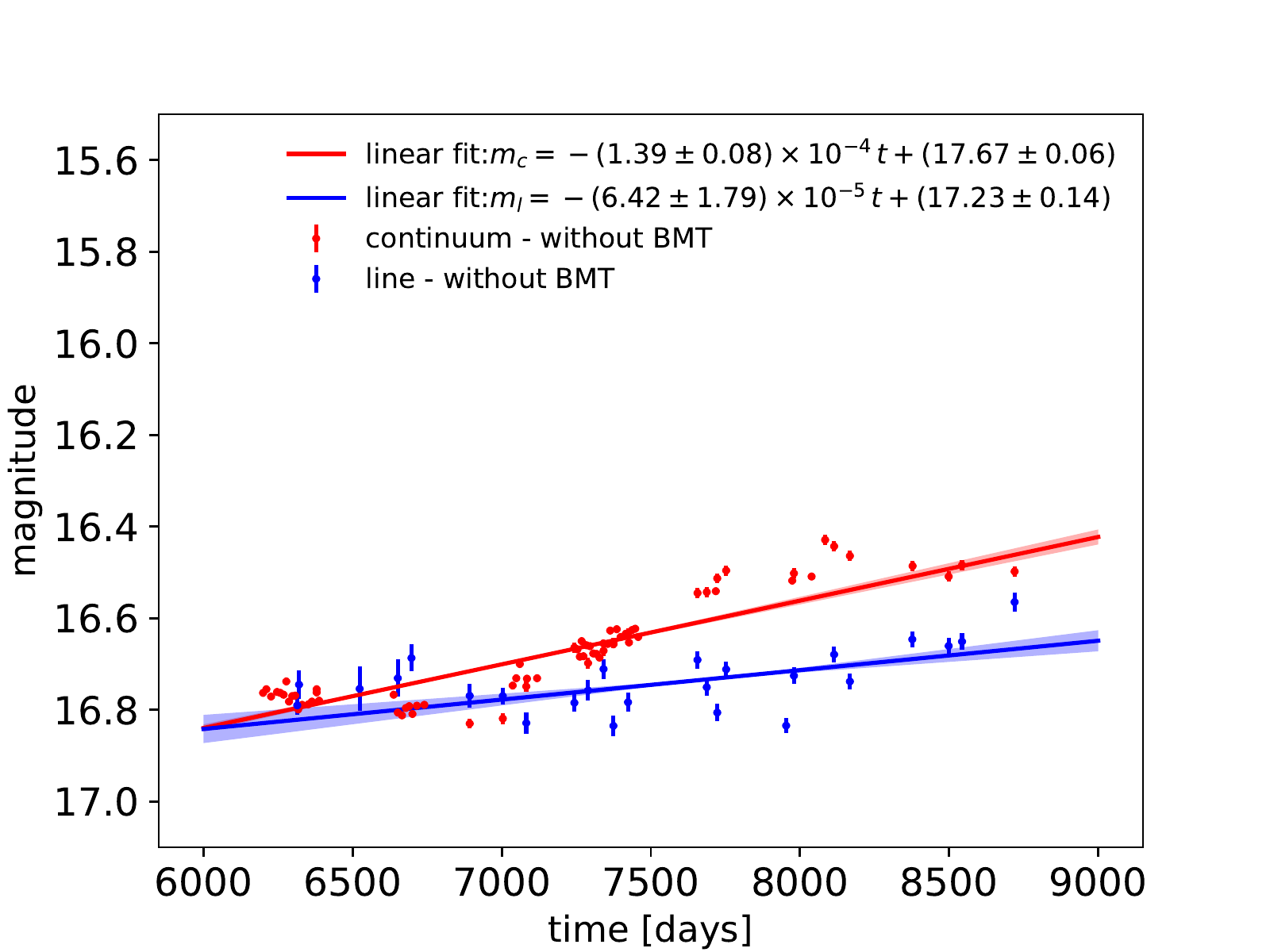}
    \caption{Linear-trend fit to both continuum and line-emission light curves in the log-space. \textbf{Left panel:} The linear trend fits with BMT data included. The BMT data were shifted by 0.171 mag to correct for the systematic shift with respect to the OGLE data. The legend includes the best-fit parameters for both the continuum and the line-emission light curves. The fit statistics are $\chi^2=7121.4$ and $\chi^2_{\rm red}=84.8$ for the continuum and $\chi^2=177.9$ and $\chi^2_{\rm red}=7.7$ for the line light curve. \textbf{Right panel:} As in the left panel, but without BMT data. The fit statistics are $\chi^2=5609.2$ and $\chi^2_{\rm red}=79.0$ for the continuum and $\chi^2=172.9$ and $\chi^2_{\rm red}=7.5$ for the line light curve.}
    \label{fig_linear_trend_fits}
\end{figure*}

The mean monochromatic luminosity at 3000\AA\, can be derived from the $V$-band magnitude of 16.5, using the extinction reported in NED with a value of 0.034, source redshift of 1.389, and standard cosmological parameters for the flat Universe \citep[$H_0=70\,{\rm km\,s^{-1}\,Mpc^{-1}}$, $\Omega_{\rm m}=0.28$, and $\Omega_{\Lambda}=0.72$, see][for details]{2015AcA....65..251K}. We obtain $\log{L_{3000}}=46.754$. The uncertainty of the monochromatic luminosity can be estimated from the minimum and the maximum points along the photometric lightcurve, 16.429 mag and 16.830 mag, respectively, which implies $\log{L_{3000}^{\rm max}}=46.782$ and $\log{L_{3000}^{\rm min}}=46.622$. Hence, for the further analysis, we consider $\log{L_{3000}}=46.754^{+0.028}_{-0.132}$.

As already pointed out, the linear trend is present in both continuum and MgII line emission light curves. In Fig.~\ref{fig_linear_trend_fits}, we show the fit of a linear function to both light curves, considering the case with and without BMT data in the left and right panels, respectively. The linear trend is towards smaller magnitudes, i.e. the continuum and line emission flux densities increase during the observational run. The slope of the linear trend is larger for the continuum than for the line-emission light curves. The continuum slope is $s_{\rm c}=0.057\,{\rm mag\,yr^{-1}}$ and the line slope is $s_{\rm l}=0.024\,{\rm mag\,yr^{-1}}$ with the BMT data included, while for the case without BMT data the continuum increase drops a little, $s_{\rm c}=0.051\,{\rm mag\,yr^{-1}}$, while the line-emission slope is comparable, $s_{\rm l}=0.023\,{\rm mag\,yr^{-1}}$. In other words, the decrease in the continuum magnitude is $2.38$ and $2.17$ larger than the decrease for the line-emission magnitude with and without BMT data, respectively.

\section{Results: Time-delay determination} \label{sec:result}

As for the intermediate-redshift quasar CTS C30.10 (z = 0.90052) \citep{czerny2019}, we apply several methods to determine the time-delay between the continuum $V$-band and MgII line emission. Apart from the standard interpolated cross-correlation function, we apply several statistically robust methods suitable for unevenly sampled, heterogeneous pairs of light curves \citep[see ][for an overview]{2019arXiv190703910Z}, namely the discrete correlation function, $z$-transformed discrete correlation function, JAVELIN, $\chi^2$ method, von Neumann, and Bartels estimator. For all seven methods, we considered the two pairs of light curves, those with and without magnitude-shifted BMT data. The detailed description of the time-delay analysis is in Appendix in Section~\ref{sec_time-delay}, with Subsections~\ref{subsec_ICCF}-\ref{subsec_chi2} describing individual methods including the corresponding plots and the tables.

\subsection{Final time-delay for MgII line}

\begin{table*}[h!]
    \centering
     \caption{Summary list of the time-delays expressed in light days in the observer's frame between the continuum and MgII line-emission light curves for the flat-spectrum radio quasar HE 0413-4031. We distinguish for all methods two cases - with and without magnitude-shifted BMT data.}
    \begin{tabular}{c|c|c}
    \hline
    \hline
    Method &  With shifted BMT data & Without BMT data\\
    \hline
    ICCF Interpolated continuum - Centroid [days]    & $1004.6^{+196.8}_{-246.2}$ & $1003.2^{+205.3}_{-235.4}$     \\
    ICCF Interpolated line - Centroid [days]    & $1008.4^{+142.2}_{-276.9}$  &  $1034.171^{+139.1}_{-248.9}$   \\
    ICCF Symmetric -Centroid [days]    &  $1009.7^{+113.6}_{-211.5}$  & $1021.7^{+114.5}_{-  207.8}$  \\
    DCF peak time-delay -- bootstrap [days] & $720.4^{+115.1}_{-147.9}$    &  $726.0^{+114.4}_{-145.7}$     \\
    zDCF Maximum Likelihood & $720.9^{+323.9}_{-527.3}$ & $720.9^{+331.3}_{-100.1}$ \\
    JAVELIN peak time-delay [days]  &   $1053.7^{+79.8}_{-163.6}$  &  $1058.5^{+77.1}_{-150.7}$   \\
    Von Neumann peak -- bootstrap [days] & $498.9^{+170.9}_{-125.9}$        &  $711.3^{+149.0}_{-139.5}$  \\
    Bartels peak -- bootstrap [days] & $710.9^{+172.3}_{-173.0}$   &  $714.6^{+176.1}_{-164.6}$  \\
    $\chi^2$ peak -- bootstrap [days] &  $720.4^{+145.6}_{-102.2}$   &   $727.7^{+160.0}_{-85.2}$      \\
    \hline
    Average of the most frequent peak - observer's frame [days] &    $718.2^{+102.8}_{-145.8}$                &   $720.1^{+89.8}_{-58.3}$           \\
    Average of the most frequent peak - rest frame &  $302.2^{+43.3}_{-61.4}$                  &  $303.0^{+37.8}_{-24.5}$            \\ 
    \hline
    Average of the secondary peak - observer's frame [days] & $1019.1^{+69.9}_{-114.2}$  & $1029.4^{+71.0}_{-107.0}$\\
    Average of the secondary peak - rest frame [days] & $428.8^{+29.4}_{-48.1}$  & $433.2^{+29.9}_{-45.0}$\\
    \hline
    \hline
    \end{tabular}
    \label{tab_summary_delay}
\end{table*}

Due to the systematic offset of the BMT data in the continuum light curve, we decided to distinguish two cases for all time-delay analysis techniques. For a matter of completeness, below we summarize in Table~\ref{tab_summary_delay} the main results for all the methods, including the cases with and without magnitude-shifted BMT data. The most prominent peak in the time-delay distributions is the peak close to 700 days in the observer's frame. This peak is generally present in all seven methods. However, the ICCF analysis generally gives longer time-delays of 900-1000 days, which could be caused by the interpolation and hence by adding new points to the analysis. A noticeable difference is also for the JAVELIN method, where the time-delay peak is close to $1050$ days. Since JAVELIN uses the damped random walk for fitting the continuum light curve, which is then smoothed and time-delayed to reproduce the MgII line-emission light curve, extra points are introduced to the light curves in a similar way as for the ICCF. This can lead to biases and artefacts especially for irregular and sparse datasets. This is why we decided to prefer the peak around 700 days, which is the most prominent for all discrete methods that do not require interpolation and are model-independent (DCF, zDCF, Von Neumann). The detected time-delay of $498.9^{+170.9}_{-125.9}$ days for the Von-Neumann estimator with shifted BMT data is most likely an artefact since it is an excess given by only one point, see Fig.~\ref{fig_von_neumann} (left panel). The second minimum of the Von Neumann estimator around 700 days is then more pronounced and clearly given by more points. In addition, the minimum around 500 days is not present for the case without BMT data, see Fig.~\ref{fig_von_neumann} (right panel).

Given the arguments above, we focus on the observed time-delay around 700 days. Concerning the average value, we obtain the rest-frame time-delay of $\tau_1=302.2^{+43.3}_{-61.4}$ days for the case with the shifted BMT data, and $\tau_2=303.0^{+37.8}_{-24.5}$ days for the case without them. The final average value then is $\overline{\tau}=302.6^{+28.7}_{-33.1}$ days, which corresponds to the light-travel distance of $R_{\rm MgII}=c\overline{\tau}=0.254^{+0.024}_{-0.028}\,{\rm pc}\sim 10^{17.9}\,{\rm cm}$. The inferred value of the light-travel distance $R_{\rm MgII}$ is larger than typical BLR length-scales inferred from other RM campaigns with time-delays of the order 10-100 light days for AGN with a broad range of black hole masses \citep{2000ApJ...536..284K,2004ApJ...606..749K,2016ApJ...818...30S}. The length-scale of the BLR has implications for the line variability as was shown by \citet{guo2019} and we will specifically discuss the MgII line-continuum variability relation in Section~\ref{subsec_response_mgII}.   

The results provided by the ICCF and the JAVELIN analyses provide a time-delay that we treat as secondary for the reasons of interpolation and the model-dependence. In the rest frame, this secondary time-delay is $428.8^{+29.4}_{-48.1}$ days for the case with the shifted BMT data and $433.2^{+29.9}_{-45.0}$ days for the case without the BMT data. The average rest frame value is $431.0^{+21.0}_{-32.9}$ days. This secondary time-delay peak should be reevaluated when more continuum and line-emission data is available to assess if it is just an artefact of data sampling irregularity.

\section{Results: MgII-based radius-luminosity relation}
\label{sec_MgII_RL_relation}

\subsection{Preliminary virial black hole mass and Eddington ratio}
\label{sec_final_time_delay_virial_mass}

The virial black-hole mass can be determined from the virial relation for the BLR, $M_{\bullet}=f_{\rm vir}c\tau_{\rm BLR} \text{FWHM}^2/G=(1.134^{+0.089}_{-0.072})\times 10^9\,M_{\odot}$, which was calculated assuming the virial factor equal to unity, the average time-delay for MgII inferred earlier, and the best fit FWHM of $4380^{+14}_{-15}$ km s$^{-1}$. In general, however, the virial factor may deviate from unity, which is indicated by the study of \citet{mejia2018}, which implies the anticorrelation between the virial factor and the line FWHM, which is in our case the main source of uncertainty. According to \citet{mejia2018}, we have
\begin{equation}
    f_{\rm vir,MgII}=\left(\frac{\text{FWHM}_{\rm obs}(\text{MgII})}{3200 \pm 800\,{\rm km\,s^{-1}}}\right)^{-1.21 \pm 0.24}\,,
        \label{eq_virial_factor}
\end{equation}
which for FWHM$=4380^{+14}_{-15}$ km s$^{-1}$ leads to the virial factor less than unity, $f_{\rm vir, MgII}=0.42-0.92$, and the virial black hole mass in the range of $M_{\bullet}=4.8 \times 10^8-1.0\times 10^{9}\,M_{\odot}$, hence we have a factor of 2 uncertainty in the virial black hole mass. The Eddington luminosity can be estimated as,
\begin{equation}
    L_{\rm Edd}=1.256\times 10^{47}\left(\frac{M_{\bullet}}{10^9\,M_{\odot}} \right)\,{\rm erg\,s^{-1}}\,,
    \label{eq_edd_ratio_he0413}
\end{equation}
while the bolometric luminosity may be calculated using the bolometric correction with respect to $\lambda=3000\angstrom$, $L_{\rm bol}=(5.62 \pm 1.14)L_{3000}$ \citep{2006ApJS..166..470R}, which leads to the Eddington ratio of $\eta_{\rm Edd}=L_{\rm bol}/L_{\rm Edd}\approx 2.18$. Using the power-law calibration of the bolometric correction by \citet{2019MNRAS.488.5185N}, we obtain $L_{\rm bol}\simeq 2.8 L_{3000}$, which gives $\eta_{\rm Edd}\approx 1.27$. Hence, these values imply close to the Eddington- or even the super-Eddington accretion mode.

\subsection{Position in the radius-luminosity plane}

By combining the rest-frame time-delay and the monochromatic luminosity of HE 0413-4031, we can position the source on the radius-luminosity plane alongside the other quasars to check for the potential deviation of HE 0413-4031 due to its high accretion rate, as was previously detected for super-Eddington sources monitored in broad H$\beta$ line \citep{Wang_2014a, Wang_2014b, Mart_nez_Aldama_2019}.

\begin{figure*}[tbh]
    \centering
    \includegraphics[width=0.49\textwidth]{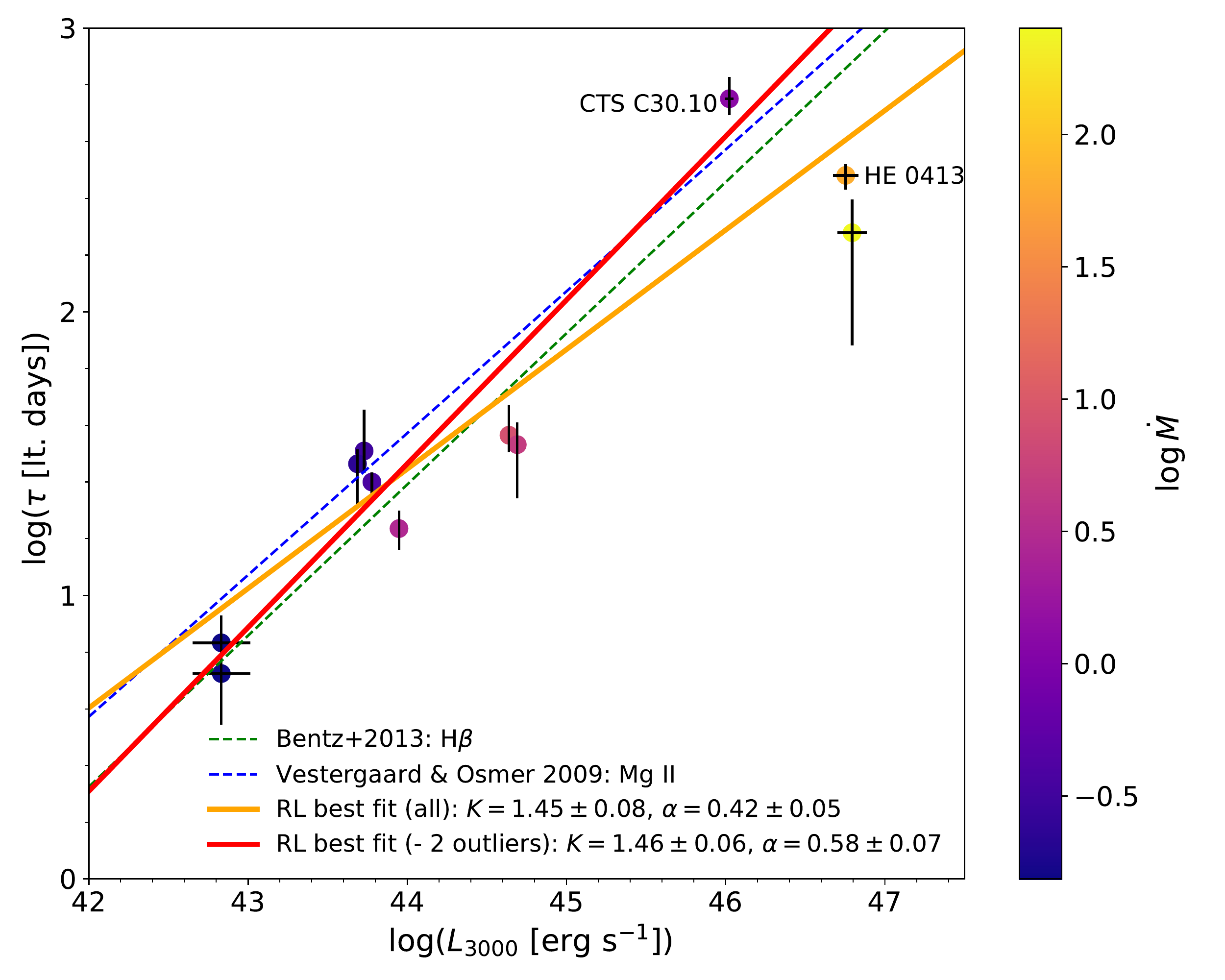}
    \includegraphics[width=0.49\textwidth]{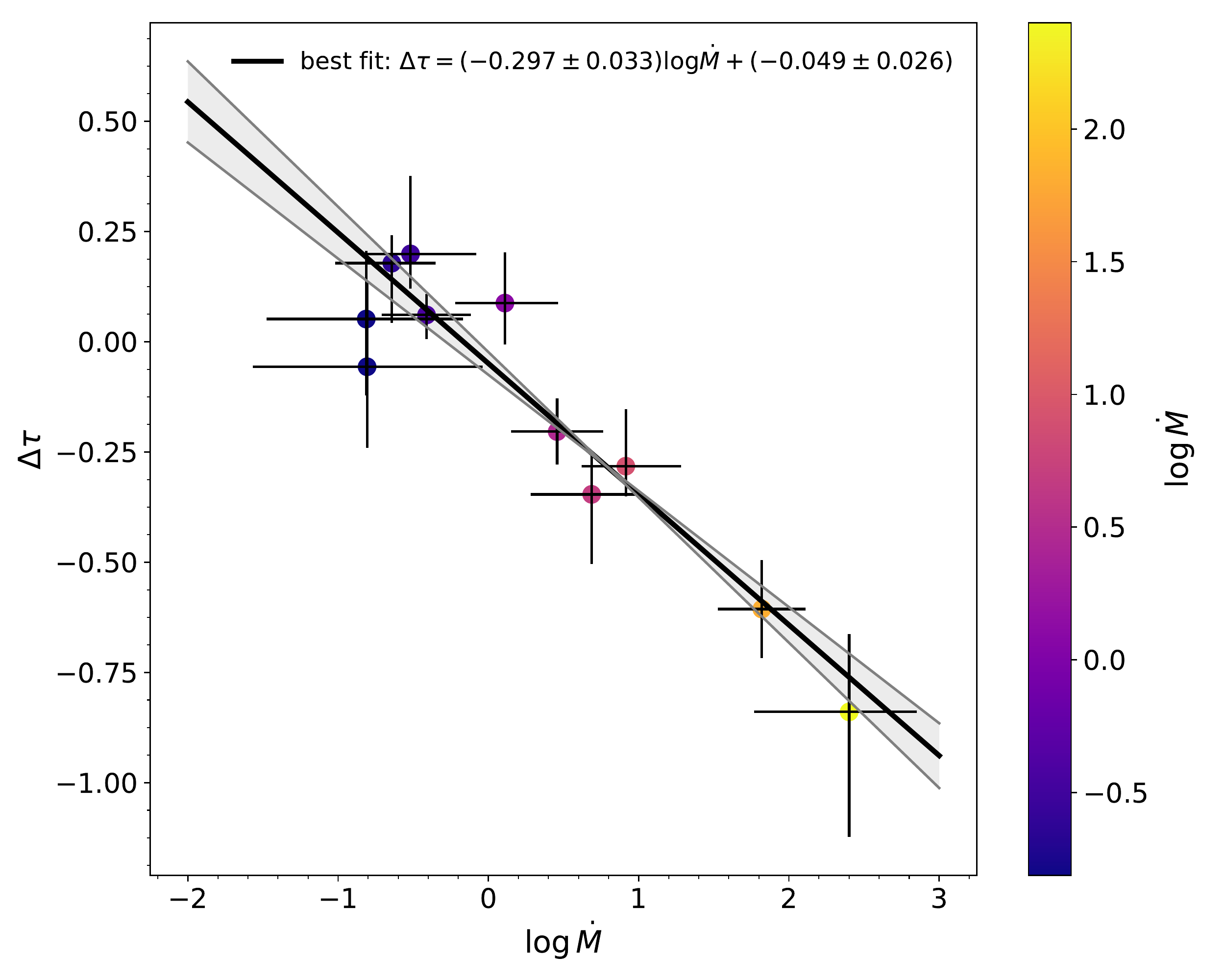}
    \caption{Radius-luminosity relation for the reverberation-mapped sources in broad MgII line resembles the radius-luminosity relation for H$\beta$ line. \textbf{Left panel:} Radius-luminosity relation for the RM quasars monitored in the broad MgII line. Clearly, the sources follow within uncertainties the scaling relationship previously derived by \citet{2009ApJ...699..800V} for MgII line (dashed blue line) as well as the H$\beta$ radius-luminosity relationship of \citet{bentz2013} (dashed green line). The best-fit relationships are also displayed, both for the case when all the sources are included in the fitting procedure (orange solid line)) and for the case when two outliers that are below the radius-luminosity relations are removed (CTS252 and HE 0413-4031) towards the higher luminosities (solid red line). Individual sources are colour-coded to show the logarithm of the dimensionless accretion rate parameter $\dot{M}$, see Eq.~(\ref{equ:mdot}), according to the colour bar to the right. \textbf{Right panel:} The strong anticorrelation (with the Pearson correlation coefficient of $p=-0.940$) of the parameter $\Delta \tau$, which expresses the rate of departure from the radius-luminosity relation, see also Eq.~(\ref{eq_departure_parameter}), with respect to the dimensionless accretion-rate parameter $\dot{M}$ expressed by Eq.~(\ref{equ:mdot}).}
    \label{fig_radius_luminosity_mgII}
\end{figure*}

With the rest-frame time-delay of $\overline{\tau}=302.6^{+28.7}_{-33.1}$ days and the monochromatic luminosity of $\log{L_{3000}}=46.754^{+0.028}_{-0.132}$, the source HE 0413-4031 lies below the expected radius-luminosity relation, $R(MgII)-L_{3000}$ \citep{2009ApJ...699..800V}. We demonstrate this in Fig.~\ref{fig_radius_luminosity_mgII}, where we compiled all the sources whose time-delay was determined for MgII line \citep[10 sources, see Table 3 in ][]{czerny2019}, including CTS C30.10 and the new source HE 0413-4031. The list of all the sources with measured time-delays and determined monochromatic luminosities is in Table~\ref{tab_sources_mgII}. With a large scatter ( $\sigma=0.246$ dex), the sources approximately follow the radius-luminosity relationship previously derived for MgII line \citep{2009ApJ...699..800V},

\begin{equation}
    \log{\left[\frac{\tau(\text{MgII})}{1 {\rm lt. day}}\right]}=1.572 + 0.5\log{\left(\frac{L_{3000}}{10^{44}\,{\rm erg\,s^{-1}}}\right)}\,,
    \label{eq_tau_luminosity_vestergaard_osmer}
\end{equation}
as well as the radius-luminosity relationship derived for H$\beta$ line for lower redshift sources by \citet{bentz2013},
\begin{equation}
    \log{\left[\frac{\tau(H\beta)}{1 {\rm lt. day}}\right]}=1.391 + 0.533\log{\left(\frac{L_{3000}}{10^{44}\,{\rm erg\,s^{-1}}}\right)}\,,
    \label{eq_tau_luminosity_bentz}
\end{equation}
where we replaced $L_{5100}$ monochromatic luminosity by $L_{3100}$ using $L_{5100}\simeq 0.556\,L_{3000}$ using the power-law relations for the bolometric corrections derived by \citet{2019MNRAS.488.5185N}. The scatter of the sources around the relation by \citet{bentz2013} is $\sigma=0.269$ dex, which is comparable to the scatter with respect to the relation by \citet{2009ApJ...699..800V}.

\begin{table*}[tbh]
\centering
 \caption{Characteristics of reverberation-mapped sources monitored using broad MgII line. From the left to the right column, the table lists the source name, redshift, measured time-delay in light days in the rest frame, the logarithm of the monochromatic luminosity at 3000 \AA, FWHM of MgII in ${\rm km\,s^{-1}}$ , the dimensionless accretion rate as defined in Eq.~(\ref{equ:mdot}), departure parameter $\Delta \tau$ defined by Eq.~\ref{eq_departure_parameter}, and the corrected time-delay expressed in light days in the rest frame, see also Eq.~\ref{eq_tau_corr}. The superscripts to sources names indicate the sources, from which we obtained the measured time-delay (first source) and the monochromatic luminosity at 3000 \AA (second source): 1) \citet{2016ApJ...818...30S}, 2) \citet{2019ApJS..241...34S}, 3) \citet{lira2018}, 4)  NED, NUV, GALEX, 5) \citet{metzroth2006},  6) \citet{1982ApJ...256....1C}, 7) \citet{czerny2019}, 8) this work, 9) a script of \citet{2010ApJ...708..927K}.}
       \resizebox{2.2\columnwidth}{!}{
    \begin{tabular}{c|c|c|c|c|c|c|c}
    \hline
    \hline 
     Source & $z$ & $\tau$ [days] & $\log{(L_{3000} [{\rm erg\,s^{-1}}])}$ & FWHM(MgII) [$\rm km\,s^{-1}$] & $\dot{M}$ & $\Delta \tau$ & $\tau_{\rm corr}$ [days]\\
     \hline
      141214.20+532546.7$^{1,2}$ &  0.45810 & $36.7^{+10.4}_{-4.8}$ &   $44.63882 \pm 0.00043$ & $2391 \pm 46$ & $8.21^{+6.94}_{-5.58}$ & $-0.28^{+0.13}_{-0.07}$ &  $80.7^{+22.9}_{10.6}$ \\
      141018.04+532937.5$^{1,2}$ &  0.46960 & $32.3^{+12.9}_{-5.3}$ &  $43.7288 \pm 0.0051$  & $3101 \pm 76$ & $0.30^{+0.31}_{-0.21}$ & $0.20^{+0.18}_{-0.08}$ & $27.1^{+10.8}_{-4.4}$  \\ 
      141417.13+515722.6$^{1,2}$ &  0.60370 & $29.1^{+3.6}_{-8.8}$ &  $43.6874 \pm 0.0029$ & $3874 \pm 86$ & $0.23^{+0.15}_{-0.20}$ & $0.18^{+0.06}_{-0.14}$ & $22.4^{+2.8}_{6.8}$ \\ 
      142049.28+521053.3$^{1,2}$ &  0.75100 & $34.0^{+6.7}_{-12.0}$ &  $44.6909 \pm 0.0009$ & $4108 \pm 39$ & $4.87^{+3.57}_{-4.57}$ & $-0.35^{+0.09}_{-0.16}$ & $64.2^{+12.6}_{-22.6}$ \\ 
      141650.93+535157.0$^{1,2}$ &  0.52660 & $25.1^{+2.0}_{-2.6}$ &  $43.778 \pm 0.002$ & $4066 \pm 202$ & $0.39^{+0.26}_{-0.27}$ & $0.06^{+0.05}_{-0.06}$ & $22.6^{+1.8}_{-2.3}$ \\ 
      141644.17+532556.1$^{1,2}$  &  0.42530 & $17.2^{+2.7}_{-2.7}$ &  $43.9480 \pm 0.0011$ & $2681 \pm 96$ & $2.87^{+2.03}_{-2.03}$ & $-0.20^{+0.07}_{-0.07}$ & $27.8^{+4.4}_{-4.4}$ \\ 
      CTS252$^{3,4}$ &  1.89000 & $190.0^{+59.0}_{-114.0}$ & $46.79 \pm 0.09$ & $3800 \pm 380$ & $251.55^{+261.03}_{-367.22}$ & $-0.84^{+0.18}_{-0.28}$ & $1136.2^{+353.1}_{-682.0}$ \\ 
      NGC4151$^{5,6}$ &  0.00332 & $6.8^{+1.7}_{-2.1}$ &  $42.83 \pm 0.18$ & $4823 \pm 1105$ & $0.15^{+0.23}_{-0.23}$ & $0.05^{+0.15}_{-0.17}$ & $4.7^{+1.2}_{-1.4}$ \\ 
      NGC4151$^{5,6}$ &  0.00332 & $5.3^{+1.9}_{-1.8}$ & $42.83 \pm 0.18$ & $6558 \pm 1850$ & $0.16^{+0.28}_{-0.27}$ & $-0.06^{0.19}_{-0.18}$ & $3.7^{+1.3}_{-1.2}$ \\ 
     CTS C30.10$^{7,7}$  &  0.90052 & $564^{+109}_{-71}$ &  $46.023 \pm  0.026$ & $5009 \pm 325$ & $1.29^{+1.05}_{-0.98}$ & $+0.09^{+0.11}_{-0.09}$ &  $721.4^{+139.4}_{-90.8}$ \\
     HE 0413-4031$^{8,9}$ &  1.37648 & $302.6^{+28.7}_{-33.1}$ & $46.754 \pm 0.080$ & $4380 \pm 14$ & $66.04^{+44.17}_{-44.76}$ & $-0.61^{+0.11}_{-0.11}$ & $1223.9^{+116.4}_{-134.1}$ \\
     \hline
    \end{tabular}}
    \label{tab_sources_mgII}
\end{table*}

We also fitted the general radius-luminosity relationship $\log{(\tau/1\,\text{lt. day})}=K+\alpha\log{(L_{3000}/10^{44}\,{\rm erg\,s^{-1}})}$ to all available data. We obtained the best-fit parameters of $K=1.45 \pm 0.08$ and $\alpha=0.42 \pm 0.05$ with $\chi^2=76.6$ and $\chi^2_{\rm red}=8.5$. Subsequently, we removed 2 outliers -- CTS252 and HE 0413-4031 --  that are significantly below RL relations in Eq.~\ref{eq_tau_luminosity_vestergaard_osmer} and Eq.~\ref{eq_tau_luminosity_bentz}. This helped to improve the fit, with the best-fit with $\chi^2=36.7$ and $\chi^2_{\rm red}=5.2$, and the final relation based on MgII data can be expressed as,
\begin{equation}
    \log{\left[\frac{\tau(\text{MgII})}{1 {\rm lt. day}}\right]}=(1.46 \pm 0.06) + (0.58 \pm 0.07)\log{\left(\frac{L_{3000}}{10^{44}\,{\rm erg\,s^{-1}}}\right)}\,.
    \label{eq_tau_luminosity_our data}
\end{equation}
Both best-fit relations are depicted in Fig.~\ref{fig_radius_luminosity_mgII} with solid lines. 

The time-delay offset can be explained by the higher accretion rate implied by the super-Eddington luminosity. The correlation of the time-delay offset with the accretion rate was shown previously for reverberation-mapped sources in H$\beta$ line \citep{2018ApJ...856....6D,Mart_nez_Aldama_2019} and we will demonstrate it in the following section for broad MgII line. By moving the source HE 0413-4031 back onto the radius-luminosity relation, we can estimate the corrected black hole mass using  \citep{2009ApJ...699..800V},
\begin{equation}
    M_{\bullet}^{MgII}=10^{zp(\lambda)}\left[\frac{\text{FWHM(MgII)}}{1000\,{\rm km\,s^{-1}}} \right]^2 \left(\frac{\lambda L_{\lambda}}{10^{44}\,{\rm erg\,s^{-1}}} \right)^{0.5}\,,
    \label{eq_mbh_mgii}
\end{equation}
which for FWHM(MgII)$=4380^{+14}_{-15}$ km s$^{-1}$ and $zp(\lambda)=6.86$ ($\lambda=3000\,\AA$) yields $M_{\bullet}^{MgII}=3.31\times 10^9\,M_{\odot}$ using the monochromatic luminosity of $\log{L_{3000}}=46.754$. 

Hence, the black hole mass obtained from the radius-luminosity relation is larger by a factor of at least $\sim 3$ than the maximum mass inferred from the RM time-delay, taking into account the uncertainty in the virial factor. The Eddington ratio for the higher mass then drops to $\eta_{\rm Edd}=0.77$ for the constant bolometric correction factor of BC$=5.62$ \citep{2006ApJS..166..470R}. For the more precise luminosity-dependent bolometric correction of BC$=25\times (L_{3000}/10^{42}\,{\rm erg\,s^{-1}})^{-0.2}=2.80$ for $L_{3000}=10^{46.754}\,{\rm erg\,s^{-1}}$ \citep{2019MNRAS.488.5185N}, we get even smaller Eddington ratio of $\eta_{\rm Edd}=0.38$, which is also consistent with the SED fitting presented in Section~\ref{sec_label_SED}.

\subsection{Correction of the accretion-rate effect along the radius-luminosity relation}

 In the optical range, it has been observed that the accretion rate is responsible for the departure of the radius--luminosity relation more than the intrinsic scatter. \citet[][and references therein]{2018ApJ...856....6D} showed that the sources with the highest accretion rates have time delays shorter than the expected from the optical radius-luminosity relation. However, the accretion rate effect can be corrected, recovering the standard results \citep{Mart_nez_Aldama_2019}. Following this idea, we repeated the same exercise for all MgII reverberation-mapped data (see Table~\ref{tab_sources_mgII}).

The black hole mass was estimated assuming a virial factor anticorrelated with the FWHM of the emission line (Eq.~\ref{eq_virial_factor}), which apparently corrects the orientation effect to a certain extent. Since the Eddington ratio has shown a large scatter in comparison with other expressions of the accretion rate \citep{Mart_nez_Aldama_2019}, we will use the dimensionless accretion rate \citep{du2016},
\begin{equation} \label{equ:mdot}
\dot{M}=20.1\left(\frac{l_{44}}{\cos{\theta}}\right)^{3/2} m_{7}^{-2},
\end{equation}
where $\it{l}\mathrm{_{44}}$ is the luminosity at 3000 \AA\ in units of 10$^{44}\,{\rm erg\,s^{-1}}$, $\theta$=0.75 is inclination angle of disk to the line of sight, and $m_{7}$ is the black hole mass in units of 10$^7$ $M_\odot$. In Figure~\ref{fig_radius_luminosity_mgII} (left panel), we show the variation of the dimensionless accretion rate along the radius--luminosity relation, which is similar to the observed one in the optical range.

\begin{figure}
    \centering
    \includegraphics[width=0.5\textwidth]{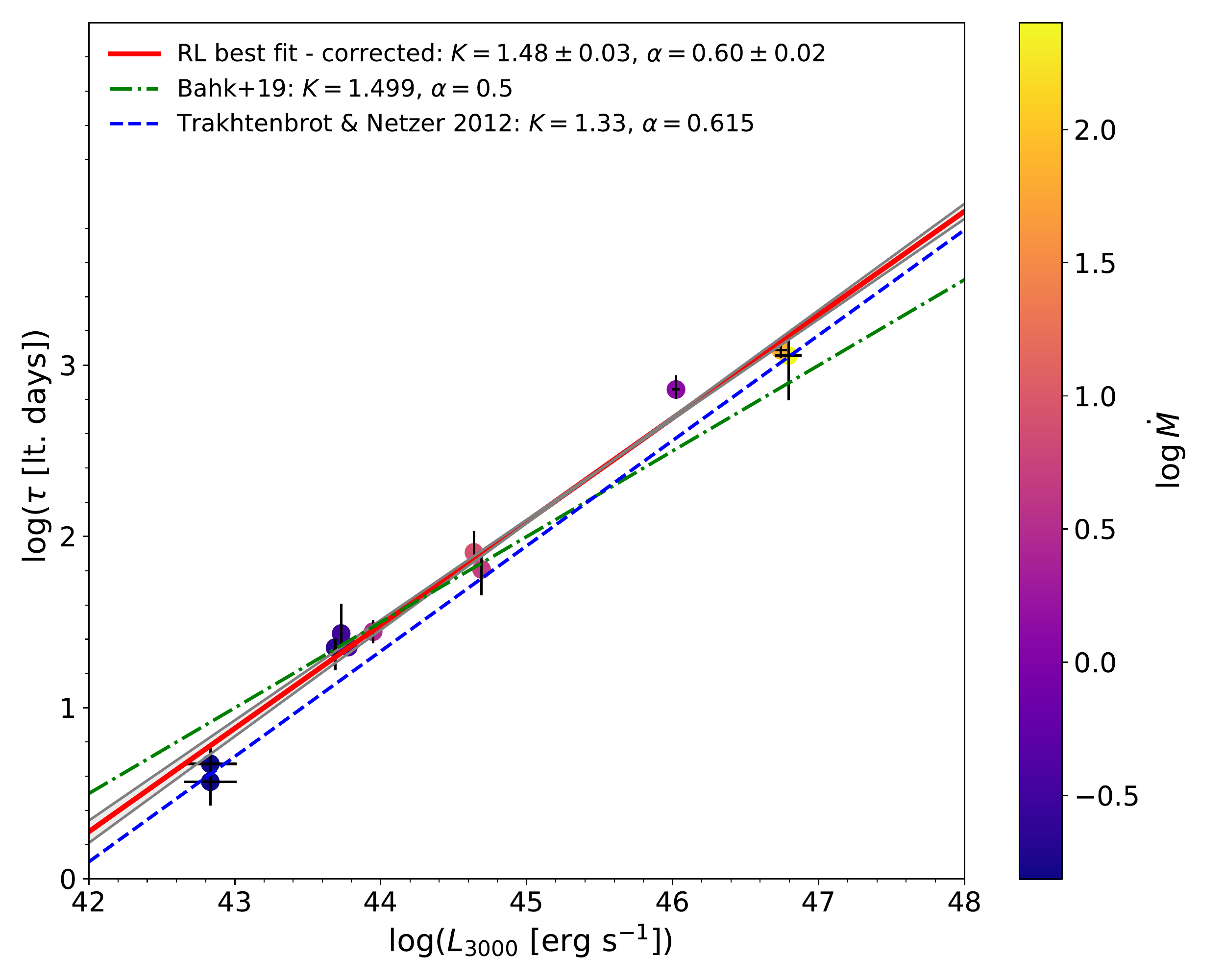}
    \caption{The radius-luminosity relation for MgII broad emission line; the rest-frame time delays were corrected for the accretion rate effect. The best-fit linear relation in the log-space is: $\log{(\tau/1\,\text{lt. day})}=(1.48 \pm 0.03)+(0.60 \pm 0.02)\log{(L_{3000}/10^{44}\,{\rm erg\,s^{-1}})}$, with $\chi^2=11.40$ and $\chi^2_{\rm red}=1.27$. For comparison, we also show the radius-luminosity relation as inferred by \citet{2019ApJ...875...50B}, which has the same normalization factor as our relation, but a smaller slope of $0.5$. In addition, we depict the radius-luminosity relation constructed by \citet{2012MNRAS.427.3081T}, which, on the other hand, has a comparable slope of $\alpha=0.615$, but a smaller normalization factor of $K=1.33$.}
    \label{fig_radius_luminosity_MgII_corr}
\end{figure}

To estimate the departure from the radius--luminosity relation, we use the parameter $\Delta \tau$, which is simply the difference between the observed time delay and the expected one from the radius--luminosity relation,
\begin{equation}
\Delta \tau=\mathrm{log}\,\left(\frac{\tau\mathrm{_{obs}}}{\tau\mathrm{_{\,R-L}}}\right)\,.
\label{eq_departure_parameter}
\end{equation}
We estimate $\tau\mathrm{_{\,R-L}}$ from the radius--luminosity relation described in Eq.~\ref{eq_tau_luminosity_our data}. Values are reported in Table~\ref{tab_sources_mgII}. The largest departure from the radius--luminosity relation is associated with the highest accretion rate sources, which is clearly evidenced in Figure~\ref{fig_radius_luminosity_mgII} (right panel). The Pearson coefficient ($p=-0.940$) also indicates a strong anticorrelation between $\Delta \tau$ and $\dot{M}$. Performing a linear fit, we get the relation 
\begin{equation}
\Delta \tau (\dot{M})= (- 0.297 \pm 0.033) \, \log{\dot{M}} + (-0.049\pm0.026)\,,    \label{equ_linearfit}
\end{equation}
for which $\chi^2=6.75$ and $\chi^2_{\rm red}=0.75$. This expression can be used to recover the expected values from the radius-luminosity relation using the relation,
\begin{equation}
\tau_{{\rm corr}}(\dot{M})\,=\,10^{-\Delta \tau{( \dot{M}})} \, \cdot \tau_{\mathrm{obs}}.
\label{eq_tau_corr}
\end{equation}
The corrected rest-frame time-delays are listed in Table~\ref{tab_sources_mgII}. Based on them, we construct a new version of the radius-luminosity relation for MgII line corrected for the accretion-rate affect (see Fig~\ref{fig_radius_luminosity_MgII_corr}). It shows a smaller scatter of $\sigma=0.104$ dex in comparison with the radius--luminosity relation before the correction, which is $\sigma=0.221$ dex when all the sources are included and $\sigma=0.186$ dex with two outliers removed. The best-fit linear relation has smaller uncertainties with $\chi^2=11.40$ and $\chi^2_{\rm red}=1.27$ and can be expressed as,
\begin{equation}
    \log{\left[\frac{\tau(\text{MgII})}{1 {\rm lt. day}}\right]}=(1.48 \pm 0.03) + (0.60 \pm 0.02)\log{\left(\frac{L_{3000}}{10^{44}\,{\rm erg\,s^{-1}}}\right)}\,.
    \label{eq_tau_luminosity_our data_corr}
\end{equation}

\textcolor{black}{The dispersion around the new relation is very small, equal to $0.104$ dex. This is smaller than the dispersion of 0.13 dex in the original radius-luminosity relation of \citet{bentz2013} after an artificial removal of outliers, despite the fact that the MgII relation covers a broad range of the luminosities, redshifts as well as Eddington ratios. It is not clear at this point whether the smaller dispersion is a property of the MgII emission or it just results from the fact that the MgII data does not come from so many different monitoring campaigns.}

The normalization coefficient in Eq.~\ref{eq_tau_luminosity_our data_corr} is within uncertainties consistent with the normalization factor inferred from the MgII-based black hole mass estimator by \citet{2019ApJ...875...50B},
\begin{equation}
    \log{\left[\frac{\tau(\text{MgII})}{1 {\rm lt. day}}\right]}\simeq 1.499 +0.5\log{\left(\frac{L_{3000}}{10^{44}\,{\rm erg\,s^{-1}}}\right)}\,,
    \label{eq_tau_luminosity_bahk19}
\end{equation}
where we adopted their fitting Scheme 4 and assumed the virial factor $f_{\rm vir}\equiv 1$ while transforming $M_{\bullet}\propto L_{3000}^{0.5}(\Delta V)^2$ relation to $\tau\propto L_{3000}^{0.5}$ relation. The relation~\ref{eq_tau_luminosity_bahk19} is also shown in Fig.~\ref{fig_radius_luminosity_MgII_corr} for comparison with our best-fit relation~\ref{eq_tau_luminosity_our data_corr}. 

However, the best-fit slope is larger than the slope of $0.5$ in Eq.~\ref{eq_tau_luminosity_bahk19}, which is also expected from the simple photoionization arguments. Currently, this may be just a systematic effect due to a small number of reverberation-mapped sources using MgII line. The larger slope currently yields a significantly small scatter, since for the relation given by Eq.~\ref{eq_tau_luminosity_bahk19} the scatter is $\sigma=0.187$ dex for the corrected time delays. For the uncorrected rest-frame time-delays, the scatter is $\sigma=0.247$ dex and $\sigma=0.189$ for the whole MgII sample (sources in Table~\ref{tab_sources_mgII}) and the MgII sample without the two outliers (CTS252 and HE 0413-4031), respectively.

On the other hand, our best fit slope is very similar to the value of $\alpha=0.615$ inferred by \citet{2012MNRAS.427.3081T}, who, on the other hand, have a smaller normalization factor $K=1.33$. Our slope value is also located between the slopes derived for the $\tau({\rm MgII})$--$L_{3000}$ relation by \citet{2002MNRAS.337..109M} ($\alpha=0.47$) and by \citet{2004MNRAS.352.1390M} ($\alpha=0.62$). However, all of the above-mentioned MgII-based radius-luminosity relations were calibrated based on the UV spectra of sources for which only H$\beta$ line reverberation mapping was performed. Certainly, more reverberation-mapped sources using MgII line are required to further constrain the $\tau({\rm MgII})$--$L_{3000}$ relation.

\section{Discussion} \label{sec:discussion}

Using the SALT data and the supplementary photometric monitoring, we were able to derive the time delay of the MgII line with respect to the continuum in $z = 1.37648$ quasar HE 0413-4031. The source is very bright in the absolute term, but the delay is formally established as $\overline{\tau}=302.6^{+28.7}_{-33.1}$ days in the comoving frame. Although the analysis of the MgII complex with the underlying power-law continuum and FeII pseudo-continuum emission is a complex task with a certain degree of degeneracy, we showed that the peak value of the time-delay distribution is not sensitive to different FeII templates, only its uncertainty may be affected due to a different number of parameters used in each model, see also Appendix~\ref{sec_appendix_FeII_templates} for a detailed discussion.

This delay is shorter than derived for CTS C30.10 \citep{czerny2019}, but similar to the delay measured for another bright quasar by \citet{lira2018}. We show that the dispersion in the measured time delay of MgII line for a given range of the monochromatic flux is related to the Eddington ratio in the source, as in H$\beta$ time delay \citep{Mart_nez_Aldama_2019}, and with the appropriate correction for this effect, the dispersion around the radius-luminosity is actually very small with $\sigma=0.104$ dex in comparison with $\sigma=0.221$ dex before the correction (when all the sources are included; $\sigma=0.186$ dex with two outliers removed), which opens up a possibility for the future applications of this relation for cosmology.

In this Section, we discuss more generally the validity and the accuracy of using MgII lines in black hole mass determination. Furthermore, we show that the intrinsic Baldwin effect is present in our source, which is another way of showing that MgII line responds to the thermal AGN continuum. To verify if the reverberating MgII line in our source is a reliable probe of its black hole mass, we performed a fit of the accretion disk model to the optical and UV continuum data of the source SED.

\subsection{Nature of MgII emission}

\citet{2013A&A...555A..89M} showed that the FWHM of MgII line is systematically narrower by $\sim 20\%$ than H$\beta$ line, which holds for all of its components as well as the full profile. The simple explanation is that MgII is emitted at larger distances than H$\beta$ from the photoionizing continuum source.
The intrinsically symmetric profile of MgII line found in this work characterized by a one-component Lorentzian is consistent with the origin of the MgII emission in the virialized BLR clouds as for H$\beta$ broad line \citep{2013A&A...555A..89M}. The Lorentzian profile may be physically explained by the turbulent motion of the emitting medium and the line broadening by its rotation \citep{2011Natur.470..366K,2012MNRAS.426.3086G,2013A&A...549A.100K,2013A&A...558A..26K}. This model of the Lorentzian line profile is also consistent with the FRADO model as such \citep{2011A&A...525L...8C}, in which the turbulence arises due to the failed outflow and the subsequent inflow and the rotation is represented by the dominant Keplerian field (see also Fig.~\ref{fig_HE0413_illustration}). For Population A sources where the MgII profile is symmetric, the MgII gas may be considered as virialized. For Population B sources, a small degree of asymmetry and the blueshift of MgII line may be related to outflows of the MgII-emitting gas \citep{2013A&A...555A..89M}.

Our results, in particular the studied intrinsic Baldwin effect in Subsection~\ref{subsec_response_mgII}, are also consistent with the work of \citet{2020MNRAS.493.5773Y}, who found for the sample of 33 extreme variability quasars that the MgII flux density responds to the variable continuum, however with a smaller amplitude. However, they also stress that the FWHM of the MgII line does not respond to the continuum as the Balmer lines do. Therefore, black hole mass estimations based on single-epoch measurements can be luminosity-biased.

Previous works also find an overall consistency between H$\beta$-based and MgII-based black hole mass estimators. \citet{2012MNRAS.427.3081T} found the scatter between these two spectral regions of $0.32$ dex in terms of the black hole mass estimation, smaller than for CIV line, for which the scatter with respect to H$\beta$ is $0.5$ dex. In addition, the same authors found FWHM(MgII)$\simeq$ FWHM(H$\beta$) up to $6000\,{\rm km\,s^{-1}}$, beyond which the FWHM of MgII seems to saturate. This is again different for CIV line, which does not show any correlations with either H$\beta$ or MgII line. Also, the FWHM(CIV)$\lesssim$ FWHM(H$\beta$) for nearly half of the studied sources \citep[see also][for a similar result]{2012ApJ...753..125S}, which contradicts reverberation mapping results. \citet{2012ApJ...754...11H} also showed that MgII-based black hole masses are comparable within uncertainties to those based on H$\alpha$, while CIV-based mass estimates differed by as much as a factor of 5. Hence, the usage of broad MgII line for black hole mass estimating is justified for sufficiently large samples, while CIV should not be applied as a reliable virial black hole mass estimator. This is in line with the overall picture where low-ionization lines (H$\alpha$, H$\beta$, MgII) originate in the bound line-emitting, photoionized clouds, which high-ionization lines (CIV) originate in the unbound outflowing gas \citep{collin1988}. 

For the $\gamma$-ray blazar 3C 454.3, \citet{2013ApJ...763L..36L} found a significant correlation between the increase in the MgII flux density and the $\gamma$-ray flaring emission (in autumn 2010), which could be related to the superluminal radio component in this source. This implies that MgII-emitting gas responds to the non-thermal continuum alongside the thermal continuum of the accretion disk. This is also in agreement with the significant correlation between the MgII flux density and the $\gamma$-ray flux increase in the blazar CTA102 \citep{2020ApJ...891...68C}, in which the superluminal radio component was also present. In addition, the MgII broad line was broader and blueshifted at the maximum of the $\gamma$-ray activity in comparison to the minimum. The BLR material in this source was inferred to be located $\sim 25\,{\rm pc}$ from the central source. \citet{2020ApJ...891...68C} conclude that the black hole mass estimation using MgII is only reliable for the sources in which UV continuum is dominated by the central accretion disk, which is also the case for our source HE0413 as we show in Subsection~\ref{sec_label_SED} based on the SED fitting based on the thermal disc emission.

In summary, based on our and previous findings of other authors, a significant fraction of the MgII emitting gas is virialized and reverberating to the variable thermal continuum as we also find in this work. For sources with a significant non-thermal emission due to the jet in the UV and the optical domain, outflowing gas at larger distances from the standard BLR region can respond to the non-thermal continuum and this contributes to the broadening and a blueshifting of the MgII line. Hence, when using MgII line in the reverberation studies, time-delay analysis should be complemented by SED modelling whenever possible to verify if photoionizing continuum is dominantly of thermal nature. 

In terms of the quasar main sequence and the four-dimensional Eigenvector 1 \citep[4DE1, ][]{sulentic2000,marziani2018}, considering the equivalent width ($27.45^{+0.12}_{-0.10}$ \AA) and the FWHM ($4380^{+14}_{-15}\,{\rm km\,s^{-1}}$) exhibited by the MgII line, HE~0413-4031 could be cataloged as a Population B1 in the 4DE1 scheme \citep[Table 2 of][]{2004ApJ...617..171B}. However, HE~0413-4031 shows a clear single-component Lorentzian profile associated with Population A sources (Sec.~\ref{sec:spec_fit}). According to the analysis presented in Appendix~\ref{sec_appendix_FeII_templates} using a different model template for FeII emission, the MgII emission could also be modelled with two kinematic components, although their nature appears to be more problematic to interpret. Moreover, the FWHM of the two Gaussian components and their relative shift with respect to the FeII emission depend strongly on the source redshift in the studied interval of $z \simeq 1.37-1.39$, see our analysis in Appendix~\ref{sec_appendix_FeII_templates}, especially Fig.~\ref{fig:mean_Serbia}.

As a high luminosity source, HE0413-4031 can be found in the population B spectral bins, being still a Population A source \citep{marziani2018}. Because of its large Eddington ratio of $\sim 0.4$, it can be further classified as an extreme Population A source (xA), with the FeII$\lambda 4570$ strength larger than unity $R_{FeII}>1$ with the FWHM(H$\beta$)$>4000\,{\rm km\,s^{-1}}$ since MgII line is generally narrower than $H\beta$ line. The difficult spectral-type classification of HE~0413-4031 stems from the fact Population A sources are typically highly accreting sources with smaller black hole masses and Population B sources have larger black hole masses and low Eddington ratios \citep{2009ApJ...698L.103M,2017FrASS...4....1F}. In this sense, HE~0413-4031 has mixed properties: a large black holes mass of a few $10^9\,M_{\odot}$ and a high Eddington ration of $\sim 0.4$. However, these general distinctions are based on the analyses of lower-luminosity low-redshift sources, while our source is at the intermediate redshift of $z\sim 1.4$ and of a high luminosity of $10^{47}\,{\rm erg\,s^{-1}}$, hence the apparent discrepancy may be solved by the cosmological argumentation that the current massive black holes with low accretion rates were highly accreting sources at higher redshifts. With a black hole mass of a few billion Solar masses, HE0413 falls into the expected mass range for type 1 AGN between redshifts of 1 and 2 \citep[see Fig. 15 of][where HE0413 is located at the age of the Universe of $4.66$ Gyr for z=1.37]{2012MNRAS.427.3081T}. On the other hand, HE0413 is still an outlier in terms of the accretion rate close to the Eddington limit for a black hole mass of a few billion Solar masses. \citet{2012MNRAS.427.3081T} suggest that a majority of such massive AGN do not accrete close to their Eddington limits even at $z\simeq 2$.    

Since HE 0413-4031 can be classified as a radio-loud AGN given its luminosity at 1.4 GHz, $L_{1.4}\approx 2.5 \times 10^{26}\,{\rm W\,Hz^{-1}}>10^{24}\,{\rm W\,Hz^{-1}}$ \citep{2016A&ARv..24...10T}, its radio-optical properties can be studied in the broader context. \citet{Ganci2019} studied the radio properties of type-1 AGNs across all main spectral types along the quasar main sequence, in particular for three classes of Kellermann's radio-loudness criterion, which is defined as the ratio of the radio and optical flux densities, $R_{\rm K}=S_{\rm radio}/S_{\rm optical}$. We follow \citet{Ganci2019}, who use 1.4 GHz flux density for $S_{\rm radio}$ and $g$-band flux density for $S_{\rm optical}$ and divide sources into three radio classes: radio detected (RD, $R_{\rm K}<10$), radio intermediate (RI, $10\leq R_{\rm K}<70$), and radio loud (RL, $R_{\rm K}\geq 70$). We derive the corresponding 1.4 GHz and $g$-band flux densities for HE 0413-4031 by linear interpolation of the averaged SED points in the log-space (see Fig.~\ref{fig:SED}), $S_{1.4}=21.04\,{\rm mJy}$ and $S_{g}=0.434\,{\rm mJy}$, which yields $R_{\rm K}=48.5$. Hence, HE 0413-4031 can be classified as a radio-intermediate source with an inverted and flat radio-spectrum towards higher frequencies according to Vizier SED\footnote{\url{http://vizier.u-strasbg.fr/vizier/sed}}, since the spectral index $\alpha$, using the notation $S_{\nu}\propto \nu^{+\alpha}$, is $\alpha_{1-5} \simeq 0.7$, $\alpha_{5-8} \simeq 1.7$, $\alpha_{8-20} \simeq -0.02$ betweeen 0.843 GHz, 5 GHz, 8 GHz, and 20 GHz, respectively. Sources with inverted to flat radio spectral indices are characterized by a compact, optically thick radio core or a core-jet system \citep{2019A&A...630A..83Z,2019arXiv191112901Z}.

\citet{Ganci2019} found that the occurrence of RD, RI, and RL sources differs along the main sequence. The classification of our source as RI with inverted-flat spectrum is consistent with its location in the extreme A population according to \citet{Ganci2019} since core-dominated sources in A3 and A4 bins are mostly RI. The source of radio emission in extreme A population can be partially due to a high star-formation rate, but also the core-jet activity. In our case, the radio spectral index implies the presence of a compact core-jet system, hence the high star-formation rate is not necessarily required. On the other hand, the presence of gas material is necessary to account for the high Eddington ratio of $\sim 0.4-0.5$. The optically thick radio core could be a sign of a restarted AGN activity \citep{2009ApJ...698..840C,2017A&ARv..25....2P}, which will eventually heat up the cold gas content and/or blow it away and slow down the star-formation.

\subsection{Response of MgII emission to continuum changes - intrinsic Baldwin effect}
\label{subsec_response_mgII}

The expected properties of the MgII line were recently modeled by \citet{guo2019}, where the authors using the CLOUDY code and the Locally Optimally Emitting Cloud (LOC) scenario showed that at the high Eddington ratio of $\sim 0.4$ the MgII line flux saturates and does not further increase with the rise of the continuum.

\begin{figure*}
    \centering
    \includegraphics[width=0.48\textwidth]{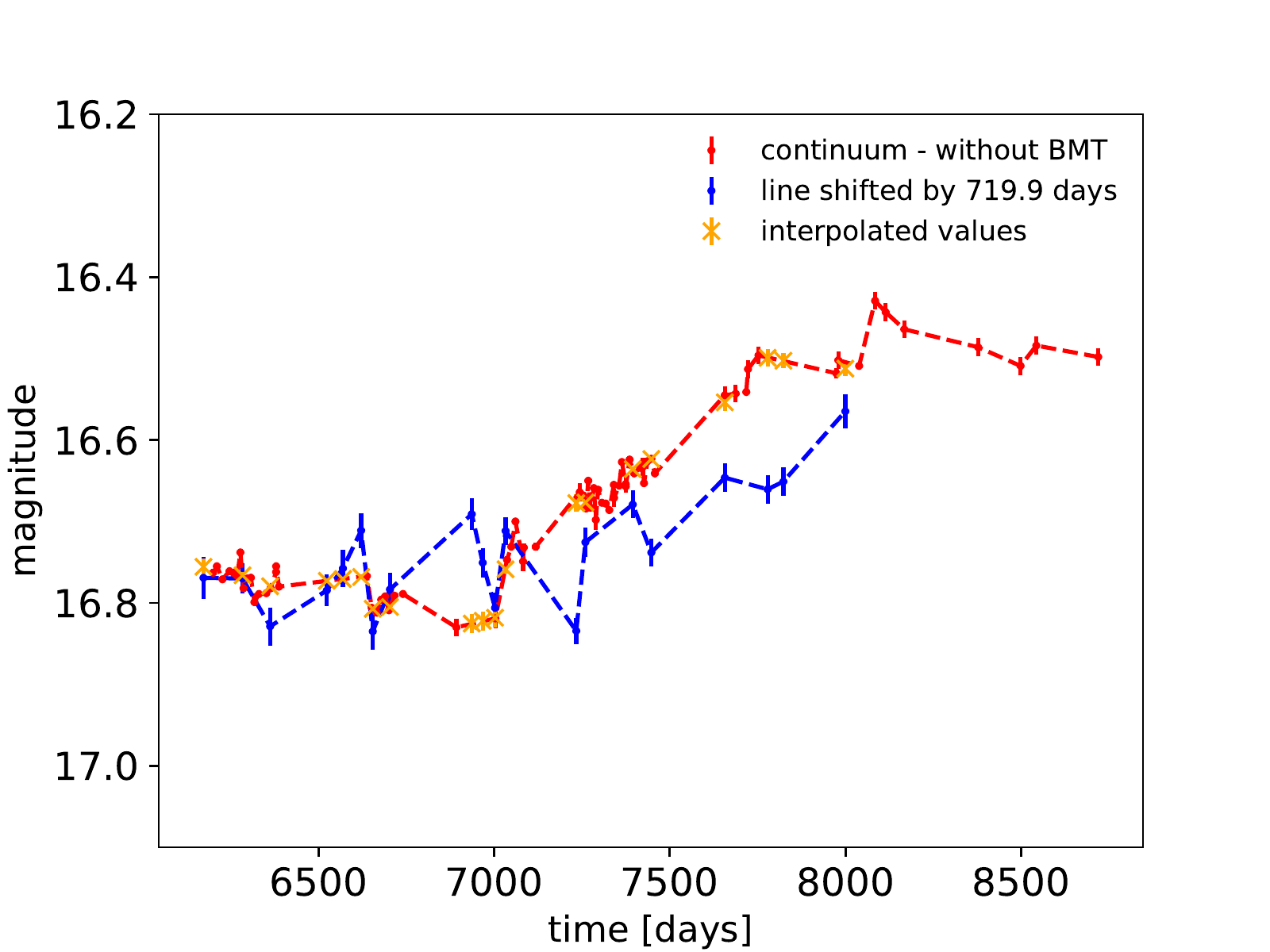}
    \includegraphics[width=0.48\textwidth]{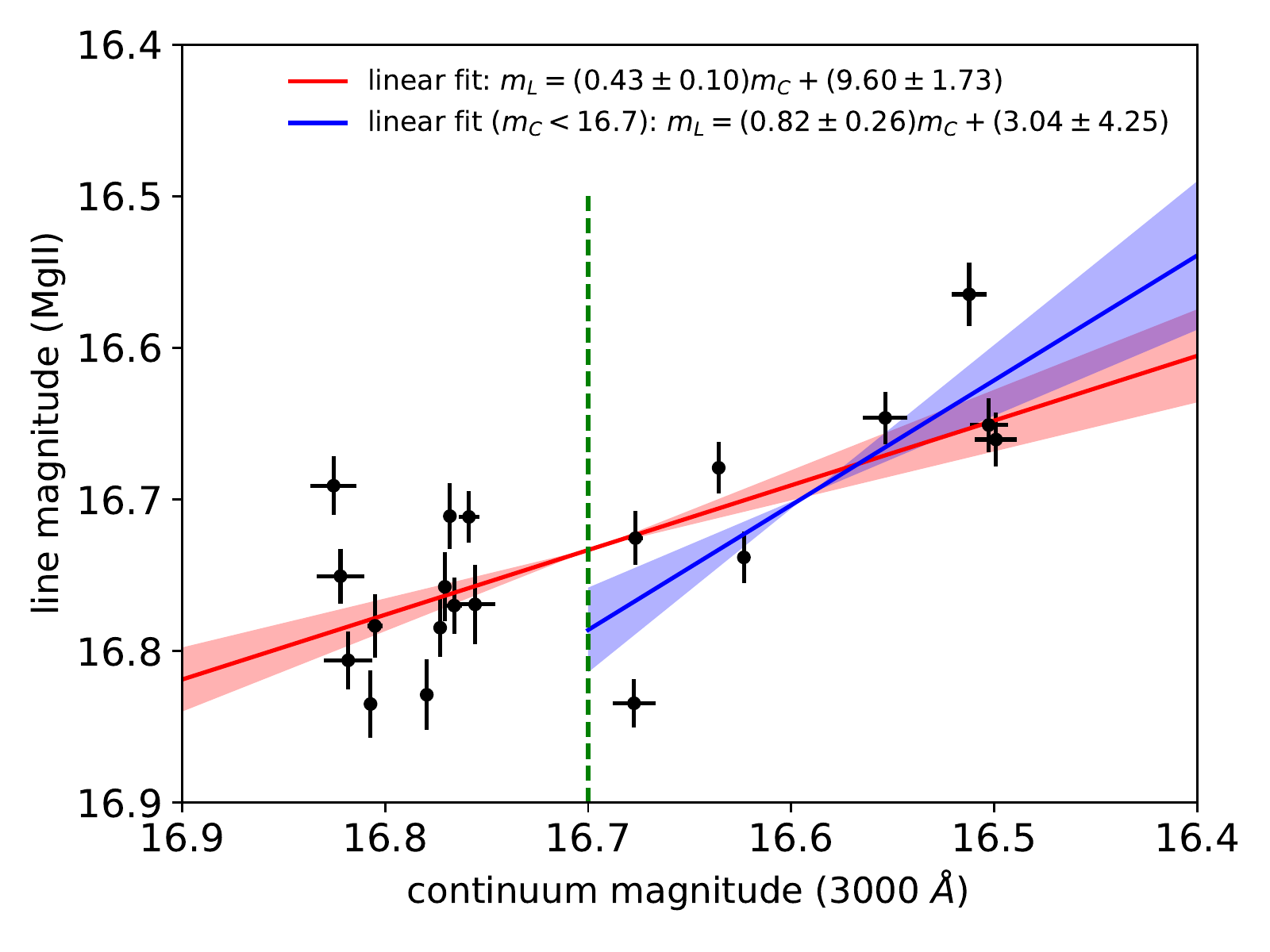}
    \caption{Determination of the continuum--line magnitude (luminosity) relation. \textbf{Left panel:} Superposition of the continuum and the time-shifted line-emission light curves. Interpolated photometry points are also shown. \textbf{Right panel:} MgII line emission magnitude versus continuum magnitude has a clear linear correlation in the logarithmic scale with the correlation coefficient of $r=0.73$ ($R^2=0.53$) with the best-fit relation of $m_{\rm l}=(0.43 \pm 0.10)m_{\rm c}+(9.60 \pm 1.73)$. The blue line represents the linear fit to the data with the continuum magnitude less than $16.7$. In this case, the correlation coefficient is higher ($r=0.79$, $R^2=0.63$) and the best-fit relation has a larger slope:  $m_{\rm l}=(0.82 \pm 0.26)m_{\rm c}+(3.04 \pm 4.25)$.}
   \label{fig_mag_line_powerlaw}
\end{figure*}

We confront this theoretical prediction with our observations of the quasar HE 0413-4031. We used the logarithm of both the continuum and MgII line-emission flux densities, i.e. magnitudes. Subsequently, we applied the determined time-delay shift to the line emission, i.e. we shifted the MgII light curve by $719.9$ days in the observer's frame. For the continuum light curve, we tried both the cases with and without BMT data, by given the fact that the BMT data are present for the epochs longer than 8000 days, they do not have a significant effect on the following analysis. As the next step, we interpolate the photometry data to the time-shifted line-emission data to have corresponding line-continuum pairs. As before, given that the photometry data come from different instruments with various uncertainties, we make use of the weighted least-squares linear B-spline interpolation with the inverse of uncertainties as weights. We show the the continuum and the time-shifted line light curves in Fig.~\ref{fig_mag_line_powerlaw} (left panel) alongside the interpolate values, which can also serve as a cross-check that the determined time-delay of $\sim 720$ days in the observer's frame represents the realistic similarity between the shapes of both light curves.    

Finally, we plot the MgII line magnitude with respect to the continuum magnitude in Fig.~\ref{fig_mag_line_powerlaw} (right panel). This relation has a significant correlation with the correlation coefficient of $r=0.73$. The best-fit linear relation is $m_{\rm l}=(0.43 \pm 0.10)m_{\rm c}+(9.60 \pm 1.73)$, which is displayed in Fig.~\ref{fig_mag_line_powerlaw} with the corresponding uncertainties. Our linear fit implies directly the power-law relation between the MgII and continuum luminosities, $L_{\rm MgII}\propto L_{3000}^{0.43 \pm 0.10}$. In combination with the measured time-delay of 303 days in the rest frame, we can conclude that the MgII line responds to the continuum variability even for the source which is highly-accreting with the Eddington ratio of $\sim 0.4$ (see also Subsection~\ref{sec_label_SED} for a detailed SED modelling). Hence, our source does not exhibit a non-responsive MgII line with a rather constant dependency on the continuum luminosity, as was analysed and shown by \citet{guo2019} (see also their Fig. 4). Moreover, from Fig.~\ref{fig_mag_line_powerlaw} (right panel) it is apparent that the line and the continuum magnitudes consist of an uncorrelated part for continuum magnitudes of more than $16.7$ mag (with the correlation coefficient of $r=0.10$, $R^2=0.01$). The part of the dependency with continuum magnitudes less than $16.7$ mag is strongly correlated with the correlation coefficient of $r=0.79$ ($R^2=0.63$) and the best-fit linear fit is $m_{\rm l}=(0.82 \pm 0.26)m_{\rm c}+(3.04 \pm 4.25)$, hence the line luminosity responds even stronger to the continuum luminosity in this part with the relation $L_{\rm MgII}\propto L_{3000}^{0.82 \pm 0.26}$, which is marginally consistent with the linear dependency within the uncertainty. 

\citet{guo2019} analyse the LOC model and MgII response for the smaller black hole mass and the 3000\AA\, luminosity ($M_{\bullet}=10^8\,M_{\odot}$ and $L_{3000}=10^{44-45}\,{\rm erg\,s^{-1}}$). However, their upper limit for the Eddington ratio, $\eta_{\rm Edd}=0.4$, is comparable to our estimated Eddington ratio and hence their flattening of MgII luminosity close to $L_{3000}=10^{45}\,{\rm erg\,s^{-1}}$ is not confirmed for HE 0413-4031. On the other hand, we observe a similar dependency of the MgII line luminosity on the continuum luminosity as \citet{guo2019} inferred for hydrogen recombination broad lines (H$\alpha$, H$\beta$) at lower Eddington ratios. For the luminosity range $\log{[L_{3000}\,(\rm erg\,s^{-1})]}=42-44$, the slope for H$\beta$ is $\alpha\sim 0.45$ and $\alpha\sim 0.42$ for H$\alpha$. In addition, \citet{guo2019} show a slower rise of MgII luminosity with respect to the continuum with the slope of $\sim 0.38$, which is smaller than our value. This implies that at least for our source, the LOC model with the initial assumption of $(R_{\rm out},\Gamma)=(10^{17.5},-2)$\footnote{Locally Optimally Emitting Cloud (LOC) models assume the power-law radial distribution of clouds, $f(r)\propto r^{\Gamma}$.}  does not apply. 

\begin{figure}[tbh]
    \centering
    \includegraphics[width=0.5\textwidth]{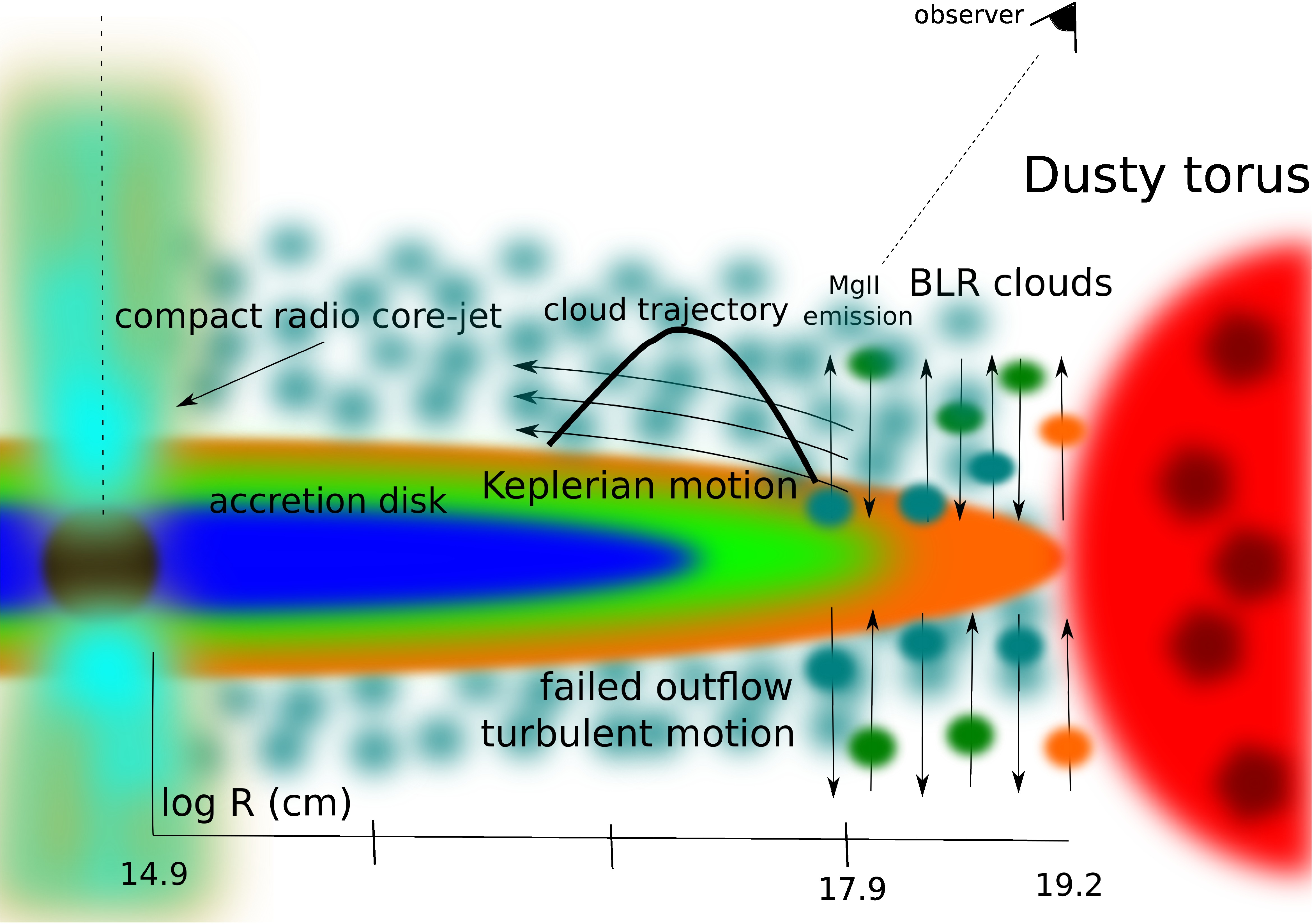}
    \caption{Illustration of the basic length-scales of the quasar HE 0413-4031. The BLR clouds are depicted with the dominantly Keplerian velocity field with a smaller outflow-inflow turbulent component according to the Failed Radiatively Accelerated Outflow Model \citep[FRADO,][]{2011A&A...525L...8C}. The axis along the bottom of the figure is expressed in the corresponding logarithms of basic radii in centimeters. From the left to the right side of the image, we include the Schwarzschild radius of $2.5\times 10^9\,M_{\odot}$ black hole, $\log{R_{\rm Schw}}=14.9$, the light-travel distance of MgII emission, $\log{R_{\rm MgII}}=17.9$, and the inner radius of the dusty torus, $\log{R_{\rm sub}}=19.2$.}
    \label{fig_HE0413_illustration}
\end{figure}

The models with the larger radial extent of the BLR with $R_{\rm out}=10^{18}\,{\rm cm}$ as shown in Fig. 9 of \citet{guo2019} seem to be more consistent with our slope of $0.43$ as they show a continuous rise of MgII luminosity even for larger continuum luminosities around the Eddington ratio of $\sim 0.4$. This is also in agreement with our inferred travel distance of $R_{\rm MgII}=c\overline{\tau}=0.254^{+0.020}_{−0.016}\,{\rm pc}\sim 10^{17.9}\,{\rm cm}$. In comparison, the location of the dusty torus is still further. Its inner radius is given by the sublimation radius, $R_{\rm sub}\sim 0.4\,{\rm pc}\,L_{45}^{0.5}\,T_{1500}^{-2.6}=5.04\,{\rm pc}\sim 10^{19.2}\,{\rm cm}$ \citep{2006ApJ...648L.101E,2008ApJ...685..147N} for our estimate of the bolometric luminosity $L_{\rm bol}=1.589 \times 10^{47}\,{\rm erg\,s^{-1}}$ and the dust sublimation temperature of $1500\,{\rm K}$. The outer radius of the dusty torus is expected to be at $R_{\rm torus}\sim YR_{\rm sub}$, where $Y\sim 5-10$ \citep{2006ApJ...648L.101E}. The light-travel distance, which can serve as a proxy for the BLR location in HE 0413-4031, is also in agreement with the model of the failed radiatively accelerated dusty outflow \citep[FRADO, ][]{2011A&A...525L...8C}. The failed dusty wind requires the existence of dust in the accretion disc, which is possible at and below $\sim 1000\,{\rm K}$. This sets the inner radius of the BLR to $r_{1000}/R_{\rm sub}=0.03 M_8^{1/6}/(\dot{m}^{1/6}\eta_{0.1}^{1/2})$, where $M_8$ is the black hole mass scaled to $10^8\,M_{\odot}$, $\dot{m}=\dot{M}/\dot{M}_{\rm Edd}$ is the dimensionless accretion rate, and $\eta$ is the accretion efficiency ($L_{\rm bol}=\eta \dot{M}c^2$). Using the best-fit SED model, see Section~\ref{sec_label_SED}, we adopt $M=2.5\times 10^9\,M_{\odot}$, $\dot{m}=0.51$, and $\eta=0.1$, which leads to $r_{1000}/R_{\rm sub}=0.0574$ or $r_{1000}=0.289\,{\rm pc}=10^{17.95}\,{\rm cm}$, which is within uncertainties consistent with the light-travel distance $R_{\rm MgII}$. We illustrate these basic length-scales of the quasar HE 0413-4031 in Fig.~\ref{fig_HE0413_illustration}.

For the continuum magnitudes smaller than $16.7$ mag the slope of the line-continuum dependency ($0.82 \pm 0.26$) is even larger than for the case when the whole range is considered. Interestingly, this slope is comparable to the exponent of the line-continuum relation as studied for the sample of flat-spectrum radio quasars \citep{2016FrASS...3...19P}, which is related to the global Baldwin effect between the equivalent width of originally broad UV lines (CIV, Ly$\alpha$) and the corresponding continuum luminosities (at 1350\,\AA), see the original works by \citet{1977ApJ...214..679B}, \citet{1978Natur.273..431B}, and \citet{1984ApJ...276..403W}. In general, the equivalent width decreases with the increasing luminosity, which can be described as a power-law relation, EW$_{\rm line}\propto L_{\rm cont}^{\gamma}$. This can be rewritten as a relation between the line and the corresponding continuum luminosities using EW$\simeq L_{\rm line}/L_{\rm cont}$, which yields $L_{\rm line} \propto L_{\rm cont}^{\gamma+1}$. The original Baldwin effect is also called global or ensemble \citep{1977ApJ...214..679B,1978MNRAS.185..381C}, which is derived based on single-epoch observations of an ensemble of AGNs, while the analogical relation studied for individual AGNs is related to as an intrinsic Baldwin effect \citep{1992AJ....103.1084P}.

\citet{2016FrASS...3...19P} analyze the line-continuum luminosity relation $L_{\rm line}-L_{\rm cont}$, including MgII line and 3000\,\AA\, continuum, for a sample of 96 FSRQ sources (core-jet blazars). For FSRQ, they found the slope of $0.796 \pm 0.153$, which is smaller than the slope of $0.909\pm 0.002$ for the control sample of RQ AGN. Within uncertainties, their slope derived for the whole sample is comparable to our slope $L_{\rm MgII}\propto L_{3000}^{0.82 \pm 0.26}$. Hence, our detected intrinsic Baldwin effect is in agreement with the global one derived for the population of FSRQ. Previously, \citet{2017A&A...603A..49R} studied the intrinsic Baldwin effect for 6 type-I AGN and they detected it for the broad recombination lines, H$\alpha$ and H$\beta$. They found that the intrinsic Baldwin effect is not related to the global one. \citet{2016FrASS...3...19P} found the difference of the global Baldwin effect between the radio-loud (blazar) and radio-quiet AGN, which could imply the importance of the non-thermal component, i.e. boosted jet emission, to the ionizing continuum for radio-loud sources. Apparently, more data for our quasar as well as more radio-loud and radio-quiet sources are needed to study in detail both the intrinsic and global Baldwin effect and their potential relation, especially taking into account the potential non-thermal contribution for radio-loud sources.  

In summary, we detect a significant correlation between MgII and 3000\AA\, continuum after the removal of the light-travel time effect. The relation $L_{\rm line}-L_{\rm cont}$ is not linear, but has a slope of $0.43\pm 0.10$ when all the corresponding line-luminosity points are combined. The slope is larger, $\gamma+1=0.82 \pm 0.26$, when only a higher correlated part of the points is selected. These results are consistent with the MgII broad-line emission being at least partially driven by the underlying continuum.       

\subsection{SED fitting}
\label{sec_label_SED}

\begin{figure}
    \centering
    \includegraphics[width=0.5\textwidth]{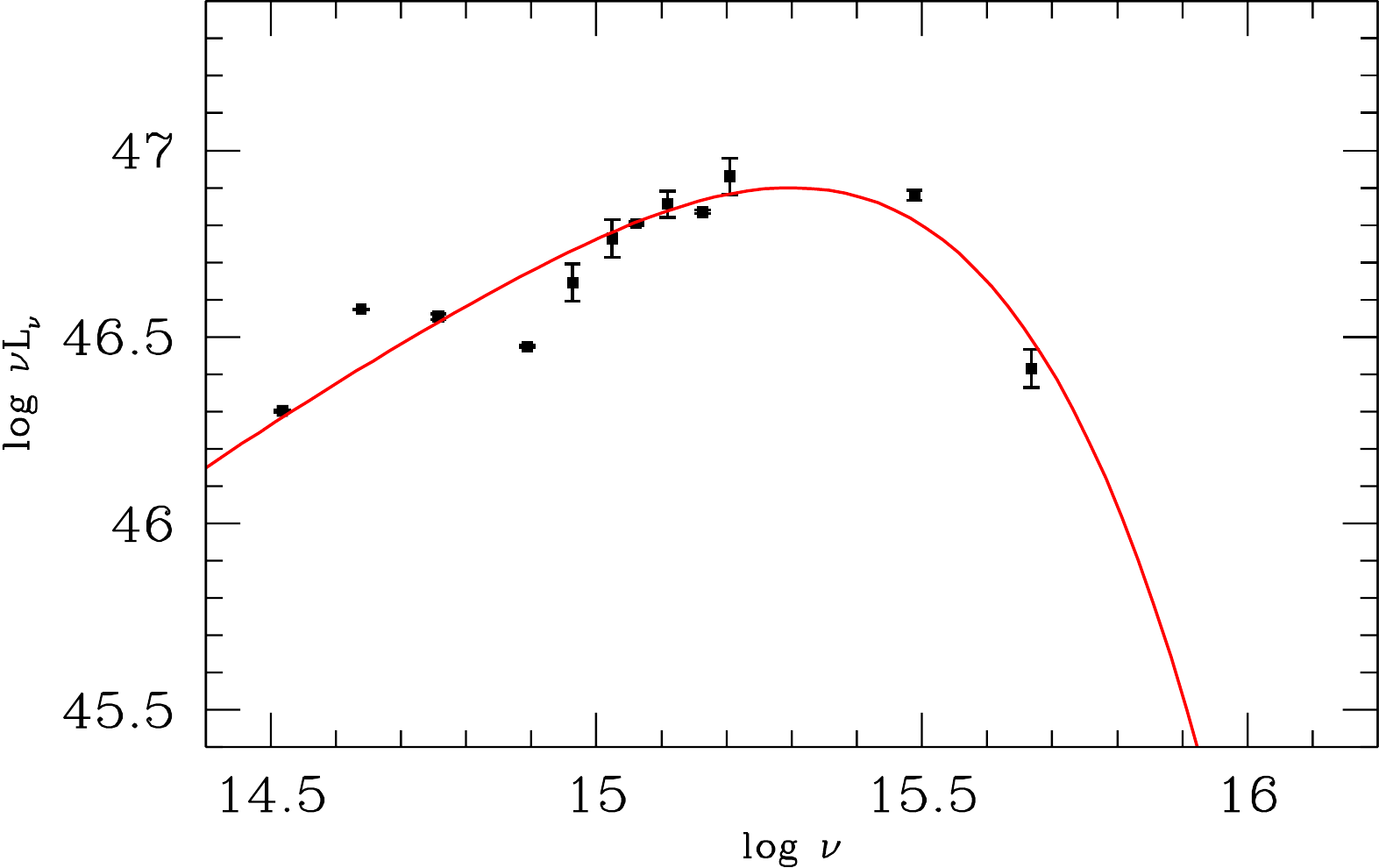}
    \caption{The SED data and the accretion disc model for HE 0413-4031 shown as the dependency of the luminosity $\nu L_{\nu}$ (in $\rm erg\,s^{-1}$) on the frequency $\nu$ (in ${\rm Hz}$) in the log-space. The data are taken from the Vizier SED data, see \url{http://vizier.u-strasbg.fr/vizier/sed}. For the fitting, we use error-weighted averages of the data. The best-fit SED model (red solid line) is represented by a red solid line and is based on the parameters $M_{\bullet} = 3.0 \times 10^9 M_{\odot}$, $\dot m = 0.35$ \textcolor{black}{(here measured in units of $1.678 \times 10^{18} (M/M_{\odot}\, g s^{-1}$)}, $a = 0.31$, and $\iota = 34$ deg.}
    \label{fig:SED}
\end{figure}

Our determination of the black hole mass in Section~\ref{sec_final_time_delay_virial_mass} is not unique since it requires additional assumptions about the virial factor. As a test of the mass range we obtained, we attempted to obtain the constraints for the black hole mass directly, from the accretion disk fitting to the continuum.

We used the data points available from Vizier SED photometric viewer\footnote{http://vizier.u-strasbg.fr/vizier/sed/}. After removal of the multiple entry and converting the measurements to the rest frame (assuming $z = 1.37648$ as determined in Section~\ref{sec_mean_spectrum}), and adopting $H_0 = 69.5, \Omega_m = 0.286, \Omega_L = 0.714$ \citep{2014ApJ...794..135B}, we obtain the IR to UV SED (see Figure~\ref{fig:SED}). We corrected the data for the Galactic extinction although the effect is not strong in the direction of HE 0413-4031. The data points come from various epochs. Therefore, we added and additional error of 0.08 (in log space) to the measurements in order to account for the variability. For disk fitting, we used only points at the frequencies above 14.5 in the log scale, rest frame, since the rest frame near IR emission in quasars come from the hot dust component. We did not assume any presence of the blazar component since the data point did not seem to suggest its need.

\begin{table}[tbh]
    \centering
      \caption{Best-fitted parameters for the black mass in the range of $1-5 \times 10^9\,M_{\odot}$. The parameters include the spin, the accretion rate $\dot m$ (here measured in units of $1.678 \times 10^{18} (M/M_{\odot}\, g s^{-1}$), and the viewing angle $\iota$. The smallest $\chi^2$ of $19.17$ is for the case with the parameters $M =2.5 \times 10^9 M_{\odot}$, $\dot m = 0.51$, $a=0.025$, and $\iota=33.8$ deg.}
    \begin{tabular}{c|c|c|c|c}
\hline
\hline
$M\,[10^9\,M_{\odot}]$  &   spin       &    $\dot m$      &  $\iota$ [deg]     &  $\chi^2$\\
\hline
$1$   & $-0.99$   &     $1.81$   &    $0.0$  &      $31.57$\\
$1.5$ & $-0.94$   &     $1.08$    &   $0.0$   &     $19.38$\\
$2$   & $-0.13$   &     $0.60$   &    $0.0$  &     $19.54$\\
\hline
$2.5$ &  $0.025$ &      $0.51$   &   $33.8$  &     $19.17$\\
\hline
$3$   &  $0.46$  &      $0.31$    &   $25.8$  &     $19.31$\\
$3.5$ &  $0.52$  &       $0.26$    &   $33.6$  &     $19.23$\\
$4$   &  $0.44$  &      $0.29$    &   $49.8$   &    $20.20$\\
$4.5$ &  $0.0062$ &     $0.54$     &  $68.2$   &     $20.31$\\
$5$   &  $0.044$ &      $0.50$     &  $69.5$   &     $20.30$\\
\hline
    \end{tabular}
    \label{table_sed_fitting}
\end{table}

We used the fully relativistic Novikov-Thorne model \citep{1973blho.conf..343N}, with all propagation effects as described in \citet{czerny_spin2011}. The model is characterized by the black hole mass, acccretion rate (in Eddington units, \textcolor{black}{assuming the fixed efficiency of 1/12 in the definition, i.e. in units of $1.678 \times 10^{18} (M/M_{\odot}) {\rm g s^{-1}}$),} spin, and viewing angle. We performed the fitting without constraints for any of those parameters. First, we performed the fitting for all the data points available and we obtained the best fit model with the parameters: $M = 3.0 \times 10^9 M_{\odot}$, $\dot m = 0.35$, $a = 0.31$, and $\iota = 34$ deg. We present this fit in Figure~\ref{fig:SED}. Second, we also performed the fitting for the error-weighted averages of the data and for the uncertainties we used error-weighted standard deviations. In this case, the best-fit solution was formally with the parameters: $M =2.5 \times 10^9 M_{\odot}$, $\dot m = 0.51$, $a=0.025$, and $\iota=33.8$ deg. However, fits are highly degenerate, so the $\chi^2$ allows for a broad mass range from $1 \times 10^9 M_{\odot}$ to masses even above $5 \times 10^9 M_{\odot}$, see Table~\ref{table_sed_fitting}. Large masses, however, do not provide an acceptable solution since they also require a very high viewing angle. A high viewing angle is not expected since the unification scheme of AGN excludes it due to the presence of the dusty/molecular torus \citep[see e.g.][for a recent review]{2017A&ARv..25....2P}, and the clear excess in the near-IR shows that the torus is present in HE 0413-4031.

If we constrain the allowed parameters to $\iota < 45 $ deg, the upper limit for the black hole mass is $M = 3.5\times 10^9 M_{\odot}$. In our fits, the black hole spin is never large, for very small black hole masses the accretion rate is super-Eddington and the spin is retrograde, the highest value of the spin we get is 0.52. However, this determination highly relies on one data point - the far-UV GALEX measurement which is in the spectral range where relativistic effects are important. If there is some internal reddening in the quasar, probably the allowed spin could be higher but the SED data quality is not good enough to attempt more complex modelling.

The obtained black hole mass range $1.5 \times 10^9 M_{\odot} - 3.5 \times 10^9 M_{\odot}$ is consistent with those presented in Section~\ref{sec_final_time_delay_virial_mass}.

By integrating the best-fit SED, we can derive the bolometric correction BC for the monochromatic luminosity at 3000\,\AA. We obtain $L_{\rm bol}=2.8 L_{3000}$, which is smaller than the mean value of $5.62 \pm 1.14$ provided by \citet{2006ApJS..166..470R} for the same wavelength. However, it is consistent with the luminosity-dependent relation for the bolometric correction derived by \citet{2019MNRAS.488.5185N}, which gives BC$=25\times (L_{3000}/10^{42}\,{\rm erg\,s^{-1}})^{-0.2}=2.80$ for $L_{3000}=10^{46.754}\,{\rm erg\,s^{-1}}$. The consistency with the power-law relation of \citet{2019MNRAS.488.5185N} stems from the fact that they used essentially the same model of an optically thick, geometrically thin accretion disc that is used in this work to fit the SED.   

In summary, the SED fitting showed that the canonical thin, optically thick accretion disc can still account for the dominant part of the continuum in our highly accreting quasar. At higher accretion rates, the inner parts of the accretion flow are expected to become geometrically and optically thick like in slim accretion discs, which can account for the reduction of the ionizing flux and shortening of time delays \citep{Wang_2014}. This is indeed supported by the existence of stable geometrically thick and optically thick ``puffy'' accretion discs in global 3D GRMHD simulations for sub-Eddington accretion rates comparable to our values of $\dot{m}=0.3-0.6$ \citep{2019ApJ...884L..37L}. However, the current computational facilities still do not allow a self-consistent treatment of the accretion disc-BLR dynamics on the scales of as much as 1000 gravitational radii, while the analytical and semi-analytical models explain the main observational features \citep{czerny_spin2011,2011A&A...525L...8C}.

\section{Conclusions} \label{sec:conclusions}

We summarize the main findings of the paper as follows:
\begin{enumerate}
    \item Using seven different methods, we found a rest-frame time-delay between the continuum and MgII line emission for the bright quasar HE 0413-4031,  $\overline{\tau}=302.6^{+28.7}_{-33.1}$ days, which was the most frequent peak in time-delay distributions.
    \item In combination with the data for 10 other sources monitored in MgII line, we construct a radius-luminosity relation, which is consistent with the theoretically expected dependency, $R\propto L^{1/2}$. The new quasar HE 0413-4031 with the monochromatic luminosity of $\log{L_{3000}}=46.754^{+0.028}_{-0.132}$ lies below the expected relation, which can be explained by its higher accretion rate. In general, for all MgII sources, the departure from the radius-luminosity relation, i.e. the shortening of their time-delays, is larger for higher-accreting sources. The same effect was previously observed for the sources monitored in H$\beta$.
    \item We determined the response of MgII line luminosity to the photoionizing continuum luminosity, $L_{\rm line}\propto L_{\rm cont}^{0.43\pm 0.10}$, which is comparable to the response of recombination emission lines H$\alpha$ and H$\beta$ according to theoretical photoionization models. This is consistent with the outer radius of the BLR at $R_{\rm out}=10^{18}\,{\rm cm}$, which is in turn in agreement with the light-travel distance inferred from the rest-frame time-delay.   
    \item The virial black hole mass determined based on the measured rest-frame time delay, $M_{\rm RM}^{\rm uncorr}(f=1)\simeq 1.1\times 10^9\,M_{\odot}$ is smaller by a factor of four than the value expected from the radius-luminosity relation, $M_{\rm RM}^{\rm corr}(f=1) \simeq 4.6\times 10^9\,M_{\odot}$. The black hole mass inferred from fitting a thin accretion disk model to the source SED, $M_{\rm SED}=1.5 \times 10^9 M_{\odot} - 3.5 \times 10^9 M_{\odot}$, is in agreement with these values within the uncertainty. Other best-fitted parameters for the source are the Eddington ratio of $0.26 \leq \dot{m} \leq 1.08$, the black hole spin of $-0.94 \leq a \leq 0.52$, and the viewing angle of $0 \leq \iota\leq 34$ degrees.
\end{enumerate}

\software{IRAF \citep{1986SPIE..627..733T,1993ASPC...52..173T},  JAVELIN \citep{2011ApJ...735...80Z,2013ApJ...765..106Z,2016ApJ...819..122Z}, PyCCF \citep{2018ascl.soft05032S}, vnrm.py \citep{2017ApJ...844..146C}, zdcf\_v2.f90 \citep{1997ASSL..218..163A}, plike\_v4.f90 \citep{1997ASSL..218..163A}, delay\_chi2.f \citep{2013A&A...556A..97C}}

\acknowledgments

We thank the referee for constructive comments that helped to improve the clarity of the manuscript. The authors acknowledge the financial support by the National Science Centre, Poland, grant No.~2017/26/A/ST9/00756 (Maestro 9), and by the Ministry of Science and Higher Education (MNiSW) grant DIR/WK/2018/12. GP acknowledges the grant MNiSW DIR/WK/2018/09. KH acknowledges support by the Polish National Science Centre grant
2015/18/E/ST9/00580. The OGLE project has received funding from the National Science
Centre, Poland, grant MAESTRO 2014/14/A/ST9/00121. The Polish participation in SALT is funded by grant No. MNiSW DIR/WK/2016/07.

%




\appendix

\setcounter{table}{0}
\renewcommand\thetable{\Alph{section}.\arabic{table}}

\section{Photometric and spectroscopic data}
\label{sec_photometry_spectroscopy}

In this section, we summarize the characteristics of FeII and MgII lines in Table~\ref{tab:spectr}, where we specifically list the FeII and MgII equivalent widths in $\AA$, velocity shift in ${\rm km\,s^{-1}}$, line width in $\AA$, and the MgII flux density in ${\rm erg\,s^{-1}\,cm^{-2}\,\AA^{-1}}$. The continnum magnitudes (V-band) from the three instruments -- OGLE, SALTICAM, and BMT -- are included in Tables~\ref{sec:phot1} and \ref{sec:phot2}. 

\begin{table*}
\caption{Table of FeII and MgII equivalent widths in \AA, velocity shift in ${\rm km\,s^{-1}}$, line width in \AA, and the MgII flux density in ${\rm erg\,s^{-1}\,cm^{-2}\,\AA^{-1}}$. The flux density was calculated for the case without the BMT data, see the text for the description.}
\label{tab:spectr}
\begin{center}
    
{\renewcommand{\arraystretch}{1.5}
\begin{tabular}{|c|c|c|c|c|c|c|}
 \hline
Obs. & JD& EW(FeII) & EW(MgII) & Shift & Width & Flux density \\ 
 No. & -2 450 000 & \AA & \AA & km/s & \AA & ${\rm erg\,s^{-1}\,cm^{-2}\,\AA^{-1}}$ \\
 \hline
 \hline
 1 & 6314.4087 & $13.71^{+0.91}_{-0.93}$ & $34.18^{+0.51}_{-0.49}$ & $1666.98^{+17.04}_{-16.65}$ & $2149.29^{+28.50}_{-40.08}$ &  $(3.836 \pm 0.074)\times 10^{-14}$  \\ 
 2 & 6320.3859 & $9.75^{+1.34}_{-1.32}$ & $35.89^{+0.79}_{-0.74}$ & $1723.70^{+23.73}_{-22.96}$ & $2196.68^{+54.72}_{-52.86}$ & $(4.000 \pm 0.114)\times 10^{-14}$\\ 
 3 & 6523.5954 & $10.73^{+2.28}_{-2.22}$ & $34.67^{+1.25}_{-1.20}$ & $1529.11^{+39.28}_{-39.12}$ & $2165.96^{+88.44}_{-81.07}$ & $(3.966 \pm 0.178) \times 10^{-14}$\\ 
 4 & 6651.4751 & $10.38^{+1.77}_{-1.72}$ & $36.59^{+1.04}_{-1.02}$ & $1598.83^{+31.86}_{-31.09}$ & $2289.73^{+81.43}_{-64.80}$ & $(4.051 \pm 0.152) \times 10^{-14}$\\ 
 5 & 6697.3600 & $14.13^{+1.32}_{-1.34}$ & $37.93^{+0.78}_{-0.75}$ & $1617.81^{+22.58}_{-22.60}$ & $2316.05^{+46.37}_{-61.37}$ & $(4.219 \pm 0.114)\times 10^{-14}$ \\ 
 6 & 6892.5678 & $17.07^{+1.11}_{-1.03}$ & $36.06^{+0.56}_{-0.58}$ & $1554.76
^{+18.02}_{-18.05}$ & $2218.54^{+41.03}_{-38.86}$ & $ (3.911 \pm 0.094) \times 10^{-14}$\\ 
 7 & 7003.5182 & $13.25^{+0.58}_{-0.64}$ & $35.64^{+0.36}_{-0.34}$ & $1608.01^{+11.37}_{-11.35}$ & $2307.95^{+24.90}_{-23.98}$ & $ (3.908 \pm 0.067)\times 10^{-14} $\\ 
 8 & 7082.2985 & $13.29^{+0.83}_{-0.82}$ & $31.20^{+0.48}_{-0.44}$ & $1615.91^{+17.11}_{-17.14}$ & $2300.84^{+46.54}_{-35.14}$ & $ (3.702 \pm 0.080)\times 10^{-14} $\\ 
 9 & 7243.6124 & $13.93^{+0.69}_{-0.72}$ & $30.65^{+0.41}_{-0.36}$ & $1597.95^{+14.64}_{-14.64}$ & $2250.33^{+34.74}_{-28.81}$ & $(3.856 \pm   0.069)\times 10^{-14}$\\ 
 10 & 7289.4741 & $13.57^{+0.89}_{-0.86}$ & $31.14^{+0.47}_{-0.46}$ & $1613.10^{+17.51}_{-17.45}$ & $2213.75^{+45.45}_{-31.14}$ &  $(3.953 \pm   0.082) \times 10^{-14}$\\ 
 11 & 7341.3298 & $12.46^{+0.82}_{-0.87}$ & $32.33^{+0.45}_{-0.49}$ & $1597.92^{+16.98}_{-17.10}$ & $2296.71^{+40.25}_{-39.59}$ & $(4.127 \pm  0.082) \times 10^{-14} $\\ 
 12 & 7374.4950 & $13.43^{+0.83}_{-0.83}$ & $28.87^{+0.43}_{-0.45}$ & $1594.19^{+19.07}_{-17.77}$ & $2200.05^{+36.00}_{-35.27}$ & $ (3.681 \pm   0.075) \times 10^{-14} $\\ 
 13 & 7423.3702 & $10.12^{+0.74}_{-0.72}$ & $29.97^{+0.41}_{-0.44}$ & $1669.13^{+16.79}_{-16.53}$ & $2331.95^{+34.35}_{-39.93}$ & $ (3.860 \pm 0.073) \times 10^{-14}$\\ 
 14 & 7656.2474 & $10.64^{+0.63}_{-0.61}$ & $28.06^{+0.36}_{-0.33}$ & $1664.97^{+14.98}_{-13.65}$ & $2290.71^{+35.96}_{-30.11}$ & $ (4.204 \pm 0.075) \times 10^{-14} $\\ 
 15 & 7687.3943 & $10.65^{+0.65}_{-0.62}$ & $28.08^{+0.36}_{-0.35}$ & $1664.77^{+15.31}_{-14.89}$ & $2288.94^{+37.67}_{-30.63}$ & $(3.979 \pm 0.067) \times 10^{-14}$ \\ 
 16 & 7722.5591 & $11.99^{+0.66}_{-0.66}$ & $26.62^{+0.38}_{-0.34}$ & $1672.60^{+16.16}_{-15.95}$ & $2254.84^{+29.84}_{-40.84}$ & $(3.781 \pm   0.066) \times 10^{-14}$\\ 
 17 & 7752.4673 & $10.26^{+0.58}_{-0.58}$ & $27.94^{+0.34}_{-0.32}$ & $1682.02^{+14.26}_{-14.03}$ & $2298.09^{+29.95}_{-33.91}$ & $(4.125 \pm 0.065)\times 10^{-14}  $\\ 
 18 & 7953.6598 & $12.10^{+0.64}_{-0.64}$ & $25.37^{+0.33}_{-0.34}$ & $1648.26^{+15.85}_{-15.69}$ & $2208.36^{+32.31}_{-31.79}$ & $ (3.683 \pm 0.054) \times 10^{-14}$\\ 
 19 & 7979.5920 & $11.83^{+0.59}_{-0.62}$ & $28.05^{+0.35}_{-0.34}$ & $1640.45^{+14.78}_{-14.73}$ & $2363.17^{+33.85}_{-35.90}$ & $(4.072 \pm 0.067) \times 10^{-14}$\\ 
 20 & 8114.4800 & $10.35^{+0.56}_{-0.65}$ & $27.40^{+0.36}_{-0.32}$ & $1681.22^{+14.14}_{-14.81}$ & $2337.25^{+33.20}_{-37.38}$ & $(4.250 \pm 0.067) \times 10^{-14}$\\ 
 21 & 8167.3265 & $9.26^{+0.53}_{-0.58}$ & $26.59^{+0.34}_{-0.27}$ & $1649.54^{+14.75}_{-14.32}$ & $2358.10^{+33.34}_{-34.15}$ & $ (4.025 \pm 0.063) \times 10^{-14}$\\ 
 22 & 8376.5021 & $13.63^{+0.65}_{-0.62}$ & $29.30^{+0.36}_{-0.35}$ & $1631.38^{+14.73}_{-14.41}$ & $2326.37^{+30.10}_{-29.91}$ & $(4.380 \pm   0.069) \times 10^{-14} $\\ 
 23 & 8498.4167 & $10.98^{+0.61}_{-0.65}$ & $29.25^{+0.37}_{-0.37}$ & $1800.13^{+15.32}_{-14.75}$ & $2343.83^{+30.65}_{-41.66}$ & $(4.323 \pm   0.070) \pm 10^{-14}$\\ 
 24 & 8543.3100 & $8.85^{+0.61}_{-0.62}$ & $29.55^{+0.38}_{-0.36}$ & $1751.82^{+15.48}_{-14.00}$ & $2368.61^{+33.36}_{-42.12}$ & $(4.361 \pm 0.071) \times 10^{-14}$\\  
 25 & 8719.5708 & $15.47^{+0.94}_{-0.91}$ & $32.19^{+0.54}_{-0.52}$ & $1667.05^{+19.55}_{-19.16}$ & $2359.25^{+52.31}_{-40.14}$ &  $(4.722 \pm 0.092)\times 10^{-14}$ \\
 \hline
 \end{tabular}} \quad
\end{center}
 \end{table*}

\begin{table}
\caption{Table of continuum magnitudes with uncertainties. The epoch is given in Julian Dates ($-2450 000$). The last column denotes three different instruments used to obtain the photometry data: 1. OGLE, 2. SALTICAM, 3. BMT. \textit{Note:} The BMT photometry points were shifted by $0.171$ mag to larger magnitudes to match the last OGLE point with the closest BMT point in the light curve.}
\label{sec:phot1}
\begin{tabular}{c|c|c|c}
\hline
\hline
JD& magnitude ($V$-band)  & Error & Instrument \\ 
-2 450 000 & mag & mag & No. \\
 \hline
6199.79634 & 16.763 & 0.005 & 1\\
6210.81464 & 16.755 & 0.003 & 1\\
6226.67656 & 16.771 & 0.004 & 1\\
6246.69516 & 16.761 & 0.004 & 1\\
6257.74660 & 16.763 & 0.005 & 1\\
6268.68051 & 16.767 & 0.004 & 1\\
6277.68239 & 16.738 & 0.003 & 1\\
6286.66584 & 16.782 & 0.004 & 1\\
6297.61482 & 16.770 & 0.004 & 1\\
6307.57245 & 16.769 & 0.004 & 1\\
6317.63928 & 16.799 & 0.004 & 1\\
6330.65489 & 16.789 & 0.003 & 1\\
6351.54598 & 16.788 & 0.004 & 1\\
6363.57130 & 16.782 & 0.003 & 1\\
6379.48424 & 16.762 & 0.004 & 1\\
6379.49181 & 16.755 & 0.004 & 1\\
6387.50984 & 16.780 & 0.003 & 1\\
6637.66923 & 16.767 & 0.003 & 1\\
6651.62009 & 16.806 & 0.003 & 1\\
6665.60325 & 16.812 & 0.004 & 1\\
6678.59717 & 16.796 & 0.003 & 1\\
6689.67132 & 16.792 & 0.003 & 1\\
6700.63473 & 16.809 & 0.004 & 1\\
6715.57393 & 16.791 & 0.003 & 1\\
6740.48864 & 16.789 & 0.004 & 1\\
6892.59242 & 16.830 & 0.011 & 2\\
7003.54330 & 16.819 & 0.012 & 2\\
7036.65108 & 16.747 & 0.004 & 1\\
7048.65280 & 16.731 & 0.003 & 1\\
7060.60356 & 16.700 & 0.004 & 1\\
7082.30016 & 16.749 & 0.012 & 2\\
7084.53369 & 16.732 & 0.005 & 1\\
7118.50567 & 16.731 & 0.005 & 1\\
7243.61293 & 16.664 & 0.011 & 2\\
7253.88913 & 16.668 & 0.003 & 1\\
7261.88037 & 16.684 & 0.004 & 1\\
7267.91217 & 16.650 & 0.004 & 1\\
7273.84457 & 16.683 & 0.004 & 1\\
7283.84655 & 16.659 & 0.004 & 1\\
7289.47056 & 16.698 & 0.012 & 2\\
7295.84011 & 16.661 & 0.004 & 1\\
7306.77839 & 16.677 & 0.004 & 1\\
\hline \label{tab:Vmag}
 \end{tabular}
 \end{table}

\begin{table}
\caption{Table of continuum magnitudes with uncertainties. The epoch is given in Julian Dates ($-2450 000$). The last column denotes three different instruments used to obtain the photometry data: 1. OGLE, 2. SALTICAM, 3. BMT. \textit{Note:} The BMT photometry points were shifted by $0.171$ mag to larger magnitudes to match the last OGLE point with the closest BMT point in the light curve.}
\label{sec:phot2}
\begin{tabular}{c|c|c|c}
\hline
\hline
JD& magnitude ($V$-band)  & Error & Instrument \\ 
-2 450 000 & mag & mag & No. \\
 \hline
7317.73770 & 16.678 & 0.004 & 1\\
7327.77204 & 16.686 & 0.004 & 1\\
7340.70393 & 16.655 & 0.003 & 1\\
7341.32488 & 16.671 & 0.011 & 2\\
7355.69184 & 16.656 & 0.004 & 1\\
7363.66368 & 16.627 & 0.003 & 1\\
7374.49091 & 16.654 & 0.011 & 2\\
7374.70619 & 16.657 & 0.003 & 1\\ 
7385.55446 & 16.624 & 0.003 & 1\\
7398.61439 & 16.641 & 0.003 & 1\\
7415.58224 & 16.634 & 0.003 & 1\\
7423.36782 & 16.633 & 0.011 & 2\\
7426.56315 & 16.653 & 0.003 & 1\\
7436.52206 & 16.626 & 0.004 & 1\\
7447.52422 & 16.623 & 0.003 & 1\\
7457.51899 & 16.641 & 0.003 & 1\\
7656.47709 & 16.545 & 0.011 & 2\\
7687.38749 & 16.543 & 0.011 & 2\\
7717.70291 & 16.541 & 0.003 & 1\\
7722.55378 & 16.513 & 0.011 & 2\\
7752.46369 & 16.496 & 0.011 & 2\\
7973.91046 & 16.518 & 0.006 & 1\\
7979.59391 & 16.502 & 0.011 & 2\\
8038.85902 & 16.509 & 0.004 & 1\\
8084.30756 & 16.429 & 0.011 & 2\\
8090.70000 & 16.338 & 0.008 & 3\\
8114.47660 & 16.193 & 0.011 & 2\\
8138.70000 & 16.305 & 0.008 & 3\\
8139.60000 & 16.311 & 0.004 & 3\\
8146.60000 & 16.284 & 0.003 & 3\\
8165.60000 & 16.296 & 0.005 & 3\\
8167.32241 & 16.464 & 0.011 & 2\\
8173.60000 & 16.277 & 0.007 & 3\\
8180.50000 & 16.261 & 0.006 & 3\\
8196.50000 & 16.291 & 0.008 & 3\\
8205.50000 & 16.274 & 0.006 & 3\\
8365.90000 & 16.285 & 0.006 & 3\\
8377.50208 & 16.486 & 0.011 & 2\\
8386.90000 & 16.302 & 0.008 & 3\\
8414.80000 & 16.319 & 0.005 & 3\\
8498.41249 &  16.509 &  0.011 & 2\\
8543.30431 & 16.484  & 0.011 & 2\\
8566.50000 & 16.319  & 0.008 & 3\\
8719.57089 & 16.498  & 0.011 & 2\\
\hline
 \end{tabular}
 \end{table}
 
\section{Overview of time-delay determination methods}
\label{sec_time-delay}
 
\subsection{Interpolated cross-correlation function (ICCF)}
\label{subsec_ICCF} 

The interpolated cross-correlation function (ICCF) is a standard method for determining the time-delay between the continuum and line-emission light curves. In general, both light curves are unevenly sampled, while the ICCF by its definition requires regular sampling with a certain timestep, which is achieved by the interpolation of the continuum light curve with respect to the line-emission light curve or vice versa (asymmetric ICCF). The definition of the ICCF between the two light curves, $x_i$ and $y_i$, with the step-size of $\Delta t=t_{i+1}-t_i$ is,
\begin{equation}
    CCF(\tau_{k})=\frac{(1/N)\sum_{i=1}^{N-k}(x_i-\overline{x})(y_{i+k}-\overline{y})}{[(1/N)
    \sum_{i}^{N}(x_i-\overline{x})^2]^{1/2}[(1/N)\sum_i^{N}(y_i-\overline{y})^2]^{1/2}}\,,
    \label{eq_CCF}
\end{equation}
where $\tau_{k}$ is the time-shift $\tau_k=k \Delta t$, where the index $k=1$,.\,.\,.,$N-1$, of the second light curve with respect to the first one, and $\overline{x}$ and $\overline{y}$ are the means of the two light curves $x_i$ and $y_i$.
The final, symmetric ICCF is obtained by averaging the ICCFs from both interpolations. 

\begin{figure*}[h!]
    \centering
    \includegraphics[width=\textwidth]{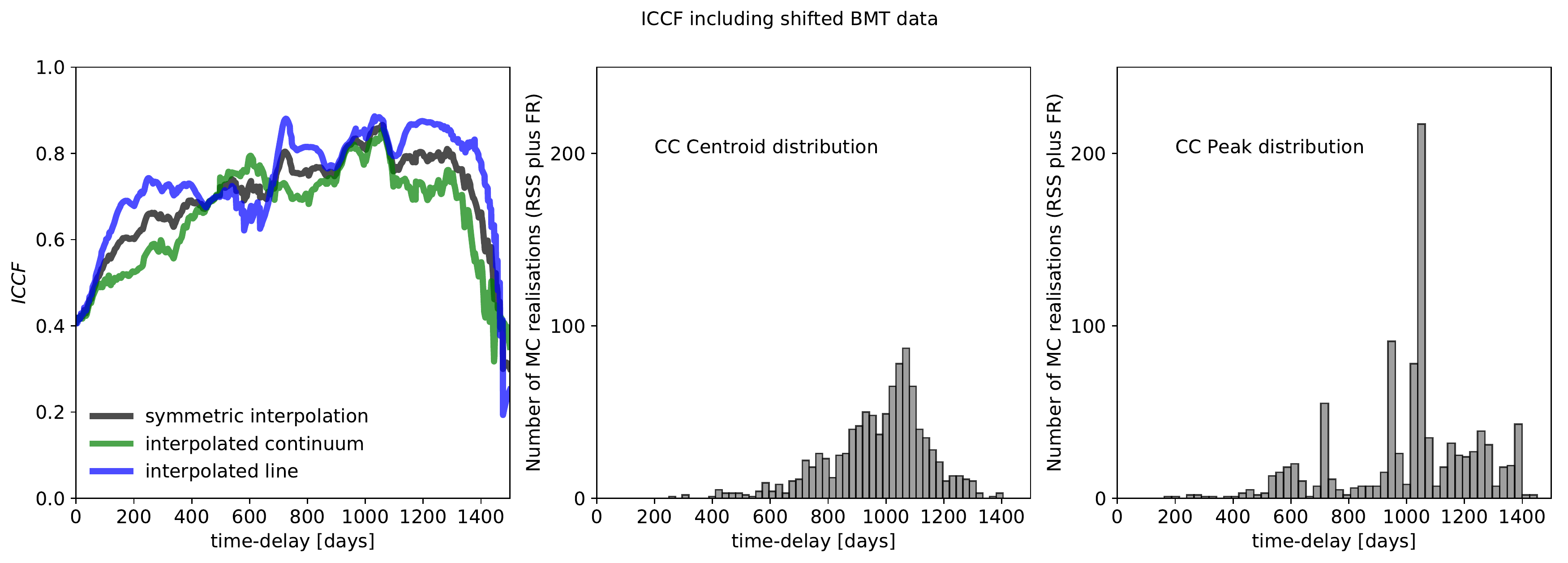}
    \includegraphics[width=\textwidth]{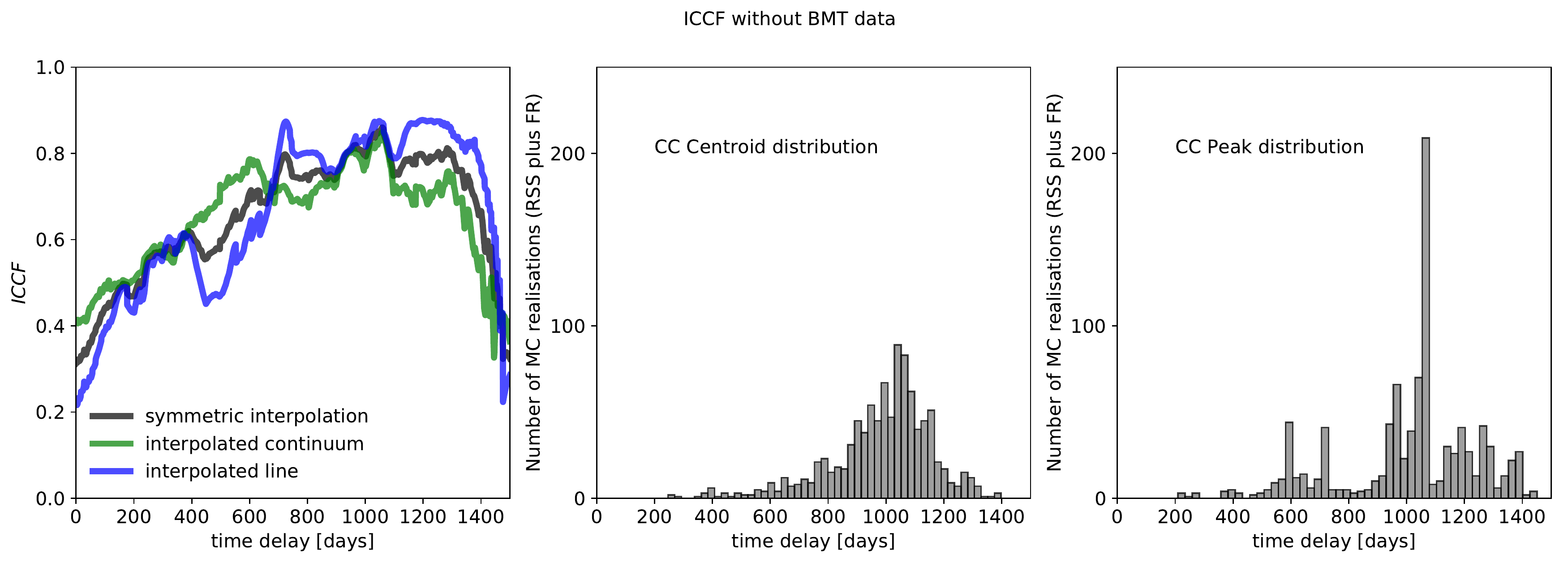}
    \caption{Interpolated cross-correlation coefficient as a function of time-delay in the observer's frame. \textbf{Top panel:} The interpolated cross-correlation function (ICCF) as a function of time-delay, including shifted BMT points. The middle panel displays the distribution of cross-correlation centroids, while the right panel shows the distribution of cross-correlation peaks. \textbf{Bottom panel:} The same as in the top panel but without BMT data.}
    \label{fig_ICCF}
\end{figure*}

We apply the PYTHON implementation of ICCF, the script PYCCF \citep{2018ascl.soft05032S} based on an earlier ICCF analysis of \citet{1998PASP..110..660P}, which calculates the ICCF including the continuum, line-emission, and symmetric interpolation. Using thousand Monte Carlo realizations of random subset selection (RSS) and flux randomization (FR), we obtained ICCF peak and centroid distributions, including their corresponding uncertainties.  

\begin{table*}[h!]
    \centering
    \caption{Results of the interpolated cross-correlation function applied to HE 0413-84031 light curves. We include centroids and peaks with uncertainties for interpolated continuum light curve, interpolated emission light curve, and symmetric ICCF. Cases with and without BMT data are separated. The time-delays are expressed in light days in the observer's frame.}
    \begin{tabular}{c|c|c}
     \hline
     \hline 
         & With shifted BMT data & Without BMT data  \\
    \hline     
    Interpolated continuum - Centroid [days]    &    $1004.6^{+196.8}_{-246.2}$ & $1003.2^{+205.3}_{-235.4}$  \\
    Interpolated continuum - Peak [days]    &  $1060.0^{+228.0}_{-342.6}$ & $1061.0^{+228.2}_{-270.8}$   \\
    Interpolated line - Centroid [days]    & $1008.4^{+142.2}_{-276.9}$  &  $1034.171^{+139.1}_{-248.9}$   \\
    Interpolated line - Peak [days]    &  $984.0^{+227.6}_{-349.0}$  &  $ 1001.0^{+252.3}_{-282.0}$   \\
    Symmetric - Centroid [days]    &  $1009.7^{+113.6}_{-211.5}$  & $1021.7^{+114.5}_{-  207.8}$  \\
    Symmetric - Peak [days]   &  $1056.0^{+197.0}_{-332.1}$ & $1057.0^{+196.0}_{-343.8}$ \\    
    \hline 
    \end{tabular}
    \label{table_ICCF}
\end{table*}

First, we cross-correlated the full continuum light curve, which included SALTICAM, OGLE, and flux-shifted BMT data, in total 86 points, with the MgII light curve (25 points). In addition, given the systematic offset of the BMT flux densities from SALTICAM points, we decided to perform the ICCF analysis also without them, which reduced the photometric light curve to 73 points. The ICCF values with respect to the time-delay, for which we separately calculated the continuum-interpolated, line-interpolated, and symmetric ICCF, with the corresponding peak and centroid distributions (for symmetric CCF) are displayed in Fig.~\ref{fig_ICCF} with and without BMT points in the top and bottom panels, respectively. We summarize the peak and centroid values of the ICCF for the interpolated continuum, interpolated line-emission, and symmetric case in Table~\ref{table_ICCF} where the cases with and without BMT datapoints are separated as well. When comparing these two cases in Table~\ref{table_ICCF}, the peak and centroid values for the case without the BMT data are generally comparable within the uncertainties, which is also visible in centroid and peak distributions in Fig.~\ref{fig_ICCF}.

The maximum values of the ICCF are about 0.8 (for the interpolation of the photometry) which for such a relatively short emission line light curve is a high value, supporting the view that the delay determination should be in general reliable since the line and continuum are well correlated. For comparison, the maximum value of the ICCF for the quasar CTS C30.10, with similar formal data quality, was only 0.65 \citep{2019arXiv190703910Z,czerny2019}.

We also tested if the linear trend present in the continuum should be eventually subtracted before the time delay is measured. However, we noticed that such a trend subtraction decreases the maximum value of the correlation in ICCF. For example, for the interpolated continuum, without BMT data points, $r_{max}$ decreases from $0.85$ down to $0.61$. Thus we conclude that the trend subtraction is not beneficial for the time-delay analysis. The presence of the trend is natural if the light curve of the red noise character like here covers the period shorter than the maximum timescale present in the system. The time delay measurement is not strongly affected anyway, we obtain for the same case the peak time-delay of $1037.0$ days instead of $1061.0$ days. Thus, in further analysis, we do not consider the trend subtraction.

\subsection{Discrete Correlation Function}
\label{subsec_DCF} 

\citet{1988ApJ...333..646E} suggested to use the Discrete Correlation Function (DCF) since the ICCF by definition introduces additional interpolated datapoints and can thus distort the time-delay determination, especially for the unevenly and sparsely sampled pairs of light curves. The basic algorithm is to search for data pairs $(x_i,y_j)$ between the two light curves that fall into the time-delay bin $\tau-\delta \tau/2\leq \Delta t_{ij} <\tau+\delta \tau/2 $, where $\tau$ is the time-delay, $\delta \tau$ is the chosen time-delay bin, and $\Delta t_{ij}=t_j-t_i$. Given $M$ such pairs, we can calculate the unbinned discrete correlation coefficient for each of them,
\begin{equation}
    UDCF_{ij}=\frac{(x_i-\overline{x})(y_j-\overline{y})}{\sqrt{(s_x-\sigma_x^2)}\sqrt{(s_y-\sigma_y^2)}}\,,
    \label{eq_udcf}
\end{equation}
where $\overline{x}$ and $\overline{y}$ are the light curve means in the given time-delay bin, $s_x$, $s_y$ are the variances and $\sigma_x$, $\sigma_y$ are the mean measurement errors for a given bin. The discrete correlation function for a given time-delay is calculated by averaging over $M$ data point pairs,
\begin{equation}
   DCF(\tau)=\frac{1}{M}\sum_{ij} UDCF_{ij}\,.
   \label{eq_dcf_tau}
\end{equation}
The error of the DCF can be formally inferred from the relation,
\begin{equation}
   \sigma_{\rm DCF}(\tau)=\frac{1}{M-1}\sqrt{\sum [UDCF_{ij}-DCF(\tau)]^2}\,.
\end{equation}

For our DCF analysis, we make use of the python code pyDCF by \citet{2015MNRAS.453.3455R} with the possibility of applying the Gaussian weighting scheme to matching pairs of both light curves. We also tested different time-delay bins as well as the searched time-delay intervals. In addition, we extended the DCF analysis by including the bootstrap technique to construct time-delay distributions and to infer the actual peaks and their uncertainties.  

We explore the correlation of the two light curves on two timescales:
\begin{itemize}
    \item between 0 and 1500 days, with a time-step of 120 days,
    \item between 200 and 1100 days, with a smaller time-step of 20 days.
\end{itemize}

\begin{table*}[h!]
    \centering
     \caption{Time-delay in light days corresponding to the peak values of DCF in the observer's frame. Two time intervals are analyzed: between 0 and 1500 days, and the narrower interval between 200 and 1100 days. The bottom two lines show the peak and the mean time-delay as inferred from 500 bootstrap realizations. \textit{Notes:} 1) This is the value for the maximum DCF for time delays less than 1300 days, the time-delay at 1315 days has the larger DCF of $0.82$, but this value can be excluded as it approaches the end of the observational run.}
    \begin{tabular}{c|c|c}
    \hline
    \hline
         &  With shifted BMT data   & Without BMT data   \\
    \hline      
    Time-delay at the DCF peak $(0,1500; 120)$     &  812.7 (DCF$=0.82$) &   812.7 (DCF$=0.80$)$^{1}$\\
    Time-delay at the DCF peak $(200,1100; 20)$    &  730.0 (DCF$=0.93$) &   730.0 (DCF$=0.92$)\\
    \hline
    Peak time-delay -- bootstrap [days] & $720.4^{+115.1}_{-147.9}$    &  $726.0^{+114.4}_{-145.7}$     \\
    Mean time-delay -- bootstrap [days] & $658.7^{+116.2}_{-139.2}$    &  $665.4^{+115.0}_{-142.8}$        \\
    \hline
    \end{tabular}
    \label{tab_DCF}
\end{table*}

\begin{figure*}[h!]
    \centering
    \begin{tabular}{cc}
    \includegraphics[width=0.5\textwidth]{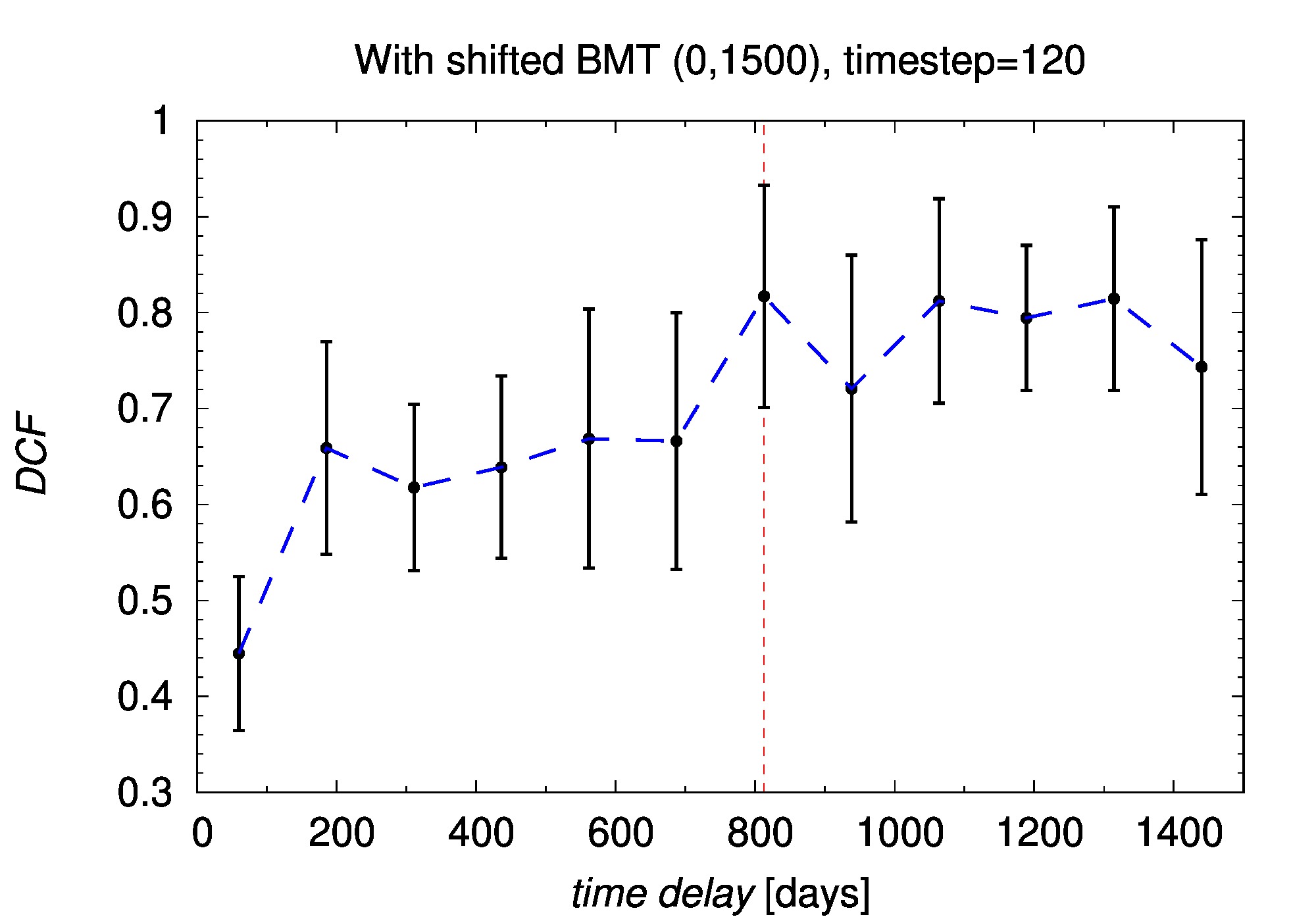} & \includegraphics[width=0.5\textwidth]{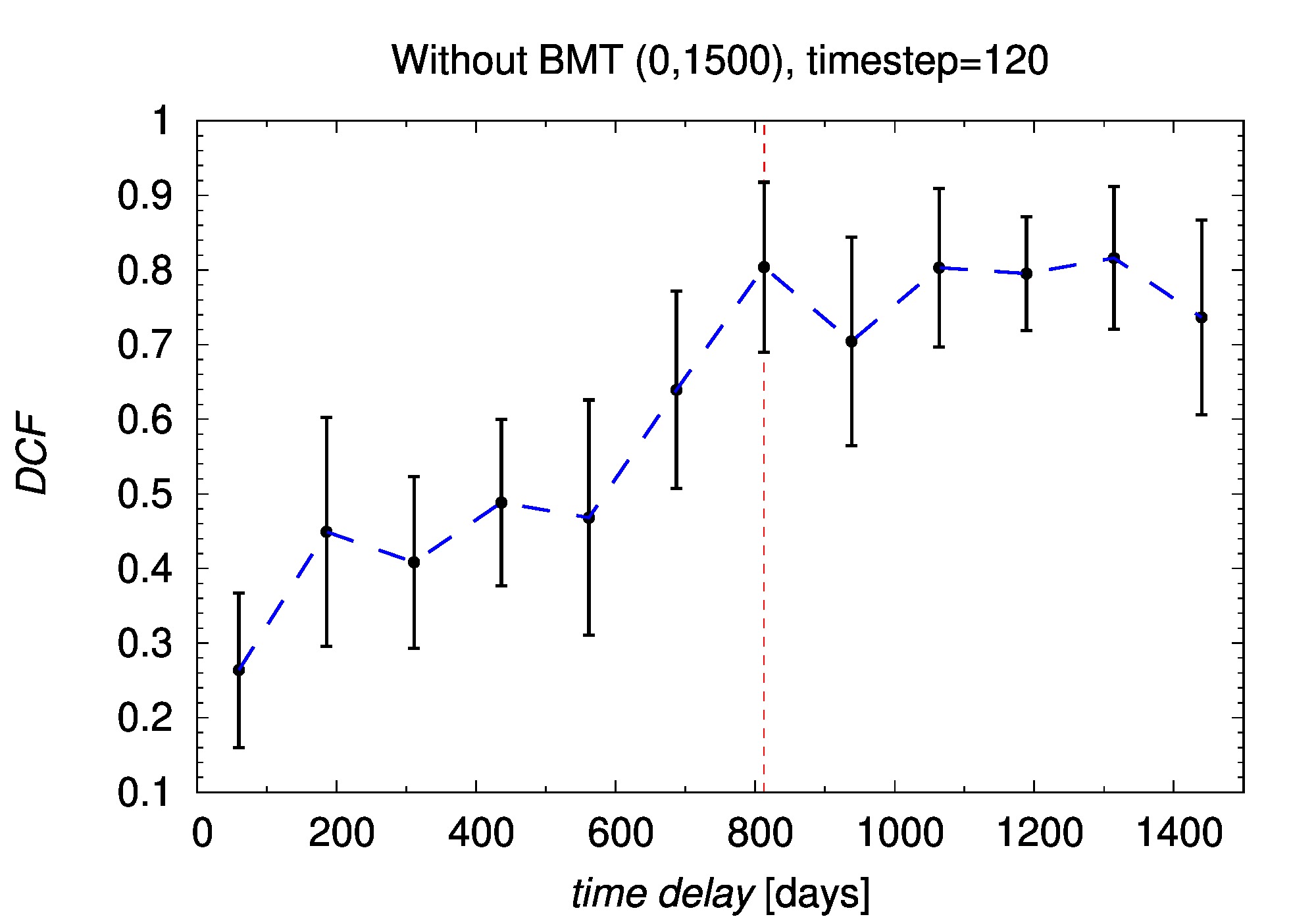}\\
    \includegraphics[width=0.5\textwidth]{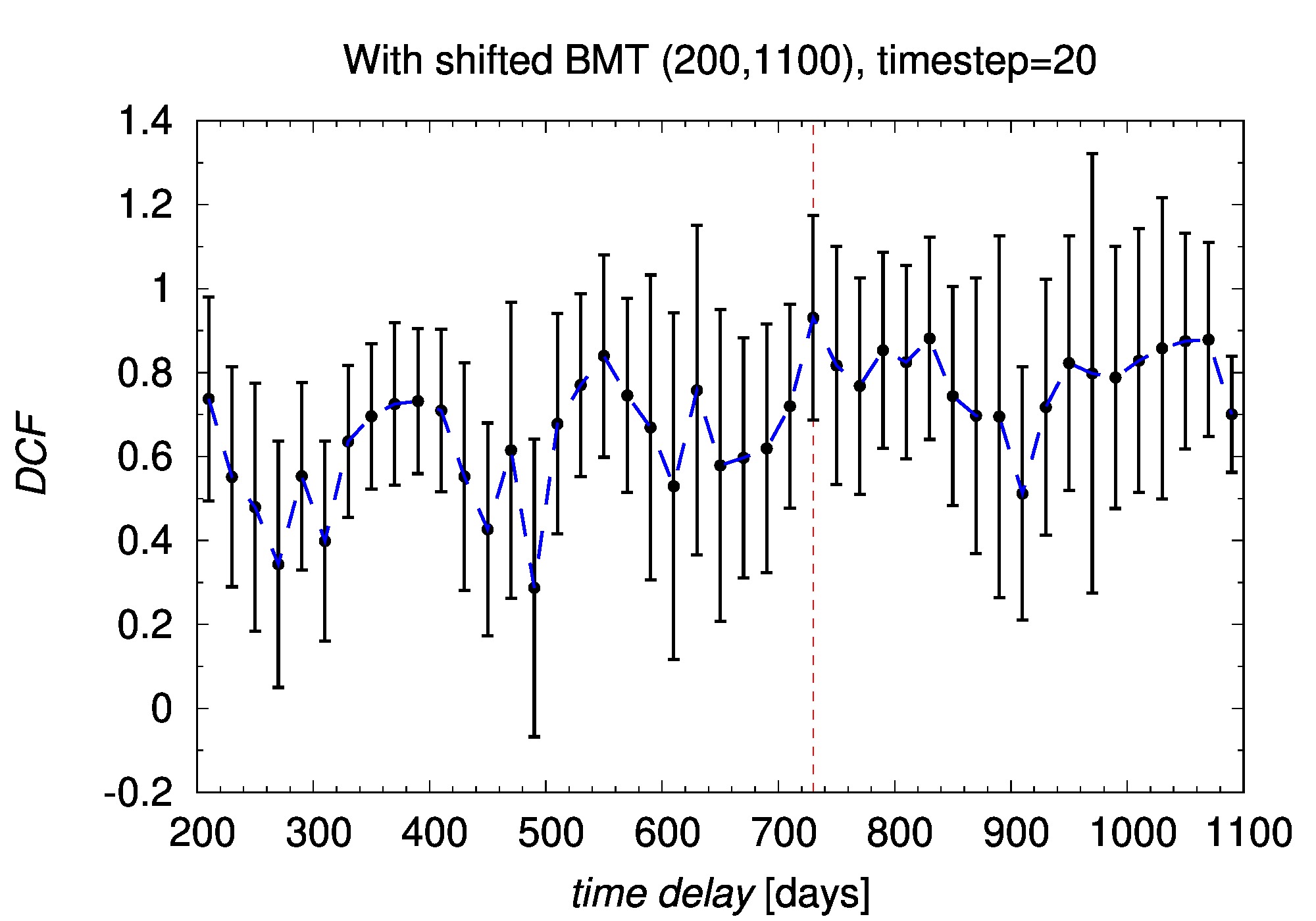} & \includegraphics[width=0.5\textwidth]{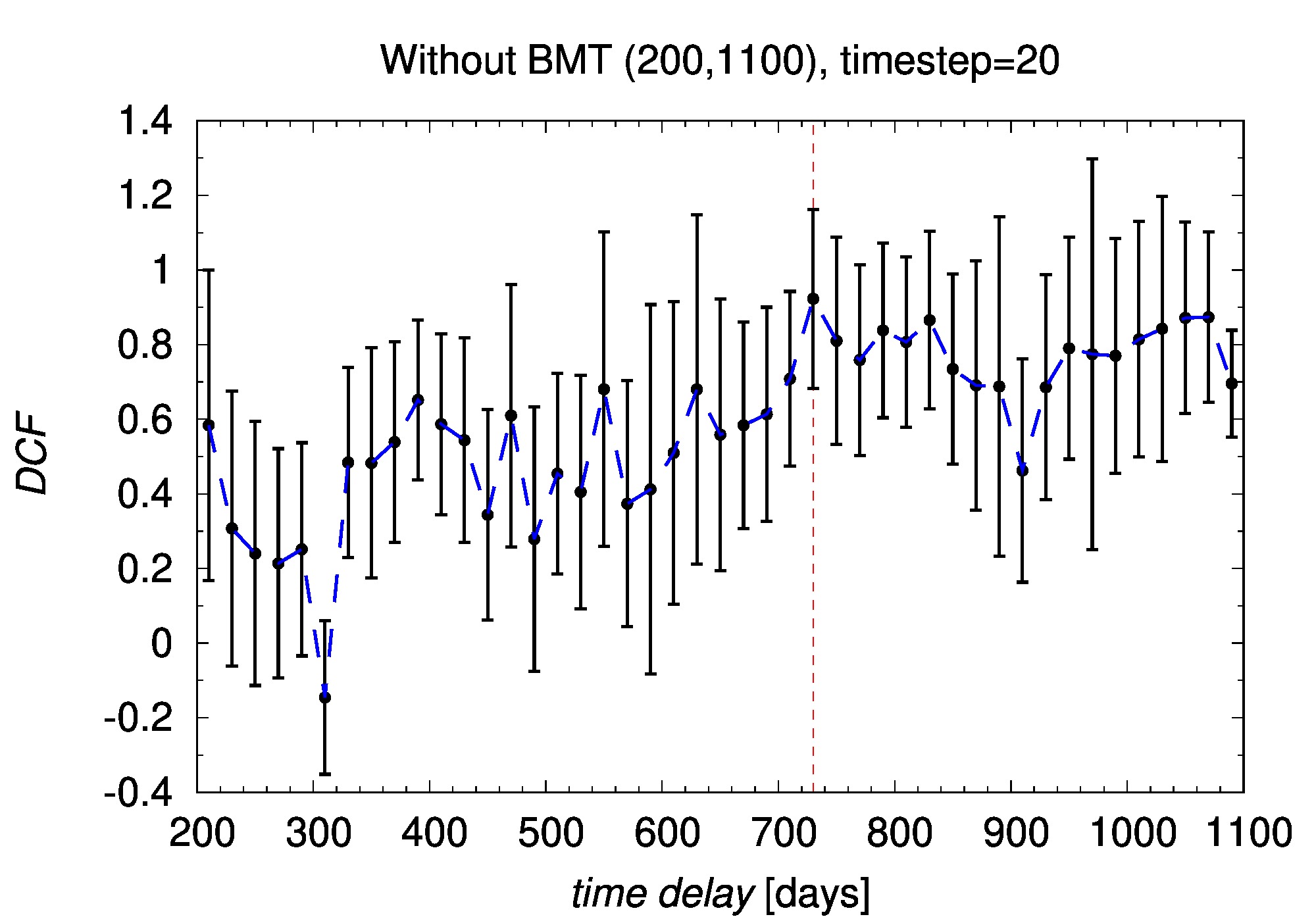}
    \end{tabular}
    \caption{The Discrete Correlation Function (DCF) as a function of the time-delay in the observer's frame. \textbf{Top panels:} The DCF determined between 0 and 1500 days, with a time-step of 120 days, to the left we include the shifted BMT flux densities for the continuum, to the right, they are omitted. The vertical dashed line denotes the time-delay for the maximum DCF. \textbf{Bottom panels:} Similar to the panel above, the DCF analysis was performed for the time-interval of 200-1100 days with a smaller time-step of 20 days. The vertical dashed line denotes the time-delay for the maximum DCF.}
    \label{fig_dcf}
\end{figure*}

As before for the ICCF analysis, we perform the DCF analysis with and without flux-shifted BMT data. The time-delays for the peak values of the DCF are shown in Table~\ref{tab_DCF}. The figures of the DCF versus the time-delay are in Fig.~\ref{fig_dcf} for the case with and without BMT datapoints in the left and right panels, respectively. In the top panels of Fig.~\ref{fig_dcf}, we show the whole explored time-range between 0 and 1500 days (with a time-step of 120 days), in the bottom panels, we display the DCF analysis in the time-range $(200, 1100)$ days with a smaller time-step of 20 days. 

\begin{figure*}[h!]
    \centering
    \begin{tabular}{cc}
        \includegraphics[width=0.5\textwidth]{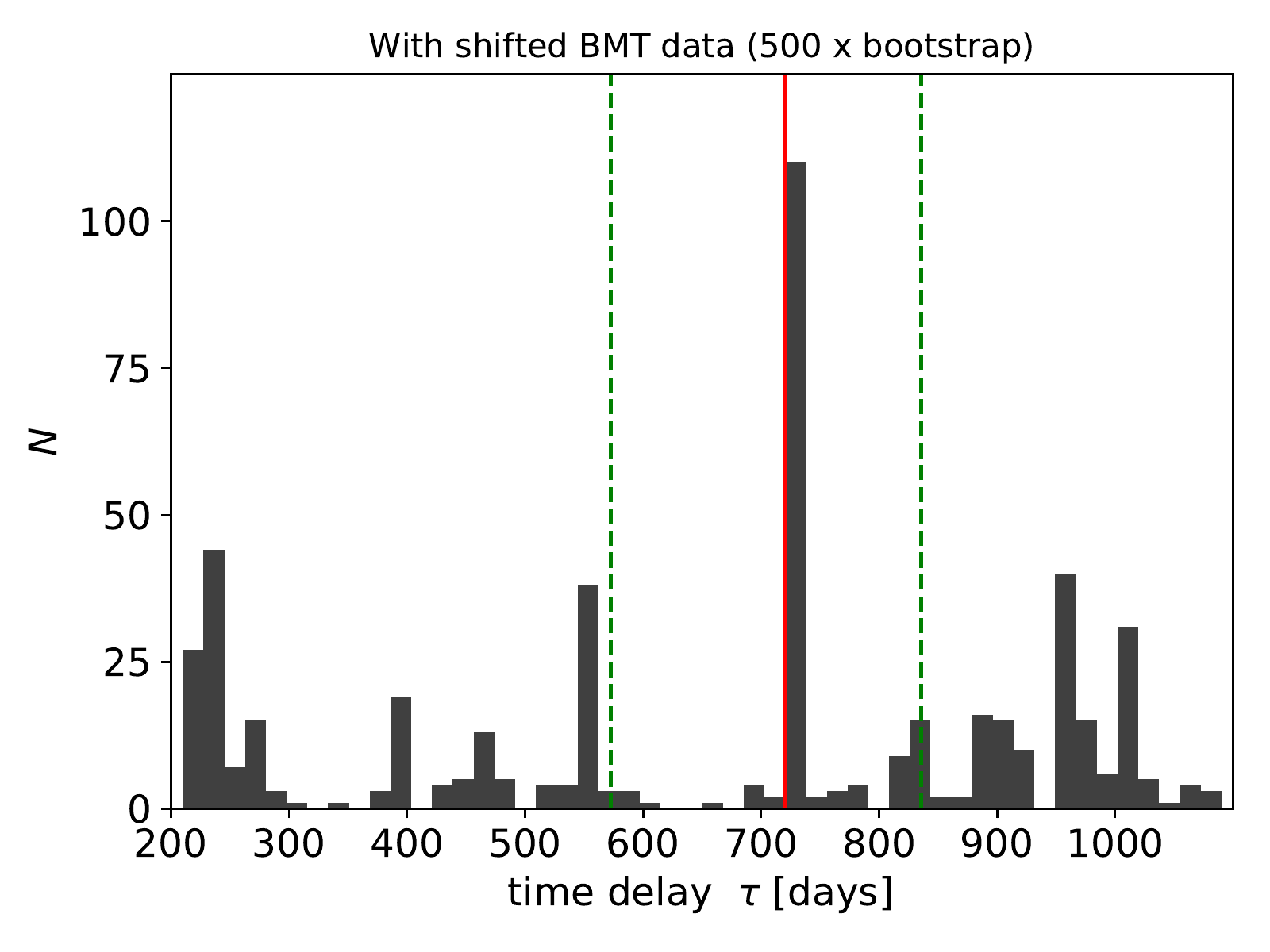} & \includegraphics[width=0.5\textwidth]{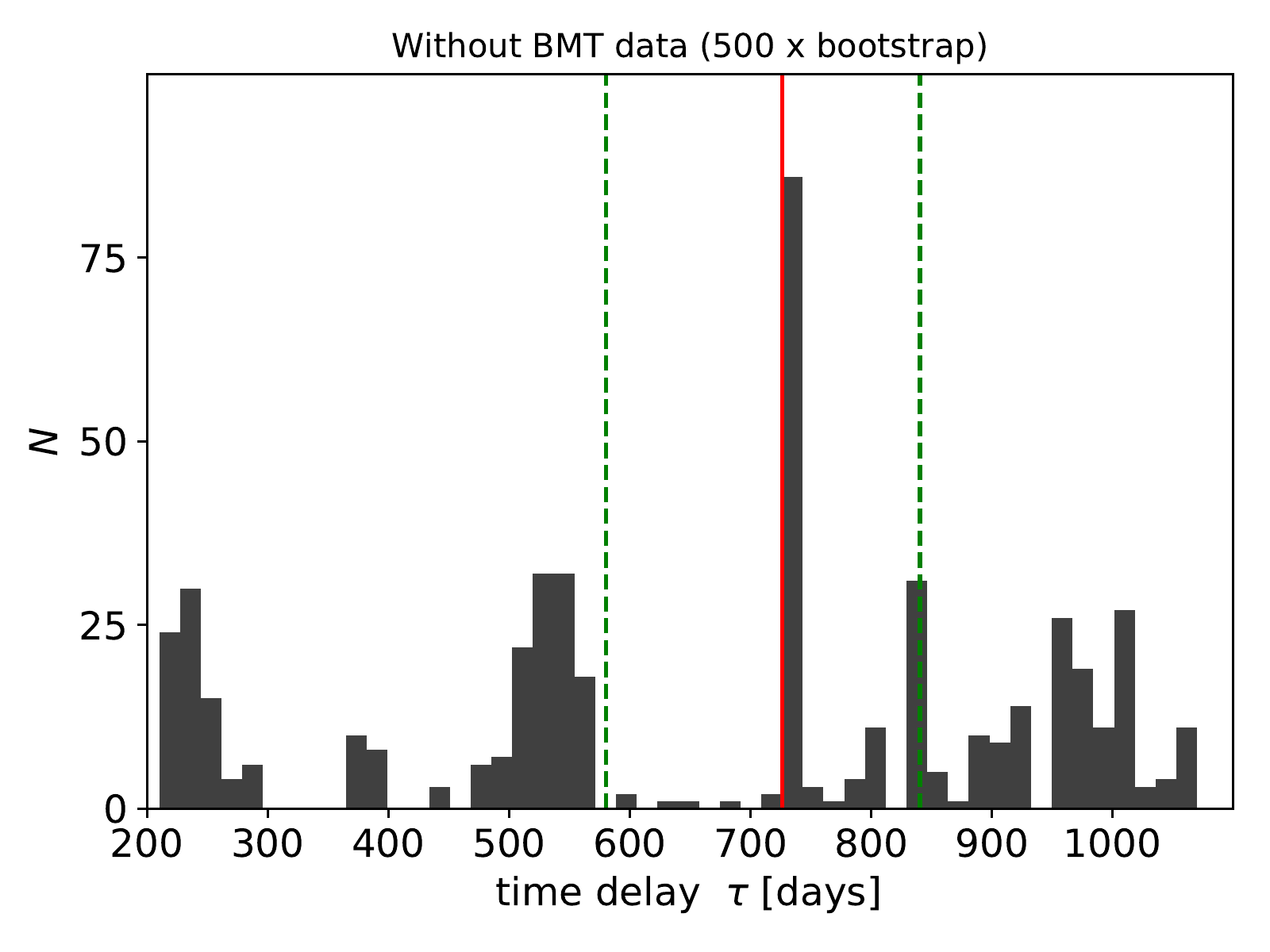}
    \end{tabular}
    \caption{Histograms of the time-delays constructed from 500 bootstrap realizations of the DCF analysis. \textbf{Left panel:} With flux-shifted BMT data included. \textbf{Right figure:} Without BMT data. In both panels, the red vertical line marks the histogram peak value and the two green horizontal lines stand for 1$\sigma$ uncertainties of the peak.}
    \label{fig_DCF_bootstrap}
\end{figure*}

To determine the uncertainty of the DCF peaks as well as the mean values for the time-delay, we perform 500 bootstrap simulations by randomly selecting subsamples of the light curves for both the cases with the flux-shifted BMT data and without them. The values of the peak and mean time-delays are in Table~\ref{tab_DCF}. The peak time-delay (for the largest DCF value) is clearly in the interval around 720-730 days as is also visible in the histograms in Fig.~\ref{fig_DCF_bootstrap} for both the cases with and without BMT data in the left and right panels, respectively.

\subsection{$z$-transformed Discrete Correlation Function  (zDCF)}
\label{subsec_zDCF} 

\citet{1997ASSL..218..163A} proposed the $z$-transformed Discrete Correlation function (zDCF) to correct several biases of the classical discrete correlation function \citep[DCF,][]{1988ApJ...333..646E}, namely it replaces equal time-lag binning with equal population binning and uses Fisher's $z$-transform. The minimum required number of observed points is 11, therefore the $z$-tranformed DCF is specially suited for undersampled, sparse and heterogeneous pairs of light curves, which is the case of our continuum and line-emission light curves as they are combined from different instruments. In addition, zDCF does not assume any light curve properties, such as smoothness, or any AGN variability process. Moreover, from Monte-Carlo generated pairs of light curves with randomized errors, it is possible to infer the uncertainty from the averaged zDCF values.

\begin{figure*}
    \centering
    \includegraphics[width=0.48\textwidth]{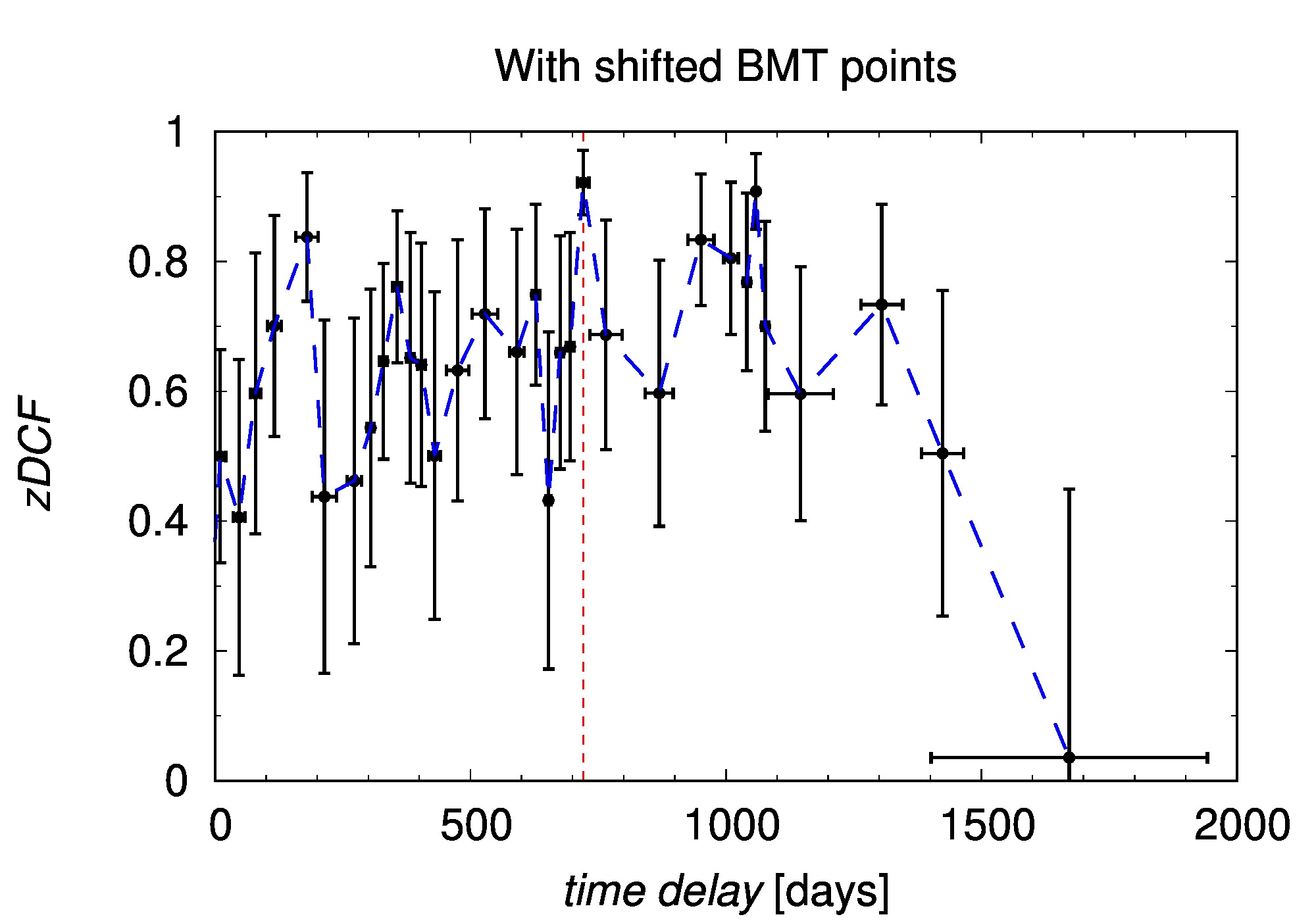}  \includegraphics[width=0.48\textwidth]{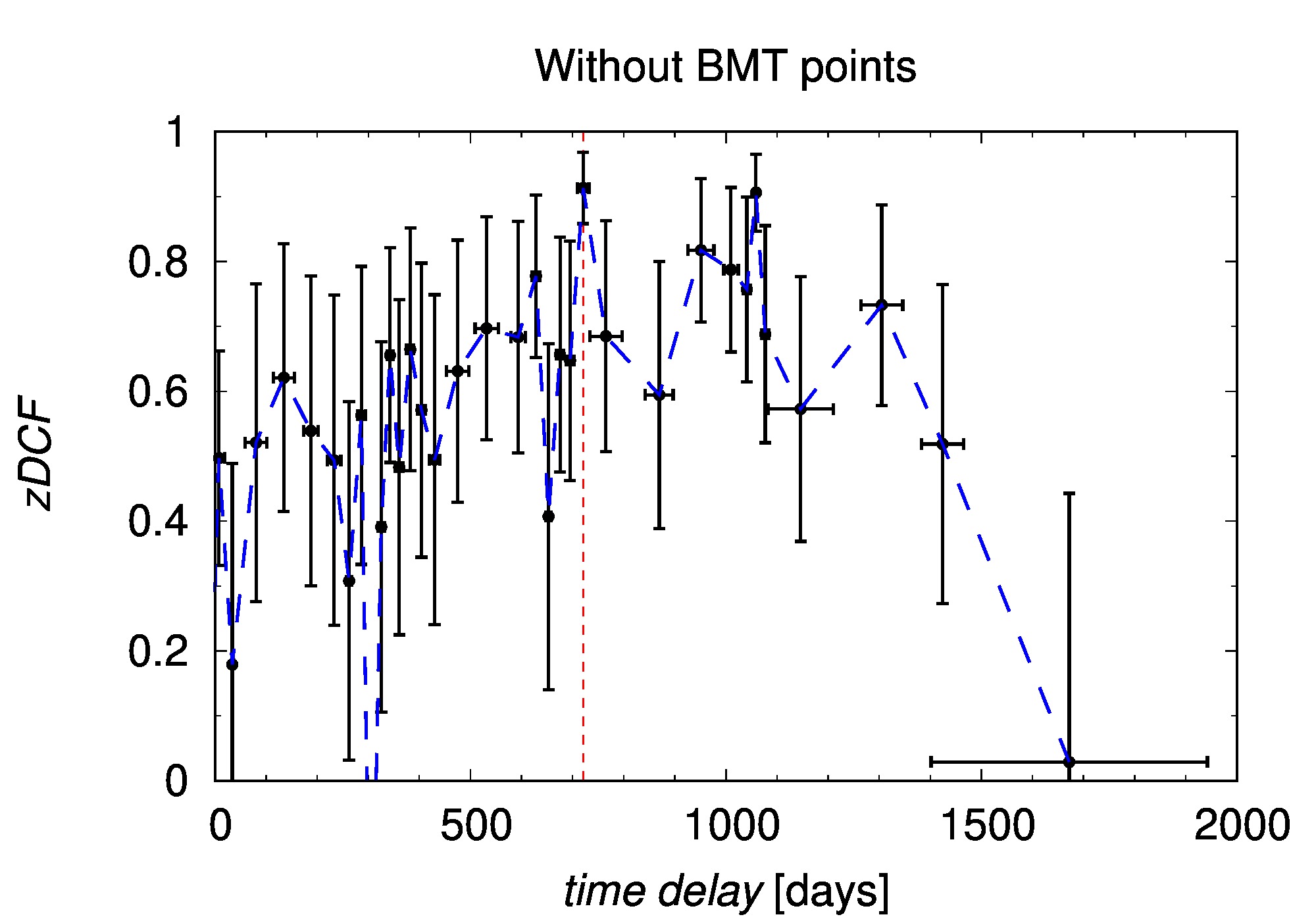}
    \caption{The zDCF values as a function of the time-delay, including the uncertainties for the time-delay and the zDCF values. \textbf{Left panel:} zDCF values as a function of the time-delay in the observer's frame based on the continuum light curve (including OGLE, SALTICAM, and flux-corrected BMT data) and MgII line-emission light curve (SALT telescope). The red dashed vertical line denotes the most prominent peak at $721$ days. \textbf{Right panel:} zDCF values versus the time-delay in the observer's frame as in the left panel, but without BMT datapoints.}
    \label{fig_zDCF}
\end{figure*}

For our zDCF analysis, we first used 86 continuum measurements (including data from OGLE, SALTICAM, and flux-corrected BMT) and 25 MgII line-emission points (from SALT spectral observations). The zDCF values as a function of the time delay are displayed in Fig.~\ref{fig_zDCF} including both errors for the time-delay and the zDCF value. Two peaks are apparent, $\tau_1=720.9$ days with zDCF$=0.92$, and $\tau_2=1059$ days with zDCF$=0.91$. To evaluate the uncertainties of these peaks, we ran the maximum-likelihood (ML) analysis for the surroundings of each peak, between 500 and 1000 days for the peak at $721$ days and 1000 and 1500 days for the peak at $1059$ days. In the next step, we performed global Maximum-Likelihood analysis of the time-delay peaks between 0 and 2000 days, with the most likely peak at $721^{+324}_{-527}$ days. The results are shown in Table~\ref{tab_zDCF} (left column).  

In the continuum light curve, the BMT points needed to be systematically shifted towards smaller flux densities to match OGLE and SALTICAM values. Therefore we also performed zDCF analysis without BMT points, with the total of 73 continuum points and 25 MgII line-emission points. The overall results concerning the time-delay peaks were not affected, see Table~\ref{tab_zDCF} (right column) and Fig.~\ref{fig_zDCF} (right panel). The global peak remained at $720.9^{+331.3}_{-100.1}$ days with a smaller lower uncertainty interval than for the case including BMT datapoints, but with a comparable likelihood value. 

\begin{table*}[h!]
    \centering
     \caption{Maximum-likelihood (ML) analysis for the zDCF time-delay values with and without flux-shifted BMT points included. The time-delays are expressed in light days in the observer's frame. The table contains results for the localized ML analysis, taking into account the surroundings of the two most prominent peaks at 721 and 1059 days. The lower part contains the peak of the global ML analysis in the searched interval between 0 and 2000 days. The actual maximum likelihood is denoted as $\mathcal{L}$ and its value is listed for the time-delay in each interval.}
    \begin{tabular}{c|c|c}
    \hline
    \hline
    Time-delay interval & With shifted BMT data & Without BMT data\\
    \hline
    500-1000 days & $720.9^{+78.8}_{-24.1}$, $\mathcal{L}=0.51$  & $720.9^{+80.6}_{-84.5}$, $\mathcal{L}=0.48$ \\
    1000-1500 days & $1059.0^{+219.5}_{-22.09}$, $\mathcal{L}=0.53$ & $1059.0^{+224.9}_{-19.6}$, $\mathcal{L}=0.54$\\
    \hline 
    0-2000 days & $720.9^{+323.9}_{-527.3}$, $\mathcal{L}=0.1434$ & $720.9^{+331.3}_{-100.1}$, $\mathcal{L}=0.12$ \\
    \hline
    \hline
    \end{tabular}
    \label{tab_zDCF}
\end{table*}

\subsection{The JAVELIN code package}
\label{subsec_javelin} 

Another way of estimating the time-delay is to model the AGN continuum variability as a stochastic process via the damped random walk process \citep[DRW;][]{2009ApJ...698..895K,2010ApJ...721.1014M,2010ApJ...708..927K,2016ApJ...826..118K}. The emission-line light curve is then modelled as a time-delayed, scaled, and smoothed response to the continuum stochastic variability. Based on this model assumptions, JAVELIN (Just Another Vehicle for Estimating Lags In Nuclei) code was developed \citep{2011ApJ...735...80Z,2013ApJ...765..106Z,2016ApJ...819..122Z}\footnote{Please visit \url{https://bitbucket.org/nye17/javelin/src/develop/} for more information on the code usage and application}. The JAVELIN package employs Markov Chain Monte Carlo (MCMC) to obtain posterior probabilities of the continuum variability timescale and amplitude. With this two parameters, distributions of three parameters -- time-delay, smoothing width of the top-hat function, scaling factor (ratio of the continuum and line-emission amplitudes $A_{\rm line}/A_{\rm cont}$) -- that describe the line-emission light curve are searched for.

\begin{figure*}[h!]
    \centering
    \begin{tabular}{cc}
    \includegraphics[width=0.5\textwidth]{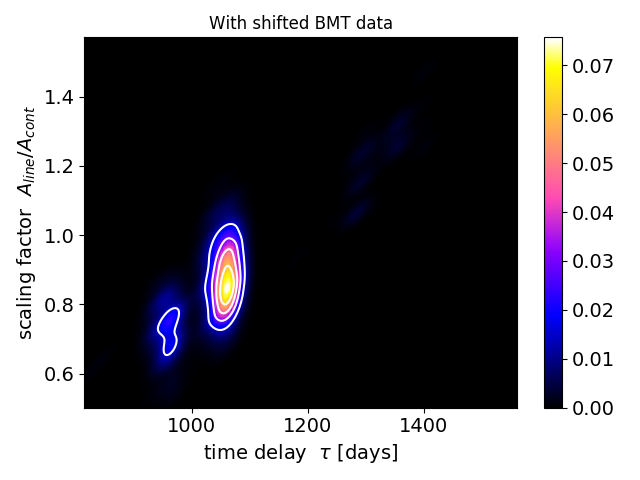} & \includegraphics[width=0.5\textwidth]{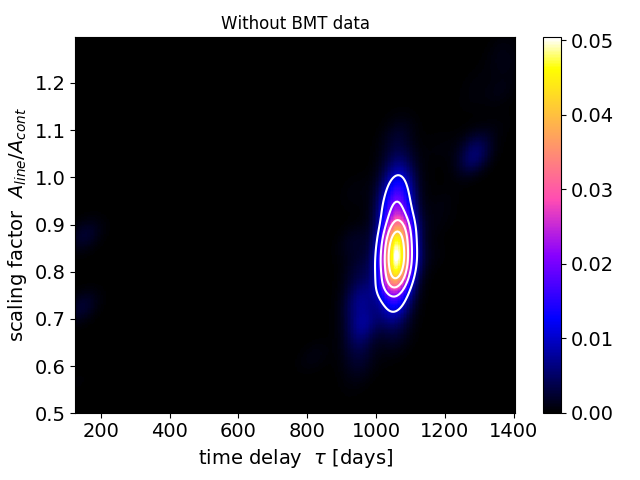}\\
    \includegraphics[width=0.5\textwidth]{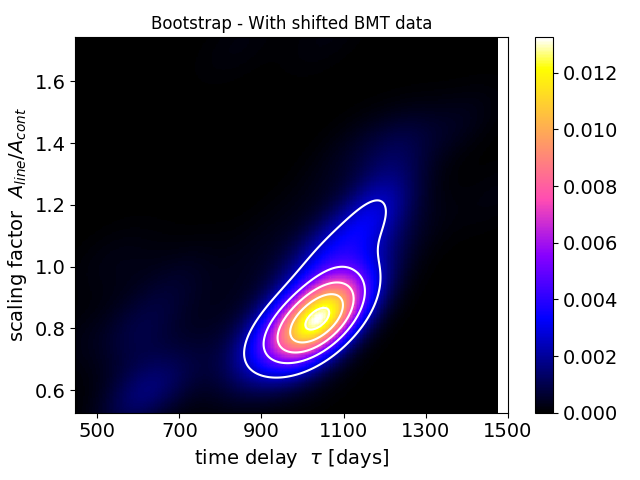} & \includegraphics[width=0.5\textwidth]{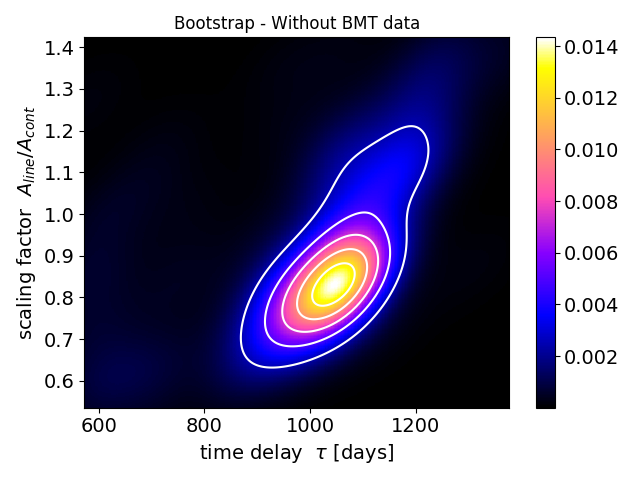}
    \end{tabular}
    \caption{Colour-coded plots of JAVELIN-code results in the time-delay/scaling factor plane. \textbf{Top row:} Time-delay distribution including magnitude-shifted BMT data (left panel) and without them (right panel). \textbf{Bottom row:} Distribution of the time-delay mean values from two hundred bootstrap realizations -- including shifted BMT data (left panel) and without them (right panel).}
    \label{fig_javelin_code}
\end{figure*}

\begin{table*}[h!]
    \centering
       \caption{The peak and mean values of the time-delay distribution (in the observer's frame) from 200 bootstrap realizations of JAVELIN. Time delays are expressed in light days.}
    \begin{tabular}{c|c|c}
    \hline
    \hline
         & With shifted BMT data & Without BMT data\\
    \hline     
   Peak time-delay [days]  &   $1053.7^{+79.8}_{-163.6}$  &  $1058.5^{+77.1}_{-150.7}$   \\
   Mean time-delay [days]  &    $1002.1^{+77.0}_{-161.8}$ &  $1016.0^{+70.5}_{-148.2}$   \\
    \hline     
    \end{tabular}
    \label{tab_JAVELIN}
\end{table*}

In Fig.~\ref{fig_javelin_code}, we show the distributions of the time-delay and scaling factor in the top panels with and without (magnitude-shifted) BMT data in the left and the right panels, respectively. Both the peak and the mean of the distributions are consistent within the uncertainties, with the peak close to $1050$ days.

To estimate the uncertainties for these time-delays, we ran 200 bootstrap realizations, generating randomly subsets of both light curves. The distributions of the means of time-delays are shown in bottom panels of Fig.~\ref{fig_javelin_code} both with shifted BMT data (left panel) and without them (right panel). The peak and mean time-delays with corresponding uncertainties are listed in Table~\ref{tab_JAVELIN}.   

\subsection{Measures of regularity/randomness -- von Neumann estimator}
\label{subsec_regularity} 

A novel technique to investigate time-delays is to measure regularity or randomness of data \citep{2017ApJ...844..146C}, which has previously been extensively applied in cryptography or electronic data compression. This method does not require interpolation of light curves as the ICCF, nor does it require binning in the correlation space as for DCF and zDCF. Moreover, the analysis is not based on any assumptions concerning the AGN variability in a way as the JAVELIN assumes for the continuum light curve. One of the most robust measures of the data regularity is an optimized von Neumann scheme, which uses the combined light curve $F(t,\tau)=\{(t_i,f_i)\}_{i=1}^{N}=F_1 \cup F_2^{\tau}$, where $F_1$ is the continuum light curve and $F_2^{\tau}$ is the time-delayed line-emission light curve. Based on the combined light curve, von Neumann estimator is defined as the mean successive difference of $F(t,\tau)$,

\begin{equation}
    E(\tau)\equiv \frac{1}{N-1}\sum_{i=1}^{N-1} [F(t_i)-F(t_{i+1})]^2\,.
    \label{eq_neumann_estimator}
\end{equation}
The minimum of the estimator $E$ is reached for a certain time-delay $\tau=\tau'$, which is expected to be close to the actual time-delay, $\tau'=\tau_0$. 

\begin{figure*}
    \centering
    \includegraphics[width=0.49\textwidth]{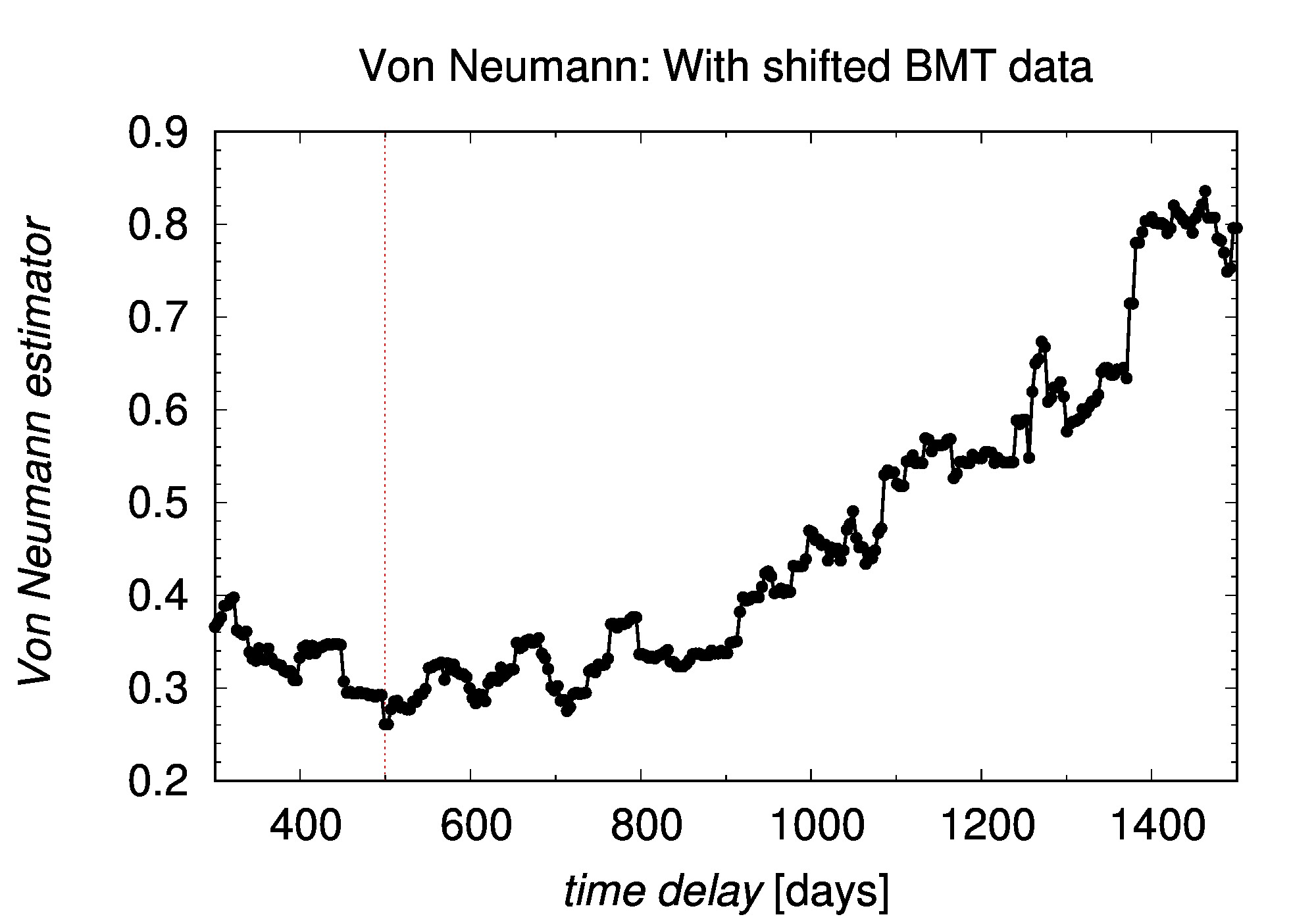}
    \includegraphics[width=0.49\textwidth]{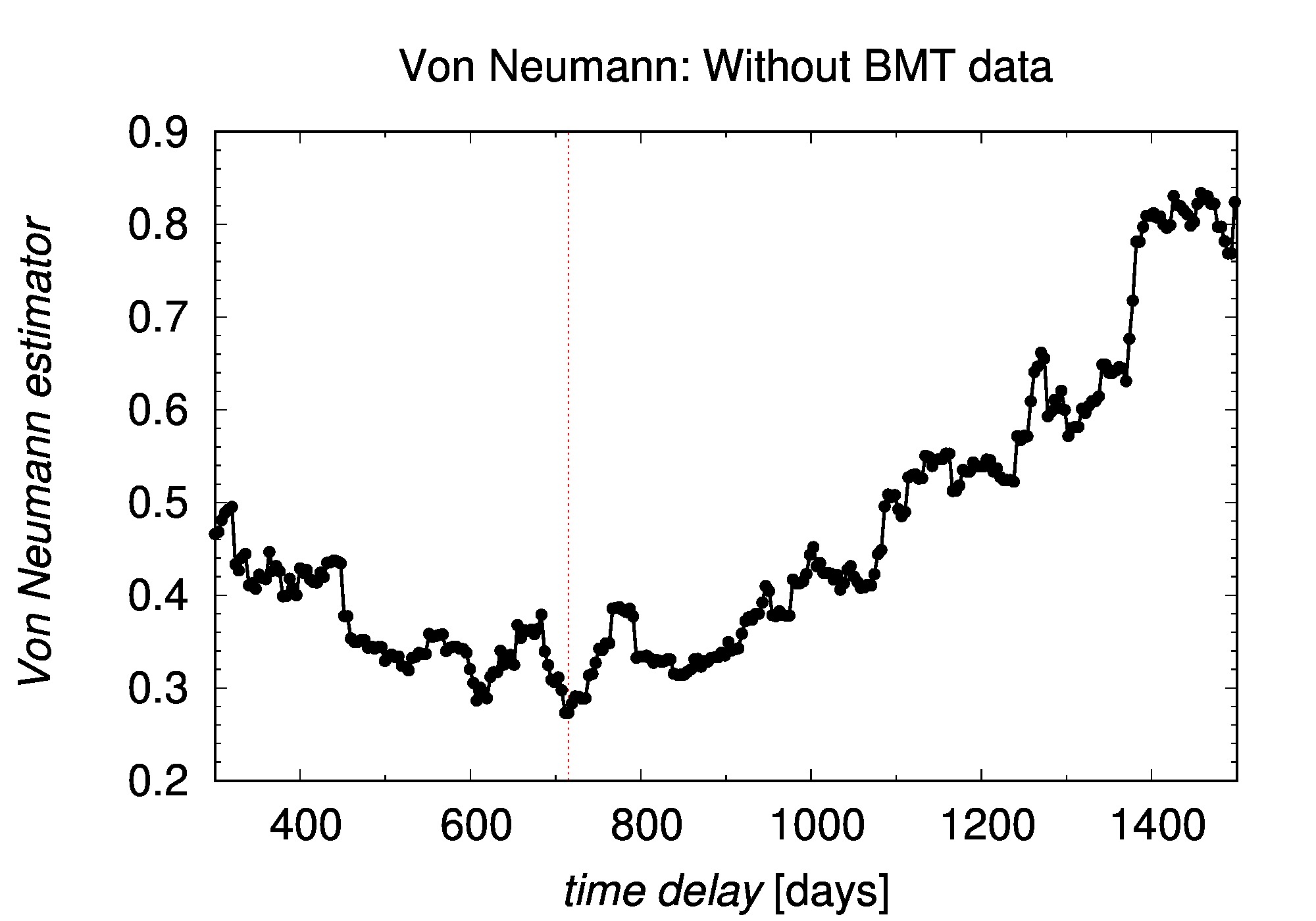}
    \caption{Von Neumann estimator as a function of the time-delay. \textbf{Left figure:} The case with magnitude-shifted BMT data, the minimum at $499.33$ days is depicted by a red vertical line. \textbf{Right panel:} The case without BMT data, the minimum at $715.18$ days is represented by a red vertical line.}
    \label{fig_von_neumann}
\end{figure*}

\begin{figure*}
    \centering
    \includegraphics[width=0.49\textwidth]{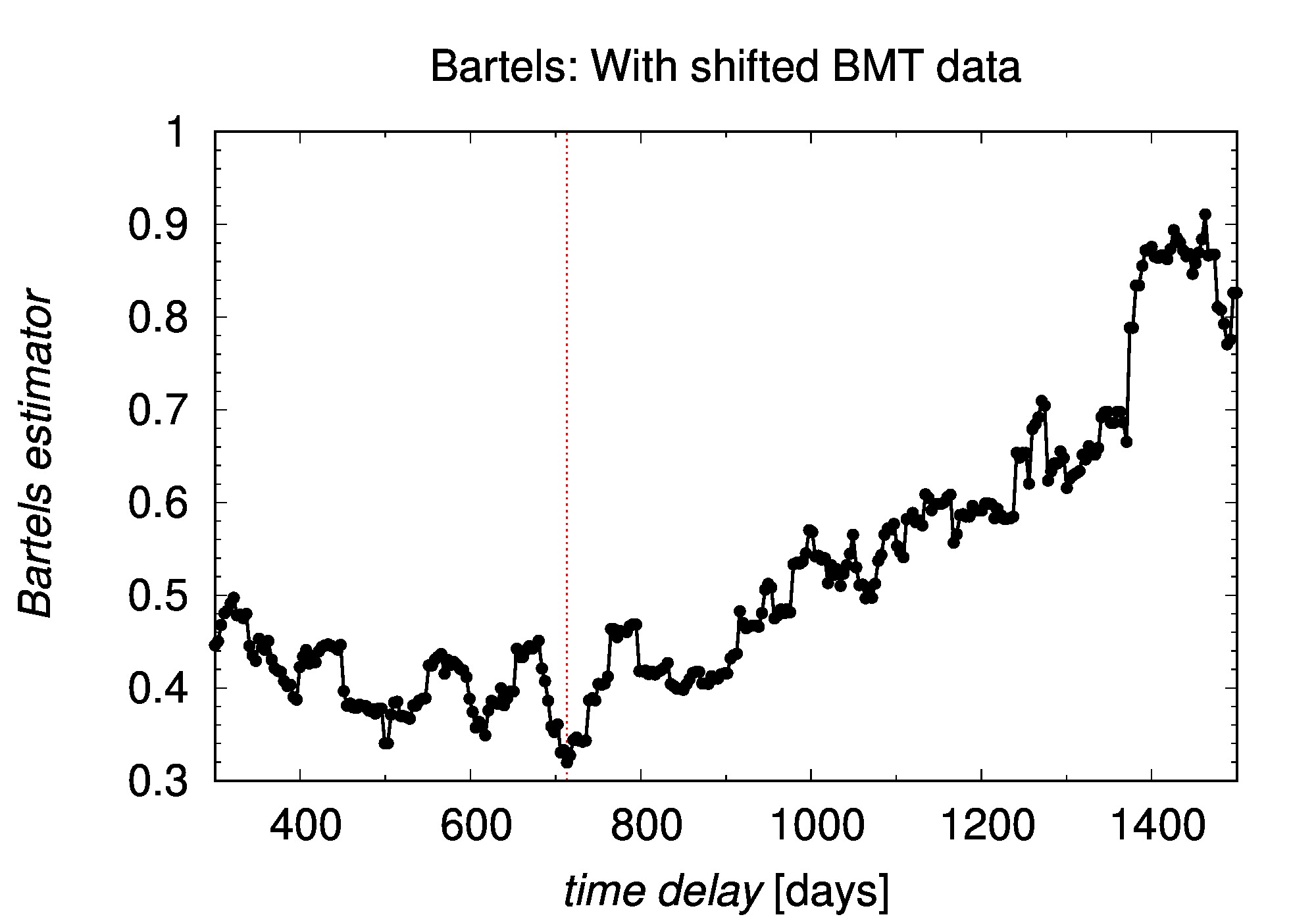}
    \includegraphics[width=0.49\textwidth]{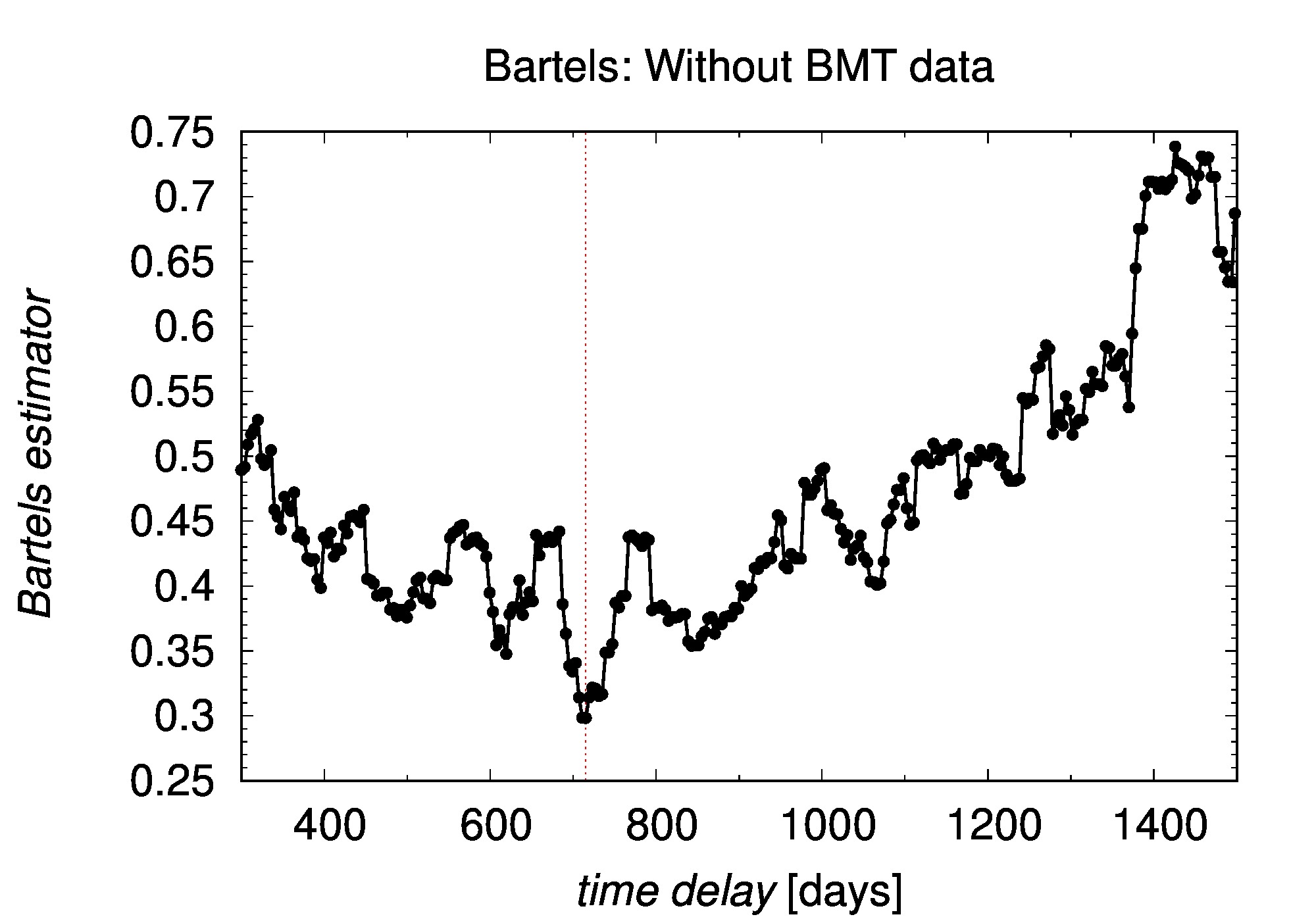}
    \caption{Bartels estimator as a function of the time-delay. \textbf{Left figure:} The case with magnitude-shifted BMT data, the minimum at $713.43$ days is depicted by a red vertical line. \textbf{Right panel:} The case without BMT data, the minimum at $715.18$ days is represented by a red vertical line.}
    \label{fig_bartels}
\end{figure*}

We applied the estimator to the data of HE 0413-4031 to estimate the time-delay between the continuum and MgII-line light curves. We made use of the python implementation of the estimator in Eq.~(\ref{eq_neumann_estimator}), which was demonstrated in \citet{2017ApJ...844..146C} \footnote{For the script, visit \url{www.pozonunez.de/astro_codes/python/vnrm.py}.}. In Fig.~\ref{fig_von_neumann}, we show the estimator value as a function of the time-delay with and without magnitude-shifted BMT data in the left and the right panels, respectively. For the case with shifted BMT data, we obtain the minimum of $E(\tau)$ at $499.3$ days, while without BMT data, the minimum is for the time-delay of $715.18$ days.

\begin{table*}[h!]
    \centering
    \caption{The upper part of the table shows the minima of the von Neumann estimator, the peak and the mean of minimum distributions for the case with and without magnitude-shifted BMT data in the left and the right columns respectively. The lower part displays the same information as above, but for the Bartels estimator, which is the modification of the Von Neumann's scheme using the ranked combined light curve.Time delays are expressed in light days in the observer's frame.}
    \begin{tabular}{c|c|c}
      \hline
      \hline
  Estimator & With shifted BMT data  & Without BMT data \\
     \hline     
Von Neumann: Minimum of $E(\tau)$ [days]  &  $499.33$   &  $715.18$ \\
Von Neumann: Bootstrap (10\,000) - peak [days] & $ 498.9^{+170.9}_{-125.9}$        &  $711.3^{+149.0}_{-139.5}$      \\
Von Neumann: Bootstrap (10\,000) - mean [days] & $588.5^{+157.3}_{-108.0}$ & $708.1^{+147.0}_{-137.9}$\\
\hline
Bartels: Minimum of $E(\tau)$ [days]  &  $713.43$   &  $715.18$ \\
Bartels: Bootstrap (10\,000) - peak [days] & $710.9^{+172.3}_{-173.0}$   &  $714.6^{+176.1}_{-164.6}$  \\
Bartels: Bootstrap (10\,000) - mean [days] & $ 634.6^{+161.9}_{-159.9}$ & $725.5^{+172.8}_{-150.3}$ \\
\hline
    \end{tabular}
    \label{tab_von_neumann}
\end{table*}

To construct distributions of the estimator minima, we perform $10\,000$ boostrap realizations for both cases with and without shifted BMT data. For both of these cases, the peak and the mean of the distributions are listed in Table~\ref{tab_von_neumann}. The minima and the peaks differ by about 200 days for the cases with and without BMT data. However, the minimum around 710 days is present for both cases, being a local minimum for the case with BMT data. This makes the peak at 710 days more robust, while we do not obtain any significant result for the peak at $1060-1070$ days, which we obtained using JAVELIN and ICCF methods.  

In addition, we perform the same analysis using the Bartels estimator, which is a modification of the Von Neumann's estimator using the ranked unified light curve $F_{\rm R}(t,\tau)$ \citep{bartels82}. In comparison with the pure Von Neumann's scheme, Bartels modification of the estimator has a consistent global minimum at $713.43$ and $715.18$ days for both cases with and without BMT photometry data, respectively, see Fig.~\ref{fig_bartels} (left and right panels, respectively). The peak values of the time-delay distribution in the observer's frame are also comparable, see Table~\ref{tab_von_neumann}. The mean value of the time-delay distribution is smaller for the case with the BMT data included, but within uncertainties the mean values of the time-delay are still comparable.

\subsection{$\chi^2$ method}
\label{subsec_chi2} 

As for the quasar CTS C30.10 \citep{czerny2019}, we also apply the $\chi^2$ method to the light curves. It was found that the $\chi^2$ method, which is frequently used in quasar lensing studies, works better than the ICCF for the AGN variability modelled as a red noise process \citep{2013A&A...556A..97C}. The light curves were prepared as for the standard ICCF, that is mean values were subtracted from them and they were normalized by their corresponding variances. Subsequently, the spectroscopic light curve was time-shifted with respect to the photometry light curve. The datapoints were linearly interpolated, but since the photometry light curve is denser than the spectroscopic light curve, we interpolated the photometry to the spectroscopy, i.e. we performed an asymmetric interpolation. Finally, we estimated the degree of similarity between shifted light curves by calculating the $\chi^2$, whose minimum may be considered as the most likely time-delay between the continuum and the line emission.

\begin{figure*}[tbh]
    \centering
    \includegraphics[width=0.49\textwidth]{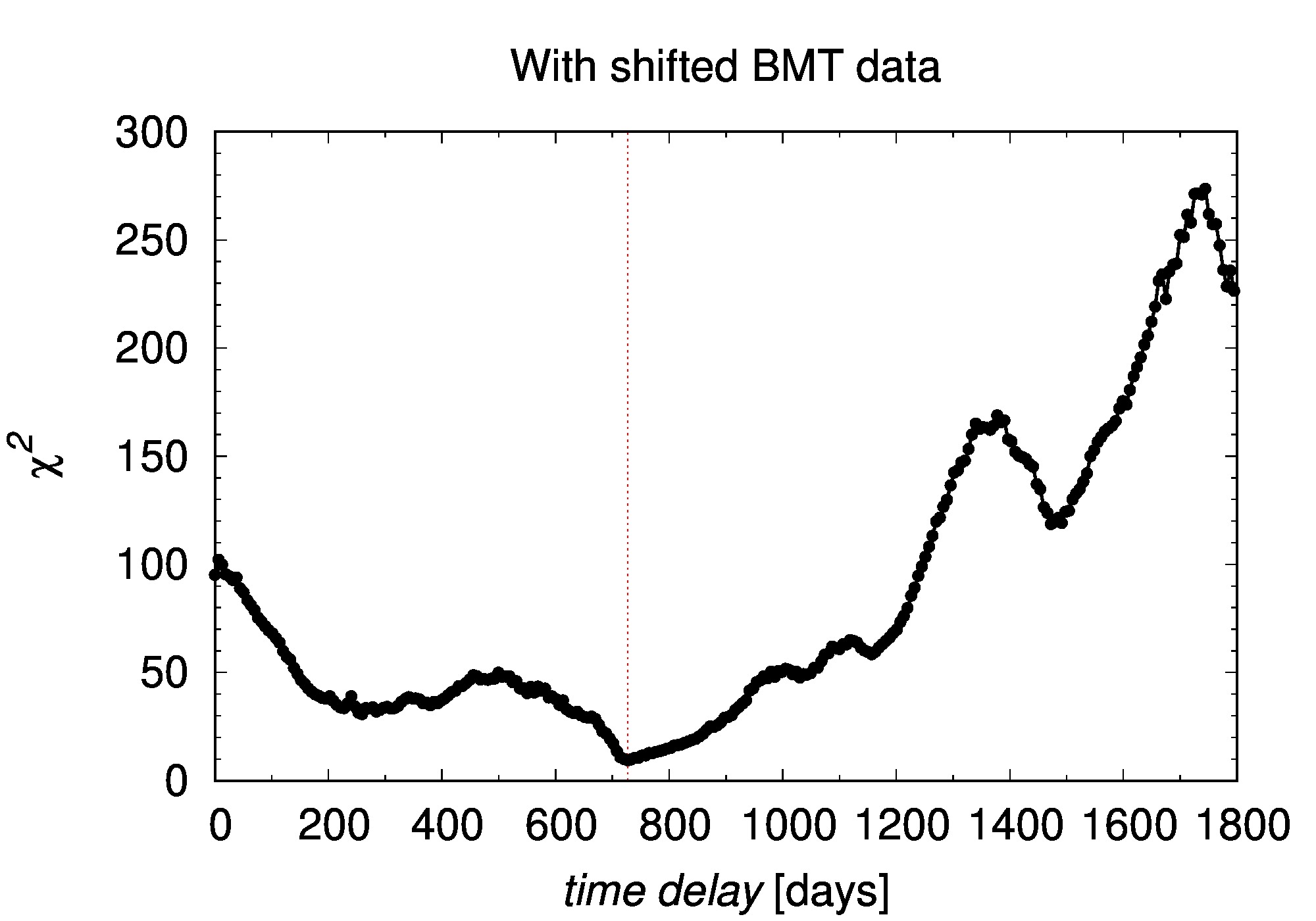}
    \includegraphics[width=0.49\textwidth]{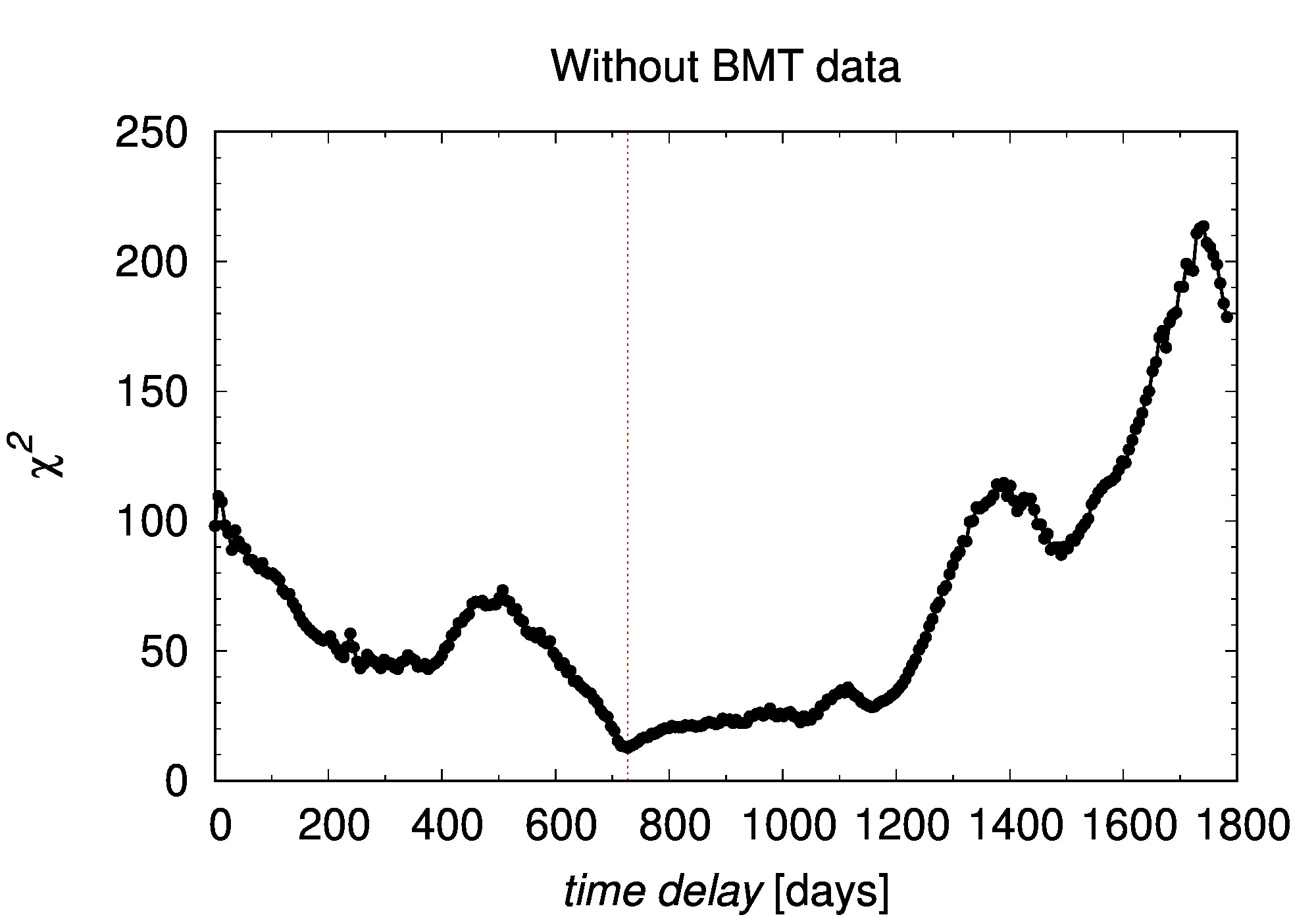}
    \caption{The values of the $\chi^2$ statistic as a function of the time-delay expressed in days with respect to the observer's frame of reference. \textbf{Left panel:} The $\chi^2$ values calculated for the case with the shifted BMT photometry data. \textbf{Right panel:} The $\chi^2$ values calculated for the case without the BMT photometry data.}
    \label{fig_chi2}
\end{figure*}

\begin{figure*}[tbh]
    \centering
    \includegraphics[width=0.49\textwidth]{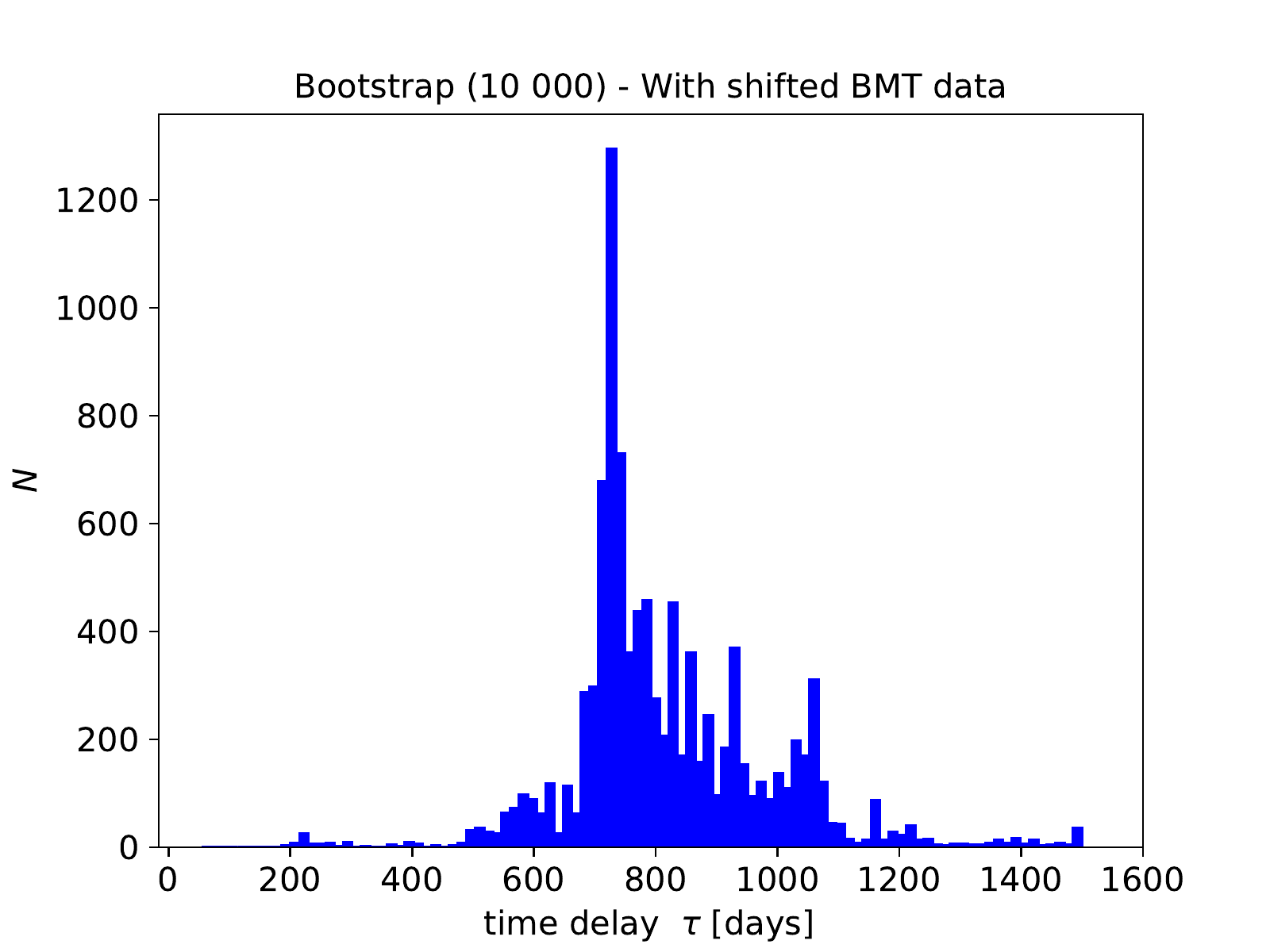}
    \includegraphics[width=0.49\textwidth]{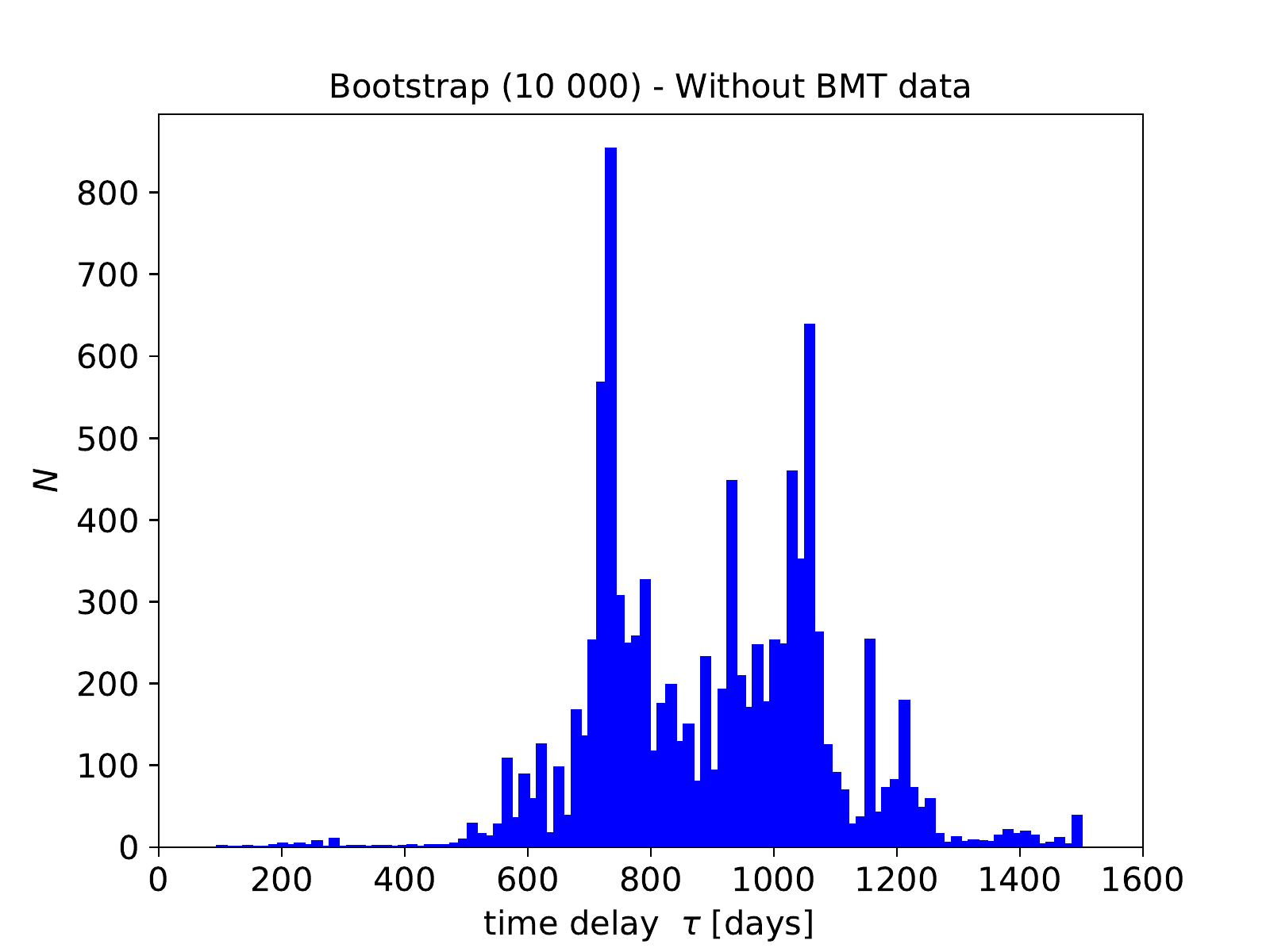}
    \caption{Distributions of the time-delays expressed in days in the observer's frame of reference for $10\,000$ bootstrap realizations. \textbf{Left panel:} The time-delay distribution based on the $\chi^2$ analysis calculated for the case with the shifted BMT photometry data. \textbf{Right panel:} The time-delay distribution based on the $\chi^2$ analysis calculated for the case without the BMT photometry data.}
    \label{fig_chi2_distribution}
\end{figure*}

In Fig.~\ref{fig_chi2}, we show the $\chi^2$ values as a function of the time-delay in the observer's frame of reference for the case with the magnitude-shifted BMT data included (left panel) and without them (right panel). In both case, the global minimum of $\chi^2$ is close to $\tau=727$ days. To determine the uncertainty of this minimum, we construct the distributions of the time-delay by performing $10\,000$ bootstrap realizations, i.e. by creating randomly selected subsets of both light curves and using the $\chi^2$ method for each new pair. The distributions are displayed in Fig.~\ref{fig_chi2_distribution}. Both cases with and without the BMT data have one main peak and secondary peaks towards longer time-delays. The peak and mean values of the distributions are listed in Table~\ref{tab_chi2}. The peak and the mean values are within uncertainties consistent, with the mean values shifted towards larger values with respect to the peak values because of the presence of secondary peaks at larger time-delays.

\begin{table*}[h!]
    \centering
     \caption{Results of the $\chi^2$ analysis of the time-delay for two cases: with and without shifted BMT data. We list the $\chi^2$ minima, the peaks, and the means of the time-delay distributions expressed for the observer's frame of reference. Time-delays are expressed in light days.}
    \begin{tabular}{c|c|c}
    \hline
    \hline
    Statistic & With shifted BMT data & Without BMT data\\
    \hline 
    $\chi^2$ minimum [days] &    $726.86$        &  $727.41$\\
    Bootstrap ($10\,000$) - peak [days]   & $720.4^{+145.6}_{-102.2}$   &   $727.7^{+160.0}_{-85.2}$        \\
    Bootstrap ($10\,000$) - mean [days]     & $818.6^{+133.0}_{-85.2}$     &  $900.0^{+110.8}_{-91.3}$        \\
    \hline
    \end{tabular}
    \label{tab_chi2}
\end{table*}

\section{Tests of complementary UV FeII templates }
\label{sec_appendix_FeII_templates}

We used the theoretical FeII templates from \citet{bruhweiler2008} (hereafter d12 template) in our basic modelling, since one of these templates,  d12-m20-20-5.dat, allowed us to get very nice and simple fits to all data sets. However, other FeII templates are also used. Therefore, for our mean spectrum, we additionally tested two other templates. The first one was a semi-empirical template of \citet{Tsuzuki2006} (hereafter T06) based on a combination of 14 low redshift quasars and CLOUDY modelling of the FeII emission to disentangle the FeII and MgII contribution. This template combined the advantage of previously used purely observational template of \citet{2001ApJS..134....1V} and a theoretical modeling. The results are given in Table~\ref{tab:Fe_templates}. The fits obtained with this template had the same number of free parameters as the fits with d12-m20-20-5.dat, if a single Lorentzian is used for MgII, but the fit quality is always lower. The fits do not depend strongly on the doublet ratio, since the MgII line is unresolved, and the change of the doublet ratio is easily compensated with the change of the redshift and the shift of the MgII line with respect to the FeII emission. Even if we add the second kinematic component, fits do not improve considerably, and final $\chi^2$ is higher that for our canonical fits.

\begin{table}
\caption{An overview of the parameters used for fitting different FeII templates to the mean spectrum as well as the inferred best-fit parameters. From the left to the right, the parameters are the template name (d12 according to \citet{bruhweiler2008}, T06 according to \citet{Tsuzuki2006}, and the KDP15 template based on \citet{kovacevic15,2019MNRAS.484.3180P}), the MgII shape (number of either Lorentzian or Gaussian profiles), the redshift (the star $^{*}$ for the KDP15 template means that the redshift was fixed in this case based on the best-fit d12 value), the MgII doublet ratio (between one and two), FeII smear velocity, the equivalent width of MgII line, and $\chi^2$ in the last column.}
\centering
{\renewcommand{\arraystretch}{1.5}
\begin{tabular}{c|c|c|c|c|c|c}
 \hline
 \hline 
FeII  & MgII Shape & Redshift & Doublet Ratio & FeII smear Velocity & EW(MgII) & $\chi^2$ \\ 
template & Shape & & & [km s$^{-1}$] & \\
 \hline
 \hline
d12     & 1 Lorentz & 1.37648 & 1.6 & 2800 & 27.44 & 2088.42 \\
T06 & 1 Lorentz &1.38205 & 1.0 & 4500 & 29.64 & 2476.92\\ 
T06 & 1 Lorentz &1.38323 & 1.7 & 4700 & 29.77 & 2482.84\\  
T06 & 2 Lorentz& 1.38323 & 1.7 & 4700 & 29.74 & 2299.37\\
KDP15 & 1 Lorentz & 1.37648$^*$ & 1.6 & 5000 & 25.45 & 2921.12 \\
KDP15 & 2 Gauss & 1.37648$^*$ & 1.6 & 4000 & 22.13 & 1749.18 \\
KDP15 & 2 Gauss & 1.37617 & 1.6 & 4000 & 22.46 & 1711.34 \\
KDP15 & 2 Gauss & 1.389 (NED) & 1.6 & 5600 & 19.02 &   1991.62 \\

 \hline
 \end{tabular}} \quad
 \label{tab:Fe_templates}
\end{table}

Next we incorporate the semi-empirical UV FeII template\footnote{\href{http://servo.aob.rs/FeII_AGN/link7.html}{http://servo.aob.rs/FeII\_AGN/link7.html}} \citep[hereafter referred to as the KDP15 template,][]{kovacevic15,2019MNRAS.484.3180P} to fit the FeII pseudocontinuum i spectral window 2700-2900 \AA~. This model includes overall 7 free parameters, which includes 5 multiplets, namely, 60 (a$^{4}D$ - z$^{6}F^o$), 61 (a$^{4}D$ - z$^{6}P^o$), 62 (a$^{4}D$ - z$^{4}F^o$), 63 (a$^{4}D$ - z$^{4}D^o$) and 78 (a$^{4}P$ - z$^{4}P^o$). Additionally, there is empirically added component  `I Zw 1 lines' that is represented with two Gaussians (at $\lambda\lambda$2720,2840 \AA). This additional empirical set of lines were included in the model as they were not identified in the emission within $\sim$ 2825-2860 \AA~ and 2690-2725 \AA~. The remaining parameter is the line width \citep[see Appendix A1 in][for more details]{2019MNRAS.484.3180P}.

The fits with this template are also given in Table~\ref{tab:Fe_templates}. In the case of a single kinematic component, the provided fits are again not better in comparison with our standard fits. However, if we allow for two kinematic components for MgII, indeed the resulting $\chi^2$ is lower, particularly if we optimize the redshift to the new template. Although during the fitting we allowed for all the six template components to vary, we noticed in the final fits that the multiplets 61, 63 and 78 converge to values close to zero, and only the multiplets 60, 62 and the `I Zw 1 lines' return non-zero values. This is consistent with the Figure A1 in the paper of \citet{2019MNRAS.484.3180P}. In their Figure, the multiplets 60 and 78 are outside our spectral window and multiplet 63 has a very weak contribution (by a factor $\sim$3 - 3.5 times with respect to multiplet 62). 

\begin{figure}[tbh]
    \centering
       \includegraphics[width=\textwidth]{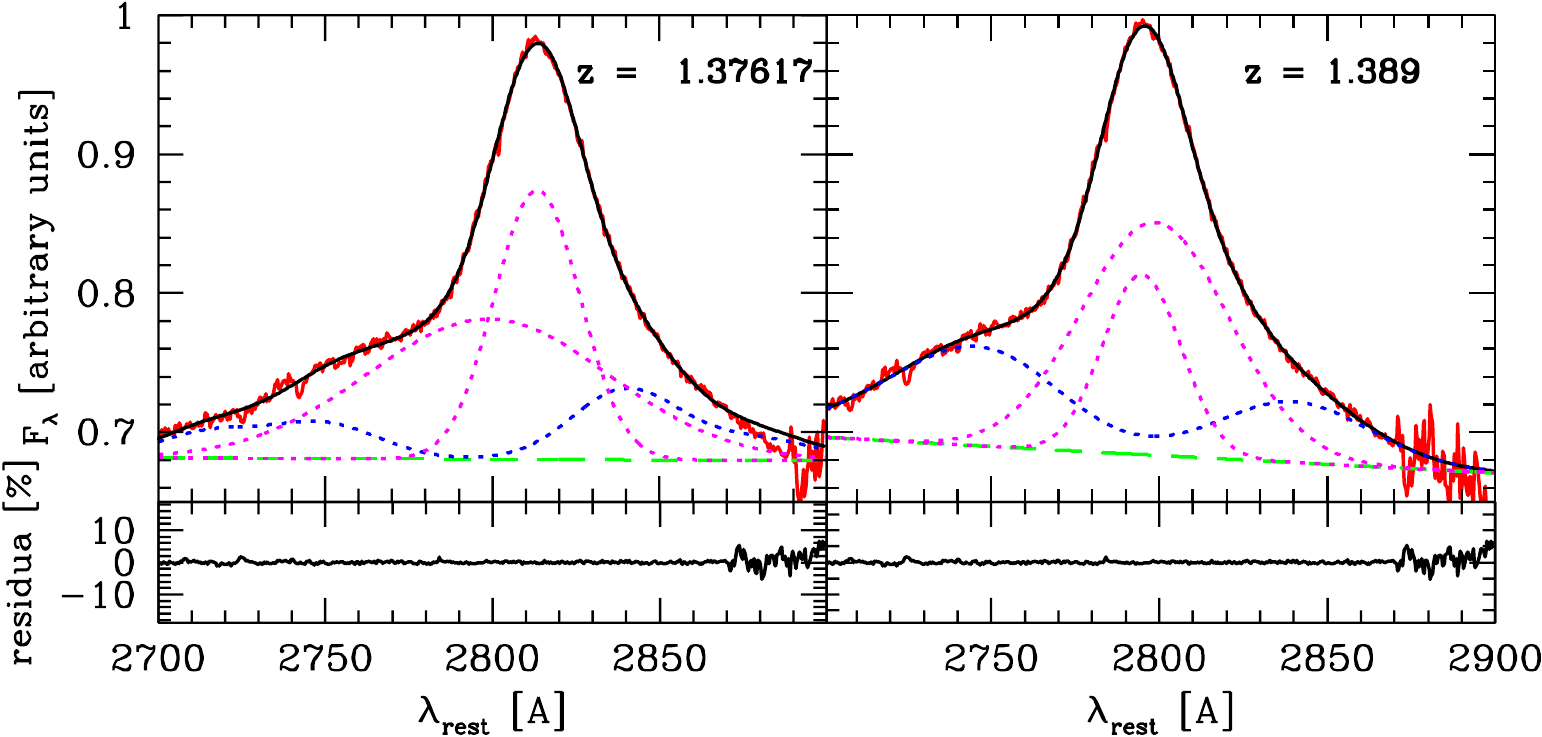}
    \caption{{\bf Left panel:} The best fit to the mean spectrum with the FeII template \citep{kovacevic15,2019MNRAS.484.3180P} and 2 Gaussian components, for the best fit redshift of $1.37617$. Visually, the residuals are similar, but the implied shape of the MgII line (dashed magenta) is very different from our standard fit shown in Figure~\ref{fig:mean}. {\bf Right panel:} Fit of the same model but for the redshift from NED, $z=1.389$.}
    \label{fig:mean_Serbia}
\end{figure}

This new best fit implies a different shape of the MgII line and different kinematics of the FeII and MgII emitting region. With d12-m20-20-5.dat, the MgII line was represented by a single Lorentzian, and FWHM of the MgII line was somewhat broader than the requested FWHM of the FeII (4380 km s$^{-1}$, and 2800 km s$^{-1}$, respectively), implying that FeII emission comes on average from a little more distant part of the BLR. In the case of the new best fit, the requested FWHM of FeII is larger, 4000 km s$^{-1}$, and the two components of MgII, if treated as separate components, have the corresponding values of FWHM of 3100  km s$^{-1}$ and  9050 km s$^{-1}$, respectively, thus considerable part of the MgII emission should originate at larger distance than FeII. If the two MgII components are treated as a single asymmetric line, then the FWHM of MgII is 4250 km s$^{-1}$ just above that for FeII emission in this model, but effectively similar to the FWHM of MgII from the basic model. However, the overall line shape is widely different, and we present the new fit in Figure~\ref{fig:mean_Serbia}. The very broad MgII component is then located at the same position as the FeII emission, but the narrower MgII component is again shifted considerably by $1545\,{\rm km s^{-1}}$ with respect to FeII, which is comparable to the shift of $1620\,{\rm km\,s^{-1}}$ in our basic fits using a single Lorentzian component discussed in Sect.~\ref{sec_mean_spectrum}.
 
In addition, we verify statistically whether the fit using the KDP15 template provides an overall  improvement. We compare the KDP15 template with the original fit using d12 FeII template, which has $p_1=8$ parameters (2 for FeII, 3 for the MgII single Lorentzian component, 2 for the power-law continuum, and 1 for the redshift). The KDP15 template uses 7 parameters for the FeII pseudo-continuum, 5 parameters for the MgII line with 2 Gaussian components, 2 for the power-law continuum, and 1 for the redshift, overall $p_2=15$ parameters. Given the $\chi^2_1=2088.42$ for the d12 fitting, $\chi^2_2=1711.34$ for the KDP15 template, and the total number of datapoints of $n=579$, we can calculate the F statistic with the null hypothesis that the apparently better fit using the KDP15 template does not lead to an improvement. The F statistic can be calculated as follows

\begin{equation}
    F=\frac{(\chi_1^2-\chi_2^2)/(p_2-p_1)}{\chi_2^2/(n-p_2)}\,,
    \label{eq_ftest}
\end{equation}
which for the values above gives $F=17.75$. When the three parameters in the KDP15 template that converge to zero are removed from the calculation, we get $F'=31.23$. Since these values are larger than the test statistic critical value, which is between 1 and 2 for our F distribution with $(7,564)$ degrees of freedom\footnote{\href{https://www.itl.nist.gov/div898/handbook/eda/section3/eda3673.htm}{https://www.itl.nist.gov/div898/handbook/eda/section3/eda3673.htm}}, the null hypothesis is rejected and formally, the fit using the KDP15 template with 2 Gaussian components for MgII line is better.

The new fit still implies a considerable shift between FeII and MgII components which is not expected according to \citet{kovacevic15} and \citet{2013A&A...555A..89M}. However, with the KDP15 FeII template the fitting results highly depend the adopted redshift. When we performed the analysis assuming that the redshift value given by NED is the right one, the decomposition of the spectrum changed significantly. Now the two Gaussians are rather similar, the dominating Gaussian coincides with the position of FeII emission and the second one is shifted only by 404 km s$^{-1}$ towards shorter wavelengths. This happened since now the FeII contribution peaks at shorter wavelengths than before, and the ratio of the multiplet 63 to multiplet 62 is 0.49 while in the previous fits the contribution from the multiplet 63 was negligible. On the other hand, formally this fit is worse, with $\chi^2$ of 1991.62 vs. 1711.34 for the redshift 1.37617. This stresses the importance of an independent and precise measurement of the redshift in this source.

\begin{figure}
    \centering
    \includegraphics[width=0.49\textwidth]{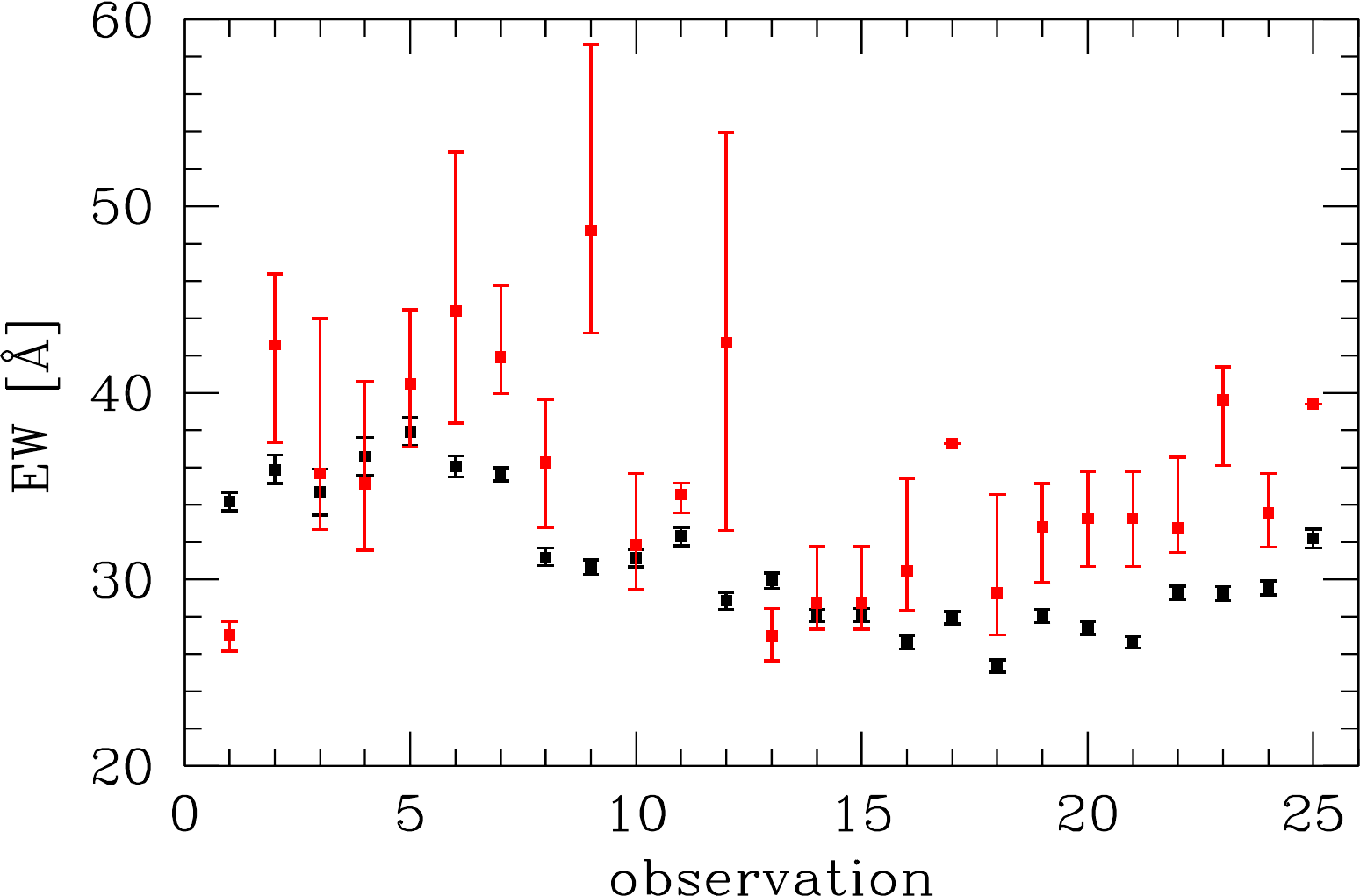}
    \caption{The equivalent width of the MgII line in each of 25 observations measured from the standard \textit{d12} model \citep[black points,][]{bruhweiler2008} and from the new model based on the FeII template \citep{kovacevic15,2019MNRAS.484.3180P} and 2 Gaussian components. The equivalent width based on the KDP15 template has noticeably larger errors.}
    \label{fig:EW_Serbia}
\end{figure}  

Since the $\chi^2$ for the new fit of the mean spectrum is better than our basic fit, we also refitted all individual 25 spectra using this model. Since three of the six parameters were unimportant for the mean spectrum fit, we optimize our model fitting and reduce the parameter space to account only for the contribution from these three non-zero FeII components. The result is shown in Figure~\ref{fig:EW_Serbia}. The overall trend of the higher values followed with the decrease in the second part of the data is still seen but the errors are much larger. This is directly related to larger number of parameters, and the MgII error is determined allowing for all the other parameters to vary (apart from the redshift, the doublet ratio and the FeII width, kept at values optimized for the mean spectrum). In particular, the parameters of the second, very broad component of the MgII line are considerably degenerate with respect to the underlying power law and FeII parameters. 

\begin{figure*}[!h]
    \centering
    \includegraphics[width=\textwidth]{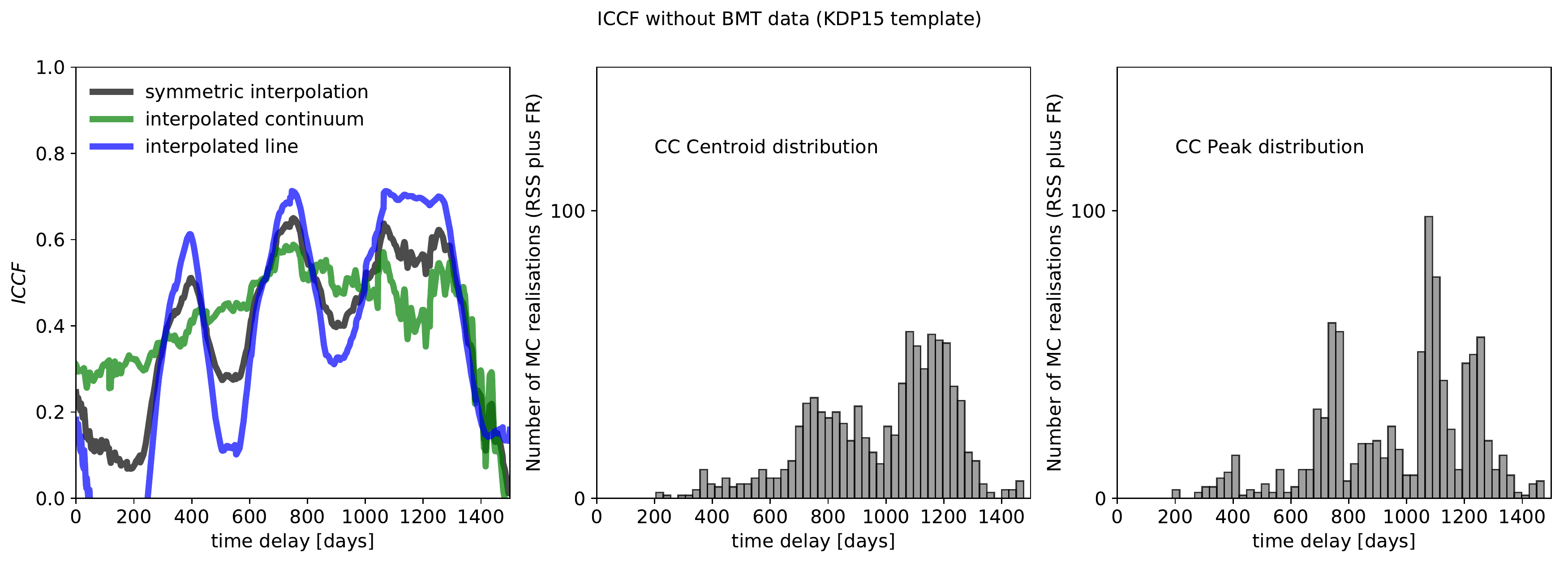}
    \caption{The ICCF values based on the KDP15 template for different interpolation cases according to the legend as a function of the time-delay in light days in the observer's frame. In the central and the right panels, we show the centroid and the peak distributions, respectively.}
    \label{fig_iccf_KDP15}
\end{figure*}

New MgII lightcurve was used again to measure the time delay using different methods. The values are comparable, but as expected the uncertainties are generally larger and the correlation coefficient between the two light curves is lower. For the ICCF, we include the centroid and the peak values for the interpolated continuum, the interpolated line emission, and the symmetric case in Table~\ref{tab_iccf_KDP15}. In Fig.~\ref{fig_iccf_KDP15}, we show the correlation coefficient as a function of the time-delay in the observer's frame with the centroid as well as the peak distributions in the central and the right panels, respectively. The peak value of the correlation coefficient is $0.65$ for the time-delay of 751 days. Whereas for the d12 template, we previously got the peak value of $0.86$ for the time-delay of $1058$ days for the case without BMT points. For the zDCF method, we obtain the peak at $\tau_{0-1000}=382.3^{+401.1}_{-59.1}$ days according to the maximum likelihood analysis in the interval of 0-1000 days. When the interval is narrowed down to $500-1000$ days, the maximum-likelihood peak is at $\tau_{500-1000}=765.3^{+70.2}_{-79.8}$ days, which is comparable within uncertainties to $\tau_{500-1000}=720.9^{+80.6}_{-84.5}$ days using the d12 template without the BMT data in the same interval. The $\chi^2$-based method gives $\tau_{\rm d12}=721^{+57}_{-45}$ days for the original d12 template, while for the MgII light curve inferred from the KDP15 template fitting, we obtained $\tau_{\rm KDP15}=751^{+104}_{-150}$ days. In summary, the basic result of our analysis -- the time-delay of the response of the MgII line -- is comparable to the previous analysis based on d12 template, only the uncertainty is larger for the KDP15 template.

\begin{table*}[!h]
  \centering
  \caption{The centroid and the peak time-delays in ligh days in the observer's frame for the MgII light curve derived based on the KDP15 template. The values are expressed in the observer's frame for the case of the interpolated continuum light curve (with respect to the line-emission), the interpolated line emission, and the symmetric interpolation.}
  \begin{tabular}{c|c|c}
  \hline
  \hline   
     & Centroid [days] & Peak [days] \\
  \hline      
  Interpolated continuum   &   $1062.9^{+228.0}_{-362.3}$  & $1057.5^{+247.9}_{-354.5}$ \\     
  Interpolated line emission     &  $1070.2^{+136.9}_{-352.9}$ & $1060.0^{+147.0}_{-345.6}$  \\
  Symmetric interpolation     &  $ 1058.9^{+149.0}_{-326.5}$ & $1063.5^{+166.5}_{-334.5}$ \\
  \hline
  \end{tabular}  
  \label{tab_iccf_KDP15}
\end{table*}

\end{document}